\documentclass[11pt,letterpaper]{article}
\usepackage{jheppub}
\usepackage{amsmath,amssymb,amsfonts,bm}
\usepackage{graphicx,wrapfig}
\usepackage{multirow}
\usepackage{verbatim}
\usepackage{appendix}
\usepackage{rotating}

\numberwithin{equation}{section}

\newcommand{\bea}{\begin{eqnarray}}
\newcommand{\eea}{\end{eqnarray}}
\newcommand{\be}{\begin{equation}}
\newcommand{\ee}{\end{equation}}
\newcommand{\mb}{\mathbf}

\newcommand{\wt}{\widetilde}

\newcommand{\ol}{\overline}
\newcommand{\ds}{\displaystyle}

\newcommand{\lead}{\mathrm{lead}}

\newcommand{\eg}{\emph{e.g.}}
\newcommand{\ie}{\emph{i.e.}}
\newcommand{\cf}{\emph{cf.}}

\newcommand{\Z}{{\mathbb Z}}
\newcommand{\R}{{\mathbb R}}
\newcommand{\C}{{\mathbb C}}

\newcommand{\Li}{{\rm Li}}

\newcommand{\Tr}{{\rm Tr \,}}
\renewcommand{\Re}{{\rm Re}}
\renewcommand{\Im}{{\rm Im}}
\newcommand{\bs}{\backslash}
\newcommand{\pd}{\partial}

\newcommand{\CA}{\mathcal{A}}
\newcommand{\CB}{\mathcal{B}}
\newcommand{\CC}{\mathcal{C}}

\newcommand{\CG}{\mathcal{G}}
\newcommand{\CH}{\mathcal{H}}
\newcommand{\CI}{\mathcal{I}}

\newcommand{\CL}{\mathcal{L}}

\newcommand{\CN}{\mathcal{N}}
\newcommand{\CO}{\mathcal{O}}
\newcommand{\CP}{\mathcal{P}}

\newcommand{\CR}{\mathcal{R}}

\newcommand{\CT}{\mathcal{T}}

\newcommand{\CW}{\mathcal{W}}

\newcommand{\CZ}{\mathcal{Z}}
\newcommand{\q}{q}

\title{3-Manifolds and 3d Indices}

\author[1,2]{Tudor Dimofte}
\author[1]{Davide Gaiotto}
\author[3,4]{Sergei Gukov}

\affiliation[1]{Institute for Advanced Study, Einstein Dr., Princeton, NJ 08540, USA}
\affiliation[2]{Trinity College, Cambridge CB2 1TQ, UK}
\affiliation[3]{California Institute of Technology, Pasadena, CA 91125, USA}
\affiliation[4]{Max-Planck-Institut f\"ur Mathematik, Vivatsgasse 7, D-53111 Bonn, Germany}

\abstract{We identify a large class $\CR$ of three-dimensional $\CN=2$ superconformal field theories. This class includes the effective theories $T_M$ of M5-branes wrapped on 3-manifolds $M$, discussed in previous work by the authors, and more generally comprises theories that admit a UV description as abelian Chern-Simons-matter theories with (possibly non-perturbative) superpotential. Mathematically, class $\CR$ might be viewed as an extreme quantum generalization of the Bloch group; in particular, the equivalence relation among theories in class $\CR$ is a quantum-field-theoretic ``2--3 move.'' We proceed to study the supersymmetric index of theories in class $\CR$, uncovering its physical and mathematical properties, including relations to algebras of line operators and to 4d indices. For 3-manifold theories $T_M$, the index is a new topological invariant, which turns out to be equivalent to non-holomorphic $SL(2,\C)$ Chern-Simons theory on $M$ with a previously unexplored ``integration cycle.''
}

\begin{document}

\maketitle

\section{Introduction}
\label{sec:intro}

The space of three-dimensional superconformal field theories is vast and only partially explored. Many superconformal field theories
can be defined as the IR fixed point of supersymmetric gauge theories coupled to matter. The inclusion of Chern-Simons couplings
gives a large variety of IR fixed points, as the CS coupling does not flow. Furthermore, in supersymmetric CS theories coupled to matter,
the CS coupling controls the strength of several interaction terms in the Lagrangian.

Theories with different  UV Lagrangian definitions can flow to the same IR SCFT, in which case they are typically called ``mirror''
descriptions of the same theory. The terminology arises from the case of gauge theories with ${\cal N}=4$ supersymmetry and no CS terms,
where supersymmetry protects the geometry of the Higgs branch, and non-trivial IR identifications typically exchange the Higgs and Coulomb branches of the theory.
In this paper we concern ourselves with ${\cal N}=2$ SCFTs. This is a rather interesting amount of supersymmetry, as it allows one to add superpotential couplings
to the theory, and the UV $U(1)_R$ R-symmetry can mix with other flavor symmetries to give interesting anomalous dimensions in the IR \cite{IW-amax, Jafferis-Zmin}.

There are several basic known ``mirror symmetries'' for ${\cal N}=2$ theories, which typically arise as a reformulation or deformation of ${\cal N}=4$ mirror symmetries \cite{IS, dBHOY, AHISS, dBHO}.
One can leverage these basic examples to build large networks of dualities, starting with a complicated Lagrangian $L$, and applying known mirror symmetries to a subsector $L'$ of the theory.
This strategy is not without danger: it relies on the assumption that in order to understand the IR behavior of $L$ it is reasonable to first have the subsector $L'$ flow to the IR, where it can be given alternative descriptions, and then to turn on the couplings between $L'$ and the rest of the theory.
Despite this danger, the strategy is still rather useful.

Given the large set of dualities constructed this way, one may seek methods for characterizing distinct IR fixed points. There are several
protected quantities that can be computed in the UV, and provide information on the IR SCFT. There are three such quantities that
are related in a surprising manner: the moduli space $\CL[T]$ of supersymmetric vacua for massive deformations of the SCFT compactified on a circle \cite{NS-I, DGH, DG-Sdual},
the ellipsoid partition function $\CZ_b[T]$ \cite{Kapustin-3dloc, HLP-wall, HHL} and the refined index $\CI[T]$ \cite{BBMR-index, Kim-index, IY-index, KSV-index, KW-index}.

The moduli space $\CL[T]$ is a useful notion if the 3d SCFT has ``enough'' flavor symmetries: to each flavor symmetry one can associate a real mass
deformation of the SCFT, and a crucial assumption is that we have enough mass deformation parameters to make the SCFT develop a mass gap.
Upon compactification on a circle, the real masses are complexified to $\C^*$ valued twisted masses $X_i$, and one can introduce the notion of an effective twisted superpotential
$\CW(X_i)$ \cite{Witten-phases}. Crucially, $\CW(X_i)$ can be usually computed in the UV.
The twisted superpotential is multivalued, both because the theory will have multiple massive vacua, and because $\CW(X_i)$ is only defined up to shifts by integer multiples of the
twisted masses. The SUSY-preserving ``effective FI parameters''
\begin{equation} \label{FIW}
P_i = \frac{\partial \CW}{\partial X_i}
\end{equation}
(really a complexification of the vev of moment map operators)
are also circle-valued. Then, upon exponentiating \eqref{FIW}, one is left with a well defined and UV-computable parameter space of supersymmetric vacua $\CL[T]$
as a Lagrangian submanifold of $(\C^*)^{2N}$. Here and in the rest of the paper, $N$ denotes the number of abelian flavor symmetry generators
available in the UV.

The ellipsoid partition function $\CZ_b[T]$ is defined by a deformation of the flat space Lagrangian for the theory $T$
that allows for some supersymmetry to be preserved upon compactification on an ellipsoid. The deformation requires a choice of R-symmetry,
which does not have to be the actual R-symmetry in the IR, and can be any linear combination of the UV R-symmetry and the other flavor symmetry generators.
The choice of R-symmetry and the real mass deformation parameters can be combined into $N$ complex parameters $\hat X_i$.
The partition function $\CZ_b(\hat X_i)$ can be computed in the UV by localization methods, and bears a close relationship with the manifold $\CL$.
Indeed, it satisfied two sets of recursion relations that are a ``quantization'' of the equations that carve out $\CL$ in $(\C^*)^{2N}$ \cite{DG-Sdual, DGG}.

This phenomenon can be explained by recasting the 3d SCFT $T$ as a boundary condition for an $U(1)^N$ free abelian four-dimensional gauge theory.
Then the equations that define $\CL$ can be promoted to Ward identities for supersymmetric 'tHooft-Wilson loops brought to the boundary.
If one realizes the 3d ellipsoid as the equator of a four-dimensional half-sphere, the supersymmetric line defects act as operators on $\CZ_b(\hat X_i)$,
and the Ward identities become operator equations for $\CZ_b(\hat X_i)$. The ``quantization'' of the operator algebra arises from a known subtlety in the
OPE of supersymmetric line defects \cite{GMNIII, Ramified}.

One of the result of this paper is to establish a similar correspondence for a third invariant of SCFTs: the refined index. The index is defined by another deformation of the Lagrangian for $T$,
which allows a supersymmetric compactification of $T$ on $S^2$. The deformation again requires a choice of R-symmetry,
which does not have to be the actual R-symmetry in the IR, and can be any linear combination of the UV R-symmetry and the other flavor symmetry generators.%
\footnote{It is important to remark that this deformation is {\it not} the same as a topological twist on $S^2$. Rather, it is akin to a superconformal transformation
from flat space to $S^2 \times \R$.} %
Then one can define a refined Witten index for the theory on the sphere, that will be a power series in a fugacity $q$ that measures the energy minus half the R-charge of the states.
The index is also a function of $N$ parameters valued in  $(S^1 \times \Z)$, associated to the flavor symmetries of the theory.
We will show that the refined index $\CI[T]$ satisfies two sets of recursion relations, which again quantize the equations which carve out $\CL$ in $(\C^*)^{2N}$.
The quantization arises in the same way as for $\CZ_b(\hat X_i)$, and the construction again uses an appropriate four-dimensional setup.

The three invariants $\CL[T]$, $\CZ_b[T]$, $\CI[T]$ provide scant information on the superpotential couplings of the theory $T$.
Mainly, the superpotential affects the calculation by breaking flavor symmetries, and thus reducing the space of available parameters.
Although this may seem trivial, it is surprisingly powerful. For example, the effect on $\CL$ of adding to $T$ a superpotential term that carries some flavor charge vector $c$
is simply to do a symplectic quotient with a moment map $C=c \cdot X$ in the ambient space $(\C^*)^{2N}$, reducing the dimension of $\CL$ by one. This accounts automatically for the effect of the superpotential on the space of vacua of the theory!
 Similar statements hold for $\CZ_b[T]$ and $\CI[T]$.

 Ultimately, we would like to identify an arena of ${\cal N}=2$ SCFTs where we can reliably label distinct SCFTs by the invariants
 $\CL[T]$, $\CZ_b[T]$, $\CI[T]$ or by some refinement of these.
For this purpose, we will impose two restrictions: first, we only consider abelian Chern-Simons matter theories.
In the absence of a superpotential we may give many different mirror abelian CSM descriptions of the same theory $T_0$.
Our second restriction is to define a theory $T$ by a restricted class of superpotential deformations of a theory $T_0$:
we  only consider a superpotential that is the linear combination of operators $\CO_I$
with the property that for each $\CO_I$ we have an abelian CSM mirror frame where $\CO_I$
is a product of elementary fields. Notice that in other mirror frames, $\CO_I$ will be a non-perturbative  monopole operator,
and there may be no mirror frame where all the $\CO_I$ are simultaneously elementary.
We will denote the resulting class of theories as ``class $\CR$.''

The invariants of the theory $T_0$ are related by a simple ``change of polarization'' to the invariants of $n$ free chiral multiplets, which we can denote
as $\CL[\Delta]^n$, $\CZ_b[\Delta]^n$, $\CI[\Delta]^n$.
In other words, any Chern-Simons levels and gauging can be undone with an appropriate $Sp(2n,\Z)$ transformation \cite{Witten-SL2} on $T_0$.
Thus the invariants  $\CL[T]$, $\CZ_b[T]$, $\CI[T]$ of a theory in the class $\CR$ come equipped with a canonical presentation as the symplectic quotient
of $\CL[\Delta]^n$, $\CZ_b[\Delta]^n$, $\CI[\Delta]^n$ by a list of $n-N$  moment maps $C_I$. It is easy to characterize the admissible sets of $C_I$
which correspond to allowed superpotentials. We can label the UV Lagrangians for theories in the class $\CR$ by
the choice of polarization $\Pi$ and the list of admissible moment maps $C_I$.

The class $\CR$ is closed under a basic abelian mirror symmetry operation, which replaces three chiral multiplets with a superpotential of the form $XYZ$
with $N_f=1$ QED, a theory involving two chiral multiplets of opposite charge under a dynamical gauge field. This ``2--3 move'' changes $n$ by one unit.
All known abelian mirror symmetries can be reduced to a combination of several 2--3 moves. The 2--3 move acts on a simple fashion on the labels $(\Pi, C_I)$.
Thus it is natural to conjecture that we can label IR fixed points of theories in the class $\CR$ by the equivalence classes of
sets of admissible $(\Pi, C_I)$ under the action of 2--3 moves.

The set of such equivalence classes is a beautiful, intricate combinatorial object, which is not fully understood. But there is at least a subset
of $\CR$ that which can be given a simple geometric interpretation, in terms of hyperbolic geometry of three-manifolds. The space $\CL[\Delta]$
can be identified with the space of ``ideal tetrahedra,'' \ie\ tetrahedra in hyperbolic space with vertices on the boundary.
There is a class of orientable 3d manifolds that can be glued together from $n$ ideal tetrahedra.
Hyperbolic geometry gives us for free a polarization \cite{NZ} and a set of linear moment maps, which have a crucial feature:
the 2--3 move relates different decompositions of the same manifold, and any two decompositions of the same manifold can be related by a sequence of 2--3 moves.%
\footnote{\label{foot:Bloch}Mathematically, it is well known that hyperbolic 3-manifolds define elements in the Bloch group $\CB(\CC)$ \cite{Bloch-regulators, DupontSah-scissors, neumann-1997}. The class $\beta(M)$ corresponding to a 3-manifold $M$ can be defined by using any ideal triangulation of $M$, and it is invariant under 2--3 moves. Our family $\CR$ of theories can be viewed as an extreme quantum generalization (and in fact refinement) of the classical Bloch group.} %
Thus for every such manifold $M$ we get a specific IR fixed point $T_M$ in the class $\CR$!

The invariants  $\CL[T_M]$, $\CZ_b[T_M]$, $\CI[T_M]$ must coincide with geometric  invariants of $M$.
Indeed, in \cite{DGH,DGG} it was proven that $\CL[T_M]$ is the space of hyperbolic metrics, or equivalently flat
$SL(2,\C)$ connections on $M$, and $\CZ_b[T_M]$ behaves as an analytically continued $SL(2,\R)$ Chern-Simons partition function on $M$ \cite{Yamazaki-3d, DG-Sdual} -- sometimes called an ``$SL(2)$'' Chern-Simons partition function \cite{gukov-2003, DGLZ, Wit-anal, Dimofte-QRS}.
In this paper, we show how $\CI[T_M]$ also admits a geometric interpretation, and behaves as a $SL(2,\C)$ Chern-Simons partition function on $M$.

Conjecturally, the theory $T_M$ has an alternative, mirror description: it can be defined by the twisted compactification of the $A_1$ 6d SCFT on $M$. This higher dimensional definition
allows a connection to the rich subject of four-dimensional theories that can be defined in terms of the $A_1$ 6d SCFT compactified on $C$ \cite{GMN, Gaiotto-dualities}.
This connection was an important inspiration behind the construction. It also featured prominently in a complementary recent construction of the theories $T_M$, based on ``R-flow'' in four-dimensional theories \cite{CCV}.
In this paper we will explore the connection of $\CI[T_M]$ to the refined indices of the four-dimensional gauge theories. We expect our dictionary to be rather useful
for four-dimensional ${\cal N}=2$ gauge theories, allowing one to compute the refined index of 4d theories in the presence of line defects and domain walls by borrowing results from the AGT correspondence \cite{AGT, AGGTV, DGOT, DGG-defects}.

Finally, we would like to remark that we expect higher rank generalizations of our construction, say based on spaces of flat $G_\C$ connections on three manifolds for any $G$,
or even more general Lagrangian submanifolds of cluster varieties, to describe larger subsets of theories in the class $\CR$. For example, for $G=SU(N)$ such theories would arise from a compactification of $K$ M5 branes on a three-manifold $M$. Mathematically, some extensions of hyperbolic invariants of 3-manifolds to higher rank have been investigated in \cite{Zickert-sl3, GTZ-slN}, following \cite{FG-Teich}, and are known to still lie in the classical Bloch group (\cf\ Footnote \ref{foot:Bloch}). The generalization of theories $T_M$ to higher rank should similarly lie in $\CR$. It would also be interesting to find an example of a 3d ${\cal N}=2$ SCFT that is
{\it not} in the class $\CR$ --- say has a parameter space $\CL$ which cannot be given as a symplectic quotient of $\CL[\Delta]^n$ by linear moment maps. We leave this problem to future work. \\

Our main goal in the remainder of the paper is to understand the index $\CI[T]$ for the theories $T_M$ defined in \cite{DGG}. We will also make statements applicable to more general theories in class $\CR$ (or even outside class $\CR$) whenever possible. Thus, we begin in Section \ref{sec:actions} by describing the general form of the index for 3d $\CN=2$ SCFT's with a $U(1)^N$ global symmetry, and defining several familiar operations on it --- for example, the descendant of the $Sp(2N,\Z)$ action on the SCFT's themselves. We introduce the line operators that will play an important role throughout the paper. In Section \ref{sec:T1}, we then specialize to the index $\CI[\Delta]$ for the tetrahedron theory $T_\Delta$, the basic building block of all theories in class $\CR$. Although $T_\Delta$ is extremely simple, consisting only of a single chiral multiplet, its index turns out to have several surprising properties that will serve as model examples for the general properties of theories in $\CR$.

We construct the indices $\CI[T_M]$ of 3-manifold theories $T_M$ in Section \ref{sec:TM}, using the building blocks $\CI[\Delta]$ and the actions of Section \ref{sec:actions}. We give a simple, combinatorial set of rules for building $\CI[T_M]$. As prefaced above, $\CI[T_M]$ is obtained as the ``symplectic reduction'' of a product index $\CI[\Delta]^n$, with the reduction realized via a certain infinite summation over broken flavor charges. We explain how the algebra of line operators acts on $\CI[T_M]$, and explore the relation between the index and the parameter space $\CL[T_M]$ of SUSY vacua. One self-consistent observation is that unconstrained chiral operators in $T_M$ --- seen as ``flat directions'' in the index --- can be easily detected by the asymptotics of the algebraic parameter space $\CL[T_M]$. In Section \ref{sec:3d4d}, we continue analyzing the quantum line operator algebra by viewing $T_M$ as a boundary condition for a 4d $\CN=2$ theory $T[\pd M]$, and considering line operators inserted into a 4d index.

Finally, in Section \ref{sec:CS}, we describe the index $\CI[T_M]$ as an intrinsic topological invariant of $M$ --- arguing that it is equivalent to an $SL(2,\C)$ Chern-Simons partition function on $M$. The fact that the index is a Chern-Simons partition function can be motivated by dualities in six dimensions, highly reminiscent of recent work on M-theory realizations of Khovanov homology \cite{Wfiveknots}. Indeed, since the index is naturally defined as an Euler characteristic of a graded vector space, both in three and six dimensions, it is perfectly ripe for categorification. We hope this topic will be explored in future work.

\section{Actions on the 3d index}
\label{sec:actions}

Let us consider a 3d $\CN=2$ SCFT $\CT$ with flavor symmetry $U(1)^N$. The generalized supersymmetric index was defined in \cite{KW-index}, following \cite{KMMR-index, BBMR-index, Kim-index, IY-index, KSV-index}, as a trace
\be \label{IH}
\CI_\CT( m;q,\zeta) \;=\;
{\rm Tr}_{\CH_{ m}} (-1)^Fe^{-\beta(E-R-j_3)}  \q^{\frac{E+j_3}{2}} \zeta^{ e}
\ee
over a superselection sector of the Hilbert space of the 3d theory on $S^2$.
Specifically, the Hilbert space $\CH_m$ is labelled by the magnetic flux
$m = (m_1,...,m_N)\in \Z^N$ on $S^2$ for $N$ background $U(1)$ gauge fields coupled to the flavor symmetry,
\be
m_i \; := \; \int_{S^2} \frac{F_i}{2 \pi} \,.
\label{mflux}
\ee
By standard arguments \cite{Witten-constraints}, the index \eqref{IH} receives contributions only from states with energy $E = R + j_3$, where $R$ is the R-charge of a state%
\footnote{Although we consider superconformal theories here, $R$ does not have to be the same as the R-charge that enters into the superconformal algebra, \cf\ \cite{IY-index}.} %
and $j_3$ is the spin on $S^2$, with respect to a fixed, chosen axis. With this in mind, we can write the index more simply as
\be \label{Izeta}
\CI_\CT( m;q,\zeta) \;=\;
{\rm Tr}_{\CH_{ m}} (-1)^F \q^{\tfrac{R}{2} + j_3} \zeta^{ e}
\ee
The fugacity $\zeta$ measures the flavor
charge $e\in \Z^N$ of states, and we use a shorthand notation $\zeta^e = \zeta_1^{e_1}\cdots \zeta_N^{e_N}$.

It is also often convenient to work with the index at fixed magnetic flux $m$ and at fixed electric charge $e$.
We therefore collect $e$ and $m$ into a symplectic charge vector
\be
\gamma = {m \choose e}
\ee
and define
\be \label{chargeindex}
\CI_\CT(\gamma;\q)\; \equiv \; \CI_\CT(m,e;\q)
 := {\rm Tr}_{\CH_{m,e}} (-1)^F \q^{\tfrac{R}{2} + j_3}.
\ee
Thus, $\ds\CI_\CT(m;q,\zeta) = \sum_{e\in \Z^N}\CI_\CT(m,e;q)\,\zeta^e$, or conversely $\ds \CI_\CT(m,e;q) = {\oint} \frac{d^N\zeta}{(2\pi i)^N
 \zeta^{e+1}} \CI_\CT(m;q,\zeta)\,,$ where $\zeta^{e+1}=\zeta_1^{e_1+1}\cdots\zeta_N^{e_N+1}$ and integration is done on the unit circle.

We note that even though the theory $\CT$ is superconformal, the R-charge used to calculate the index need not coincide with the superconformal R-charge. The index is typically computed from a UV description of the theory, by a deformed Lagrangian on $S^2$
that preserves a half of the usual superconformal symmetries no matter how $R$ is defined \cite{IY-index, FestucciaSeiberg}.
The powers of $q$ produced by such a calculation could be unconstrained.
However, if the UV theory flows to a IR SCFT whose R-symmetry is not accidental, there will be a R-symmetry redefinition that makes the powers of $q$ nonnegative. This is because the quantity $E+j_3$ in \eqref{IH} (and so $\tfrac R2+j_3$ when restricted to $E=R+j_3$) is nonnegative in superconformal theories, \cf\ \cite{Minwalla-restrictions}.
Conversely, the absence of a such redefinition --- which we have yet to encounter in class $\CR$ --- would signal a breakdown of the naive expectations for the RG flow of the UV theory.%
\footnote{This is a possibly more refined version of
the requirement of positive monopole operator dimensions used to discuss the IR behavior of $\CN=4$ gauge theories \cite{GW-Sduality}.}

 Finally, we should comment on the precise meaning of $(-1)^F$ in the trace. The most natural choice would be $(-1)^{2 j_3}$, but it is not the choice that is (implicitly) made in the
literature on the refined index: in the presence of odd magnetic flux, the angular momentum on $S^2$ of a particle of odd electric charge is shifted by a half-integral amount, but
no extra $(-1)$ sign is usually inserted compared to the index in even flux sectors. This is equivalent to a definition
\begin{equation}
(-1)^F = (-1)^{2 j_3 + e\cdot m}\,.
\end{equation}
The difference between the two conventions is fairly minimal: it is just an overall sign change $(-1)^{e\cdot m}$ of $\CI_\CT(m,e;\q)$, or a redefinition of the fugacity $\zeta \to (-1)^m \zeta$ in $\CI_\CT( m;q,\zeta)$.
In this paper we will stick to the modified form of $F$, mostly for reasons of notational convenience and backwards compatibility. We will return to some of these subtleties in later sections.

\subsection{The $Sp(2N,\Z)$ action}
\label{sec:Sp}

In \cite{Witten-SL2}, an $Sp(2N,\Z)$ action on 3d SCFT's with $N$ $U(1)$ flavor symmetries was introduced.%
\footnote{We refer the reader to \cite{Witten-SL2}, as well as \cite{DGG}, for details of $Sp(2N,\Z)$ transformations on SCFT's. Our notation here closely follows that in \cite{DGG}.} %
The index \eqref{chargeindex} in the charge basis transforms very transparently under this action. Namely, for $g\in Sp(2N,\Z)$, we have%
\footnote{Note that the $Sp(2N,\Z)$ transformation acts non-trivially on the $(-1)^{e \cdot m}$ factor we included in the definition of the fermion number. We will return to this in Section \ref{sec:aff}.} %
\be \label{Spem}
\boxed{\phantom{\int}\CI_{g\circ \CT}(g\,\gamma;q) = \CI_\CT(\gamma;q)\,.\phantom{\int}}
\ee
For simplicity, we can illustrate the action \eqref{Spem} for $N=1$ (\ie\ a single $U(1)$ flavor symmetry) by looking at the action of the two generators $T$ and $S$ of $Sp(2,\Z)\simeq SL(2,\Z)$.

A $T$ transformation simply adds one unit of Chern-Simons coupling for the background $U(1)$ gauge field.
The effect of the CS coupling for the flavor symmetry background gauge field
is simply to shift the flavor charge of a state by an amount proportional to the magnetic flux.
Therefore
\be \label{Tsim}
\CI_{T\circ \CT}(m,e+m;q) = \CI_\CT(e,m;q)\,,
\ee
which is compatible with the matrix representing $T$,
\be
T ~: \qquad
{m \choose e} \; \mapsto \;
{1 \quad 0 \choose 1 \quad 1} {m \choose e}\,.
\ee

An $S$ transformation makes the background gauge field dynamical, and adds a supersymmetric FI term for it. This effectively couples the resulting theory to a background $U(1)$ for a \emph{new} flavor symmetry,
whose conserved current is the magnetic flux of the old, now gauged, flavor symmetry. Thus, the new electric charge is the old magnetic flux. Gauging the old flavor symmetry also means that we should project onto states of zero gauge charge. However, turning on a magnetic flux $m'$ for the new flavor symmetry shifts the gauge charge of a state by $m'$, so in effect we project onto states of gauge charge $-m'$. Altogether we find
\be \label{Ssim}
\CI_{S \circ \CT}(-e,m;\q) = \CI_\CT(m,e;\q)\,,
\ee
which is compatible with the matrix representing $S$,
\be
S ~: \qquad
{m \choose e} \; \mapsto \;
{0 \; -1 \choose 1\; \quad 0}  {m \choose e}\,.
\ee
From these matrix representations, it is easy to check that $\CI_{(ST)^3\circ \CT}(\gamma;q) = \CI_\CT(\gamma;q)$, and $\CI_{S^2\circ\CT}(\gamma;q)=\CI_{\CT}(-\gamma;q)$, as expected from the group relations $(ST)^3=\rm{id.}$ and $S^2 = C,$ where $C$ acts as charge conjugation.

The $T$ and $S$ transformations \eqref{Tsim}--\eqref{Ssim} could also be written in a ``fugacity'' basis for the index \eqref{Izeta}. We find
\begin{align} \CI_{T\circ\CT}(m;q,\zeta) &= \zeta^{m}\,\CI_{\CT}(m;q,\zeta)\,, \\
 \CI_{S\circ \CT}(m';q,\zeta') &= \sum_{m\in \Z}\oint \frac{d\zeta}{2\pi i \zeta}\, \zeta'{}^{m}\zeta^{m'}\,\CI_{\CT}(m;q,\zeta)\,,\end{align}
with the integration done, as usual, on the unit circle.
In this form, it is easier to see that the transformations agree with the general rules of \cite{KW-index, IY-index} for gauging flavor symmetries and adding Chern-Simons terms to the index.

\subsection{Affine shifts}
\label{sec:aff}

As mentioned above, the R-charge used in the index does not need to coincide with the R-charge that appears in the superconformal algebra. We can take $R$ to be any R-symmetry charge. If we redefine the R-charge by a flavor symmetry, then different choices are related by $R \to R -\alpha_e \cdot e$ for some vector $\alpha_e$. Under such a shift, the index varies accordingly
as
\be
\CI_\CT(m,e;\q) \to \q^{-\tfrac{1}{2} \alpha_e \cdot e} \CI_\CT(m,e;\q)\,.
\ee
There is also a certain degree of latitude in choosing the R-charge of
the vacuum in a non-trivial flux sector of the theory. It is thus natural to
also consider redefinitions  $R \to R +\alpha_m \cdot m$, under which
\be
\CI_\CT(m,e;\q) \to \q^{\tfrac{1}{2}  \alpha_m \cdot m} \CI_\CT(m,e;\q)\,.
\ee
Altogether, we can collect
\be
\alpha = {\alpha_e \choose \alpha_m}\,,
\ee
and write
\be
\boxed{\phantom{int}
\CI_{\sigma^R(\alpha) \circ \CT}(\gamma;\q) = \q^{\frac{1}{2} \langle \alpha, \gamma \rangle} \CI_\CT(\gamma;\q) \phantom{\int}}
\ee
where $\sigma^R(\alpha) \circ \CT$ denotes the theory with the new R-charge $R+\langle \alpha, \gamma \rangle$.
Here the symplectic product $\langle \alpha, \gamma \rangle$ equals $\alpha_m \cdot m - \alpha_e \cdot e$.

Similar shifts can happen in the definition of the fermion number charge $F$.  This would not be the case if we had defined $(-1)^F = (-1)^{2 j_3}$,
as $j_3$ is a generator in a non-Abelian symmetry group. But the symplectic group acts on $(-1)^{e\cdot m}$ in a rather interesting way. Indeed, $s(\gamma) = (-1)^{e \cdot m}$ is a ``quadratic refinement of the charge lattice'',
\ie\ a $Z_2$-valued function of charges with the property
\begin{equation}
s(\gamma) s(\gamma') = s(\gamma + \gamma') (-1)^{\langle \gamma, \gamma' \rangle}\,.
\end{equation}
Any two such refinements differ by a factor of the form  $(-1)^{\langle \alpha, \gamma \rangle}$ for some $\alpha$.
Thus, we will just introduce a fermion-number shift operation
\be
\boxed{\phantom{\int} \CI_{\sigma^F(\alpha) \circ T}(\gamma;\q) = (-1)^{\langle \alpha, \gamma \rangle} \CI_\CT(\gamma;\q) \phantom{\int}}
\ee
that modifies the choice of quadratic refinement used in $F$. Note that $[\sigma^F(\alpha)]^2=1$ for any $\alpha$.
Such shifts in the definition of $R$ and $F$ are often important when comparing mirror descriptions of the same theory.
In particular, a symplectic transformation will change the choice of quadratic refinement from $s(\gamma)$ to $s(g \gamma)$,
which will have to be brought back to $s(\gamma)$ by a fermion number shift.

In the following, we will often find that $R$ and $F$ shifts happen simultaneously, with the same parameter $\alpha$.
This is ultimately due to the fact that the UV description of the theories often have a natural R-charge assignment
such that $(-1)^F = (-1)^R$, \ie\ $(-1)^{R + 2 j_3 + e \cdot m} =1$ when acting on all the chiral fields in the Lagrangian.%
\footnote{It is interesting to remark that in the context of four-dimensional ${\cal N}=2$ gauge theory, wall-crossing considerations lead
to the conjecture that $(-1)^{R + 2 j_3}$ should always be equal to a quadratic refinement of the charge lattice when acting on BPS states \cite{GMNIII}.}

Mathematically, the shifts in $R$ and $F$ generate an abelian group of translations $\Z^{2N}\times (\Z_2)^{2N}$ that acts on the index. This abelian group can naturally be combined with the symplectic group $Sp(2N,\Z)$ to generate an affine symplectic group $Sp(2N,\Z) \ltimes (\Z\times \Z_2)^{2N}$. One can check that the expected group relations are satisfied. For example, taking $N=1$, if we denote the generators of unit electric and magnetic shifts by $\sigma_e^R,\,\sigma_e^F$ and $\sigma_m^R,\,\sigma_m^F$, respectively, then
\begin{subequations} \label{affrels}
\be S\sigma_e^R = \sigma_m^R S\,, \quad\qquad
S\sigma_m^R = (\sigma_e^R){}^{-1}S\,,\ee
\be T^{-1}(\sigma_e^R)^{-1}T\sigma_e^R = \sigma_m^R\,,\quad\qquad T\sigma_m^R = \sigma_m^R T\,, \ee
\end{subequations}
and similarly for $\sigma^F_{e,m}$. The abelian $(\Z\times \Z_2)^{2N}$ subgroup of the affine symplectic group does act in the obvious way by translations on the vector $(m,e)$. However, we will see momentarily that there exists another object, a certain algebra of operators associated with the index, on which this abelian subgroup \emph{is} represented by translations.

\subsection{Adding a superpotential}
\label{sec:W}

In the framework of \cite{DGG}, another important operation on a 3d $\CN=2$ SCFT $\CT$ was the addition of a superpotential. The index is affected by superpotential terms in a simple fashion; one just has to drop, in an appropriate way, any flavor symmetry broken by the superpotential.

This has three basic consequences. To illustrate them, suppose that we add a superpotential $\CW$ charged only under the first $U(1)$ of the flavor symmetry group $U(1)^N$; in other words, $\CW$ has electric charge $c_\CW = (1,0,...,0)$. Suppose that $\CW$ also has R-charge $R_\CW$. Then:
\begin{enumerate}

\item In order for $\CW$ not to break R-symmetry, its R-charge must equal 2. This requires an electric R-charge shift by $\alpha_e = (R_\CW-2,0,...,0)$, so that $R_\CW - \alpha_e\cdot c_\CW=2$, thereby multiplying the index by $q^{(1-R_\CW/2)e_1}$, where $e = (e_1,...,e_N)$.

\item The theory cannot be coupled to flux for a broken flavor symmetry. Therefore, we must restrict $m=(0,m')$, where $m'=(m_2,...,m_N)$.

\item The Hilbert space cannot be graded by the broken flavor charge, and states whose difference in charge is a multiple of $c_\CW$ will now live in the same charge sector. This means that we must sum over $e_1$.

\end{enumerate}
Altogether, we find that the addition of $\CW$ amounts to
\be \CI_\CT(\gamma;q)\;\overset{+\CW}{\longrightarrow}\; \CI_{\CT'}(\gamma';q) = \sum_{e_1\in \Z}q^{\big(1-\frac{R_\CW}{2}\big)e_1} \CI_\CT(\gamma;q)\,\big|_{\,m_1=0}\,, \ee
with a new reduced charge vector $\gamma' = (m',e') = (m_2,...,m_N,e_2,...,e_N)$.

In general, a superpotential might contain a sum of several operators $\CO_i$, with general electric charges $c_i$ and R-charges $R_i$. We can add these operators one at a time, in any order. Each of them transforms the index as
\be \CI_\CT(\gamma;q)\;\overset{+\CO_i}{\longrightarrow}\; \CI_{\CT'}(\gamma';q) = \sum_{n\in \Z} q^{\big(1-\frac{R_i}{2}\big)n}\CI_\CT(e+n\hspace{.04cm}c_i,m;q)\,\big|_{\,m\cdot c_i = 0}\,. \ee
This can be thought of as a discrete version of a ``symplectic quotient,'' with respect to a moment map $\langle c_i, m \rangle$. After adding all the operators $\CO_i$, we will be left with a charge vector $\gamma'=(m',e')$ that obeys $m'\cdot c_i = 0$ and $e'\equiv e'+n_ic_i$\, ($n_i\in \Z$), for all $i$.

\subsection{A discrete symmetry $\rho$}
\label{sec:kappa}

We could define standard discrete symmetries $C$, $P$, and $T$ acting on theories $\CT$ and their indices, though in general the 3d SCFT's in class $\CR$ will not be invariant under any of these ``symmetries'' alone. To see which discrete symmetries stand a chance at preserving a theory $\CT$ and its index, it is useful to consider more closely the Hilbert space $\CH_m$ of $\CT$ on $S^2$ in the presence of magnetic flux $m$. Suppose further that $\CT$ is in class $\CR$, and has a $UV$ Lagrangian description as an abelian Chern-Simons-matter theory with superpotential.

The Hilbert space $\CH_m$ is graded by charges $(E,R,j_3,e)$. Since states come in complete multiplets of the $SU(2)$ Lorentz group, changing the sign of $j_3$ should preserve $\CH_m$. We could also try to flip the sign of the $U(1)$ R-charge $R$. In a Lagrangian, this is accomplished by Hermitian conjugation, switching chiral multiplets and antichirals, which  has the additional effect of flipping the flavor charge $e\to -e$. However, this overall charge conjugation still does not preserve the theory in a magnetic flux background. In order for antiparticles to behave the same way as the original particles, it is also necessary to replace $m\to -m$.

It appears that the simultaneous reversal of charges
\be \rho \;:\; (E,R,j_3,e,m)\;\to\;(E,-R,-j_3,-e,-m) \ee
is a true symmetry of the graded Hilbert spaces $\CH_m$. In fact, there exists a simple geometric operation that also realizes $\rho$ in the context of the index. If we put the Euclidean version of $\CT$ on $S^2\times S^1$, with appropriate magnetic flux and global Wilson lines so as to calculate the index, then $\rho$ corresponds to reflecting $S^2$ through its equator (along the $j_3$ axis) and reversing time. The spatial reflection flips $m$, while the time reversal flips $j_3$ and effectively $R$ and $e$. The effective negation of $R$ and $e$ happens because the fugacities in R-charge and flavor Wilson lines change sign. Altogether, this reversal of time and space is just the Euclidean analogue of $CPT$ symmetry.

The outcome is that we would expect the index to be invariant under $\rho$. Naively, from \eqref{Izeta}, this means
\begin{subequations} \label{kind}
\be \label{kzeta}
\CI_\CT(m;q,\zeta) \;\overset{\rho}{=}\; \CI_\CT(-m;q^{-1};\zeta^{-1})\,, \ee
or in a charge basis,
\be \label{kme}
\CI_\CT(m,e;q) \;\overset{\rho}{=}\; \CI_\CT(-m,-e;q^{-1})\,.\ee
\end{subequations}
However, \eqref{kind} are only true in a formal sense. In order to account for the infinite cancellations between bosons and fermions in $\CH_m$, the index should really be defined with a regulator as in \eqref{IH}. Upon applying $\rho$, the ``Hamiltonian'' appearing in the regulator changes, $H = E-R-j_3\to E+R+j_3$, so that the boson-fermion cancellations in $\CH_{-m}$ should be counted differently than would be implied by the right-hand sides of \eqref{kind}. Therefore, as written, the equalities \eqref{kind} are not actually correct. This fact is most striking if (say) we choose a superconformal R-charge assignment; then the indices on the left-hand side are series in positive powers of $q$ while the indices on the right-hand side are series in $q^{-1}$. Typically there is no analytic continuation from one to the other.%
\footnote{For example, we will see later that the index of the tetrahedron theory, as a function of $q$, has a natural boundary at $|q|=1$. We expect this to hold true for more general theories in class $\CR$.}

Making sense of $\rho$ symmetry for indices requires a reorganization of the cancellations in the spaces $\CH_m$, effectively turning the series in $q^{-1}$ on the right-hand sides of \eqref{kind} into series in $q$. One way to implement this reorganization is to separate $\CH_m$ into Fock spaces that can be ``inverted.'' For example, the reorganization in the Fock space of a free boson would correspond to rewriting the generating function of single-particle states as
\be \frac{1}{1-q^{-1}} \to \frac{-q}{1-q}\,.\ee
This reorganization will be realized in Section \ref{sec:tetkappa} for the simple tetrahedron theory $\CT_\Delta$. For more complicated theories in class $\CR$, the mathematics of plethystic logarithms and exponentials (\cf\ \cite{Hanany-pleth}) may help to convert indices $\CI_\CT(-m,-e;q^{-1})$ to series in $q$.

Although $\rho$ symmetry only holds formally for the index, it is still powerful enough to relate algebras of operators on the index (which do not feel the re-organization of Hilbert spaces). We will see this beginning in Section \ref{sec:ops} below, and find a geometric meaning for $\rho$ acting in operator algebras in Section \ref{sec:3d4d}. In Section \ref{sec:CS}, we will show that when $\CT$ is a 3-manifold theory $T_M$, $\rho$ coincides with complex conjugation in $SL(2,\C)$ Chern-Simons theory on $M$.

\subsection{An operator algebra}
\label{sec:ops}

In \cite{DGH} (see also \cite{DG-Sdual,DGG}), we encountered a certain universal structure
common to 3d $\CN=2$ theories with $U(1)^N$ flavor symmetry.
Upon compactification on $\R^2 \times S^1$,  the moduli space of SUSY vacua
maps to a Lagrangian submanifold $\CL$ of $(\C^*)^{2N}$, parameterized by
the choices of complexified twisted masses $X$ and effective complexified FI parameters $P$.
Both parameters have periodic imaginary parts (flavor Wilson lines and effective theta angles respectively),
so the true $(\C^*)^{2N}$ parameters are
\begin{equation}
x = e^X\,, \qquad p=e^P\,.
\end{equation}
The effective FI parameters are derived from the low energy effective twisted superpotential $\wt\CW$ as
\begin{equation}
p = \exp\big(\partial_X \wt\CW\big)\,.
\end{equation}

The symplectic group $Sp(2N,\Z)$ simply acts by left multiplication on the symplectic vector $(X,P)^T$. Moreover, the shifts in fermion number act as translations.
The compactification on $\R^2 \times S^1$ is done with supersymmetry-preserving boundary conditions.
Hence, an electric shift in the definition of $F$ by $\alpha_e$ is equivalent to a shift of the flavor Wilson lines by $i\pi \alpha_e$.
More generally, a shift $\alpha$ in the definition of $F$ turns into a shift of $(X,P)$ by $i \pi \alpha$.
Shifts in the definition of $R$ are not visible.
The operation of adding an operator of charge $c$ to the superpotential is equivalent to a symplectic quotient
of both the ambient space and of $\CL$ with moment map $\langle c, P \rangle$.

The classical structure of the $\R^2 \times S^1$ moduli space has a quantum counterpart in
the $S^3_b$ partition function. The $S^3_b$ partition function is a function of the complexified twisted masses
$m_X$, whose imaginary parts are now proportional to the contribution of the flavor symmetries to the choice of R-charge.
We can define operators $\hat X$ that act on the $S^3_b$ partition function as multiplication by $2 \pi b m_X$, and $\hat P$ acting as $i b \partial_{m_X}$. These operators satisfy commutation relations
\be [\hat P_j\,\hat X_{j'}]=\hbar\delta_{jj'}\,,\qquad \hbar \equiv 2\pi i b^2\,.\ee
Then the $S^3_b$ partition function behaves as a wavefunction, killed by a quantum version $\hat \CL(\hat X, \hat P, \hbar)$ of the equations that define $\CL$.

The group $Sp(2N,\Z)$ again simply acts by left multiplication on $(\hat X, \hat P)^T$.
Moreover, shifts in the definition of $R$ used in the partition function give rise to shifts of $(\hat X, \hat P)$ by $(i \pi + \hbar/2) \alpha$.
In the limit $\hbar \to 0$, the ellipsoid is degenerates into
$\R^2 \times S^1$, with a choice of fermion number that appears to depend on the choice of $R$ in $S^3_b$.
Finally, operation of adding an operator of charge $c$ to the superpotential is equivalent to a ``quantum symplectic quotient''
with moment map $\langle c, \hat P \rangle$.

We would like to carry this algebraic machinery over to index calculations.
We can actually define two commuting sets of interesting operators.
The first set is
\begin{subequations} \label{XP}
\begin{align}
\left.\begin{array}{l}
\hat X_+ = \tfrac{\hbar}{2} m - \partial_e \\[.1cm]
\hat P_+ = \tfrac{\hbar}{2} e + \partial_m
\end{array}\right.\;\;,
\qquad [\hat P_+,\hat X_+]=\hbar\,,
\end{align}
and the second set is
\begin{align}
\left.\begin{array}{l}
\hat X_- = \tfrac{\hbar}{2} m + \partial_e \\[.1cm]
\hat P_- = \tfrac{\hbar}{2} e - \partial_m
\end{array}\right.\;\;,
\qquad [\hat P_-,\hat X_-]=-\hbar\,,
\end{align}
\end{subequations}
with each $\hat X_\pm$ and $\hat P_\pm$ denoting a vector of $N$ operators, and
\be
\boxed{\hbar = \log q\,.}
\label{hlogq}
\ee
Although we write the operators in logarithmic form, as partial derivatives, the actual algebra that acts on the index is generated by the well-defined multiplication and shift operators $(\hat x_\pm,\hat p_\pm) = (e^{\hat X_\pm}, e^{\hat P_\pm})$. The exponentiated operators satisfy $q$-commutation relations
\be \hat p_j{}_+\hat x_{j'}{}_+ = q^{\delta_{jj'}}\hat x_{j'}{}_+\hat p_j{}_+\,,\qquad \hat p_j{}_-\hat x_{j'}{}_- = q^{-\delta_{jj'}}\hat x_{j'}{}_-\hat p_j{}_-\,, \ee
with all other pairs of operators commuting.

We could also consider the action of these operators on the standard index $\CI_\CT(m;q,\zeta)$ in a fugacity basis. If we set $\zeta \equiv e^{i\theta}$, we find that $\pd_e \to i\theta$ and $e\to -i\pd_\theta$, so that
\be \hat X_\pm  = \frac\hbar2 m \pm i\theta\,,\qquad \hat P_\pm = \pm\pd_m-\frac{i\hbar}2\pd_\theta\,. \label{XPz} \ee
Therefore, in a fugacity basis, the index is simply an eigenfunction of the $\hat X_\pm$ operators. In \emph{either} basis, the $\rho$ transformation of Section \ref{sec:kappa} conjugates one set of operators to the other:
\be \rho^{-1}(\hat X_\pm,\hat P_\pm)\,\rho = (\hat X_\mp,\hat P_\mp)\,. \ee

Just as in the cases of $S^3_b$ partition functions and $\R^2\times S^1$ moduli spaces, reviewed above, there is a natural action of the affine symplectic group on the operator algebra generated by \eqref{XP}. This action intertwines the affine symplectic action on the index. In particular, $g\in Sp(2N,\Z)$ acts as matrix multiplication on $(\hat X_\pm,\hat P_\pm)$, so that
\be \label{intertwine}
 \boxed{\begin{pmatrix} \hat X_\pm' \\ \hat P_\pm '\end{pmatrix}
\cdot \CI_{g\circ \CT}(\gamma') = \left[ g\begin{pmatrix} \hat X_\pm \\ \hat P_\pm \end{pmatrix}\cdot I_\CT(\gamma)\right]_{\gamma\,\to\, g^{-1}\gamma'}\,.}
\ee
(It may help to recall from \eqref{Spem} that $\CI_{g\circ \CT}(\gamma') = \CI(g^{-1}\gamma')$.) For example, if $N=1$ and $g=T$ is the $T$ element of $SL(2,\Z)$, this intertwining property would imply that
\be \hat P' \cdot \CI_\CT(m',e'-m') = \left[(\hat X+\hat P)\cdot \CI_\CT(m,e)\right]_{(m,e)\to (m',e'-m')}\,.\ee
In a similar way, the translations
\begin{align} \sigma^R(\alpha)\,:\; (\hat X_\pm,\hat P_\pm) &\;\mapsto\; (\hat X_\pm',\hat P_\pm') = (\hat X_\pm,\hat P_\pm) \pm \frac\hbar2 \alpha\,, \\
\sigma^F(\alpha)\,:\; (\hat X_\pm,\hat P_\pm) &\;\mapsto\; (\hat X_\pm',\hat P_\pm') = (\hat X_\pm,\hat P_\pm) \pm i\pi \alpha\,
\end{align}
intertwine the action of R-charge and fermion-number shifts in the index.
If we always combine an R-symmetry shift with an equal F-symmetry shift,
we recover the familiar shifts by multiples of $i \pi + \tfrac{\hbar}{2}$ that were encountered in the $S^3_b$ partition functions of $T_M$ theories \cite{DGG}.
Finally, the operation of adding an operator of charge $c$ to the superpotential is again equivalent to a ``quantum symplectic quotient''
with moment map $\langle c, \hat P_\pm \rangle$.

As we take the radius of the circle used in the definition of index to zero, {\it i.e.} $q \to 1$, or equivalently
the radius of the $S^2$ to infinity,
we expect to be able to connect back to the problem on $\R^2 \times S^1$.
One may hope that in the $q \to 1$ limit both sets of operators defined above
may go to the classical $(X,P)$ coordinates. In concrete examples,
we find a striking result: the index of the 3d SCFTs
is annihilated by the same set of equations
as the $S^3_b$ partition function is, written in terms of either set of operators:
$\hat \CL(\hat X_\pm, \hat P_\pm, \hbar_\pm) \cdot \CI_\CT = 0$. The fact that both `$\pm$' sets of equations annihilate the index is consistent with the claim of Section \ref{sec:kappa} that the index of $T_M$ theories enjoys a formal $\rho$ symmetry.  We expect this statement to have a universal validity, beyond
the examples considered in this paper. In Section \ref{sec:3d4d} we will sketch a proof of this statement.

\section{The tetrahedron index}
\label{sec:T1}

The basic building block used to construct the theory $T_M$ for any 3-manifold $M$ is the theory $T_\Delta$ associated to a single ideal tetrahedron. More precisely, we should call the theory
\be T_{\Delta,\Pi_Z}\,, \label{tetthy} \ee
since it depends on a polarization for the boundary of the tetrahedron. As discussed in Sections 2 and 4 of \cite{DGG}, we make a specific choice of polarization $\Pi_Z$ (see also Section \ref{sec:tetdiff} below). Then $T_\Delta$ consists of a single free $\CN=2$ chiral multiplet $\phi$, coupled to a background $U(1)$ gauge multiplet with a level $-\tfrac12$ Chern-Simons interaction.

In order to calculate the index of $T_\Delta$, we must specify the R-charge and fermion number assignments for the theory on $S^2\times S^1$. We take $R_\phi=F_\phi=0$, and also require that the vacuum in a flux sector of any negative charge $m$ has $R_{\rm vac}=F_{\rm vac}=0$. Then, by following the rules for constructing indices in \cite{KW-index, IY-index}, we find
\be \CI_\Delta(m;q,\zeta) \,\equiv\, \CI_{T_\Delta,\Pi_Z}(m;q,\zeta) \,=\,  (-\q^{\frac12})^{\frac12(m+|m|)}\zeta^{-\frac12(m+|m|)} \prod_{r=0}^\infty \frac{1- \q^{r+\frac12|m|+1}\zeta^{-1}}{1-\q^{r+\frac12|m|}\zeta}\,. \label{Iztet} \ee
It turns out that this expression simplifies nicely to
\be \CI_\Delta(m;q,\zeta) =\prod_{r=0}^\infty \frac{1- \q^{r-\frac12m+1}\zeta^{-1}}{1-\q^{r-\frac12m}\zeta}\,. \label{Iztet2} \ee

The various ingredients in \eqref{Iztet} can be given an intuitive explanation. The denominator in the product arises from bosonic creation operators of flavor charge $1$ and spin $r+\tfrac{|m|}{2}$ acting on the vacuum. The spin starts from $\tfrac{|m|}{2}$ due to the nontrivial magnetic flux on $S^2$. The numerator arises similarly from fermionic creation operators of flavor charge $-1$ and spin $r+\tfrac{|m|}{2}+1$.

In addition to the product, the prefactor $(-\q^{\frac12})^{\frac12(|m|+m)}\zeta^{-\frac12(m+|m|)}$ in \eqref{Iztet} turns out to be both subtle and important. It depends on the energy, R-charge, and flavor charge of the ground state in a non-trivial flux sector, which have been calculated carefully in \eg\ \cite{BKW-monopoles}. First, the flavor charge of the flux vacuum is affected by the quantization of (anti)fermions in the chiral multiplet, and receives from that a contribution $-\tfrac12|m|$. This is corrected by the Chern-Simons coupling $k=-\tfrac12$ to $-\tfrac12(m+|m|)$. In order to determine the R-charge $R_{\rm vac}$ and fermion number $F_{\rm vac}$, we can shift conventions so that $R_\phi=F_\phi=1$ for the free chiral boson. Then $R=F=0$ for the fermion zero-mode whose quantization determines the quantum numbers of the vacuum, implying $R_{\rm vac}=F_{\rm vac}=0$. In terms of the index \eqref{Iztet}, we can easily perform the shift to $R_\phi=F_\phi=1$ by setting $\zeta\to (-\q^{\frac12})\zeta$. Then it is easy to see that the vacuum acquires $R_{\rm vac}=F_{\rm vac}=0$ as expected.

It is important to remark a subtle sign difference between our prescription for the index of a chiral multiplet and the prescription that can be found in the literature \cite{IY-index,KW-index}.
After aligning the choices of R-charge and CS terms, the difference boils down to the prefactor $(-1)^{\frac12(|m|+m)}$.
This prefactor affects in a minimal way the checks of mirror symmetry that have already been done in the literature, as it only affects the overall sign of the index in a given charge sector. However, only with our definition of the chiral index can such signs be matched universally via an appropriate shift of $F$. This deeply affects the calculation of the index in theories whose superpotentials have monopole operators --- which is standard in class $\CR$.

Let us also write the index in an electric-magnetic charge basis, as in Section \ref{sec:actions}. By using several standard identities for $q$-series, we find
\be \CI_\Delta(m;q,\zeta) \,=\, \sum_{e\in \Z} \CI_\Delta(m,e;q)\zeta^e\,,\ee
with
\be \CI_\Delta(m,e;q) \,=\, \sum_{n=\lfloor e\rfloor}^\infty
 \frac{(-1)^n q^{\frac12 n(n+1)-\big( n+\frac 12e\big) m}}{(q)_n\,(q)_{n+e}}\,, \label{Imetet}\ee
where $\lfloor e \rfloor \equiv \tfrac12(|e|-e)$ and $(q)_n \equiv (1-q)(1-q^2)\cdots (1-q^n)$. The sum in \eqref{Imetet} can be thought of as defining a formal power series in $q$; for fixed charge $\gamma=(m,e)$, only finitely many terms in the sum are necessary for calculating $\CI_\Delta(m,e;q)$ to a desired order in $q$. The series \eqref{Imetet} also appears to converge to a well-defined analytic function of $q$ for $|q|<1$.

In anticipation of the interpretation of the tetrahedron index as a geometric invariant of an ideal tetrahedron $\Delta$ itself (and connections to the classical Bloch group), we can take the ``classical limit'' $q =e^\hbar \to 1$ and $m\to \infty$ with $q^m$ fixed. Let us set $q^{\frac m2}\zeta=z$, remembering that $\zeta$ is a pure phase, so that $z$ becomes a complex number. Then
\be \label{tetasymp}
\CI_\Delta(m;q,\zeta) \;=\; \prod_{r=0}^\infty \frac{1-q^{r+1}z^{-1}}{1-q^r \,\ol z^{-1}}\;\;\overset{\hbar\to 0}{\sim}\;\; \exp\left( \frac{2V_\Delta(z)}{i\hbar}+\ldots\right)\,,
\ee
where the function
\be V_\Delta(z) \equiv -\Im\,\Li_2(z^{-1}) \ee
is closely related to the hyperbolic volume of a tetrahedron with shape parameter $z$. In fact, the actual volume can be written as ${\rm Vol}_\Delta(z) = V_\Delta(z) + \arg(1-z^{-1})\log|z|$\,, \cf\ \cite{Thurston-1980}.

\subsection{Parity and $\rho$ symmetry}
\label{sec:tetkappa}

Several identities satisfied by the tetrahedron index can be understood via discrete symmetries acting on $T_\Delta$.
Let's first consider the action of parity $\mb{P}$ on $T_\Delta$. In is not an exact symmetry because it inverts the sign of the Chern-Simons level, from $k=-\tfrac12$ to $k=\tfrac12$. However, this can be compensated for by subtracting a Chern-Simons term, \ie\ applying $T^{-1}\in Sp(2,\Z)$. In the $S^2\times S^1$ background, it actually turns out that the R-charge and fermion number of the vacuum must also be shifted, so that altogether we get
\be \label{PonT}
T_\Delta\;\simeq\; \sigma_m^R\sigma_m^F T^{-1}\circ \mb{P}\,T_\Delta\,. \ee
Acting on the index, $\mb P$ simply sends $m\to -m$, so \eqref{PonT} implies
\be \CI_\Delta(m,e;q) = \big(-q^{\frac12}\big)^m\CI_\Delta(-m,e+m;q)\,.
\label{tetPem} \ee
This identity can be checked explicitly (see Appendix \ref{app:ST}), most easily after converting to the fugacity basis.

The $\rho$ symmetry of Section \ref{sec:kappa} is more universal than parity, but also more subtle to implement. Naively, it sends $\CI_\Delta(m;q,\zeta)$ to $\CI_\Delta(-m;q^{-1},\zeta^{-1})$; but the latter does not make sense for $|q|<1$. In order to properly apply the symmetry, we need to reorganize the Hilbert space $\CH_{-m}$, and to reinterpret $\CI_\Delta(-m;q^{-1},\zeta^{-1})$ as a series in $q$. For the tetrahedron, this is actually straightforward.

Let us set $z=q^{\frac m2}\zeta$, $\ol z=q^{\frac m2}\zeta^{-1}$, noting that $\rho$ acts on $z$ by complex conjugation. Then
\begin{align} \CI_\Delta(-m;q^{-1},\zeta^{-1}) &\,=\, \prod_{r=0}^\infty \frac{1-q^{-r-1}\,\ol z^{-1}}{1-q^{-r} z^{-1}}  \,=\, \exp\left(\, \sum_{n=1}^\infty \frac{-q^{-n}\ol z^{-n}+z^{-n}}{n(1-q^{-n})} \right) \notag \\
 &\simeq\; \exp\left(\, \sum_{n=1}^\infty \frac{-q^n z^{-n}+z^{-n}}{n(1-q^n)}  \right)\label{tetreorg} \\
 &\,=\, \prod_{r=0}^\infty \frac{1-q^{r+1}z^{-1}}{1-q^r \,\ol z^{-1}} = \CI_\Delta(m;q,\zeta)\,. \notag
\end{align}
The necessary reorganization of cancellations happened in the middle step \eqref{tetreorg}, and established the $\rho$ symmetry. Note that this is \emph{not} analytic continuation, since neither of the exponentials here make sense on the unit circle $|q|=1$ --- the expressions diverge at every root of unity.

\subsection{Triality}
\label{sec:tri}

\begin{figure}[htb]
\centering
\includegraphics[width=3.8in]{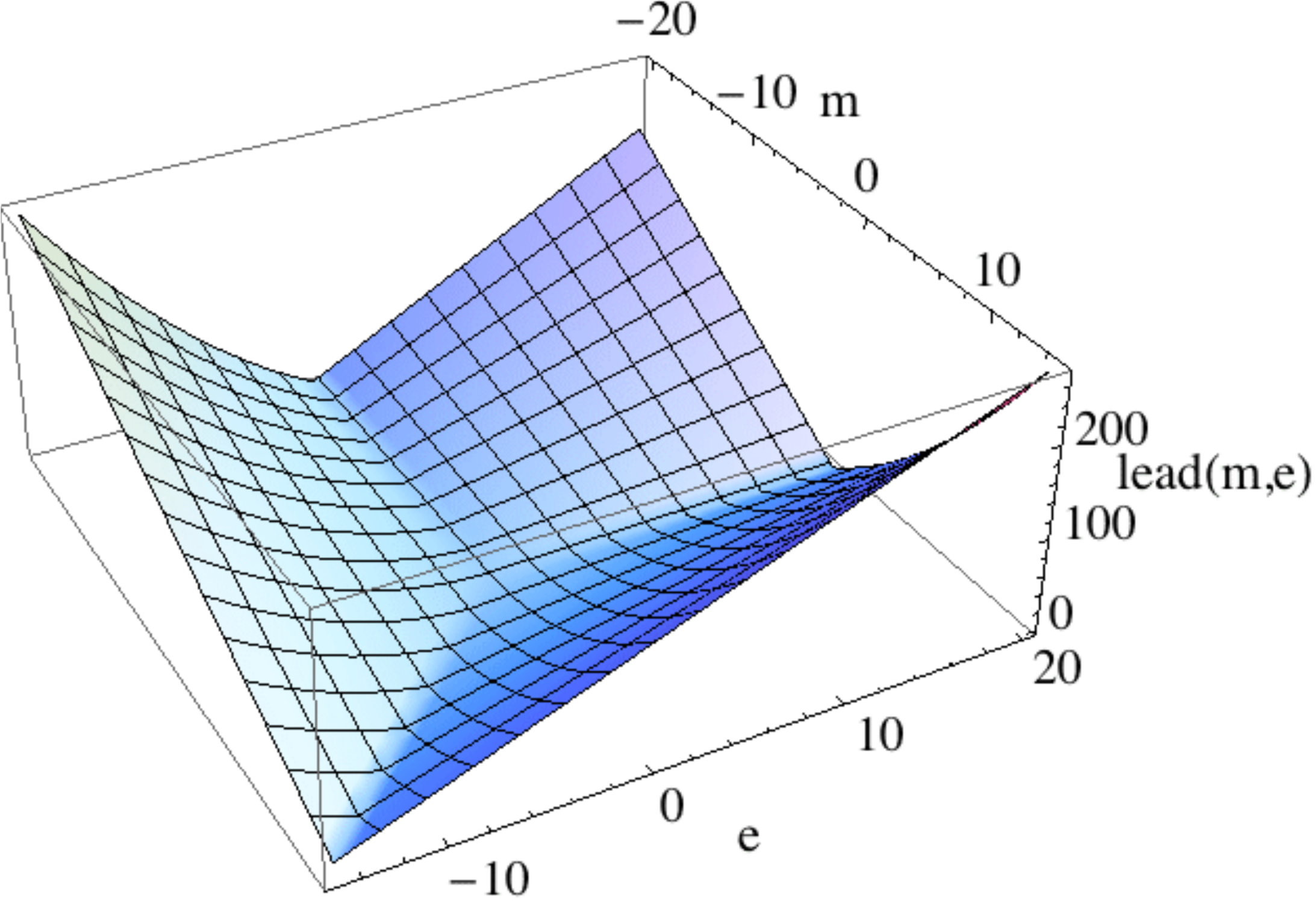}
\caption{Leading powers of $q$ in the index. The graph interpolates smoothly between integer points.}
\label{fig:lead}
\end{figure}

Experimentally, we observe that the tetrahedron index enjoys yet another more interesting discrete symmetry of order three:
\be
\CI_\Delta(m,e;q) = \big(-q^{\frac12}\big)^{-e}\CI_\Delta(e,-e-m;q)=\big(-q^{\frac12}\big)^{m}\CI_\Delta(-e-m,m;q)\,. \label{tettri} \ee
For a visual demonstration of this symmetry, let us define ${\rm lead}(m,e)$ to be the leading power of $q$ that appears in $\CI_\Delta(m,e;q)$ when written as a series in $q$. For example,
\be
\begin{array}{ll}
 \CI_\Delta(0,0;q) = 1-q-2q^2-2q^3-2q^4+q^6+5q^7+...\qquad &\Rightarrow\quad {\rm lead}(0,0)=0\,, \\[.1cm]
 \CI_\Delta(1,0;q) = -q-q^2+q^4+3q^5+4q^6+6q^7+6q^8+...\qquad &\Rightarrow \quad {\rm lead}(1,0)=1\,, \\[.1cm]
 \CI_\Delta(1,2;q) = -q^2-q^3-q^4-q^5+q^7+3q^8+5q^9+... \qquad &\Rightarrow\quad {\rm lead}(1,2) = 2\,,
\end{array}
\ee
{\it etc.} It can be shown from \eqref{Imetet} that ${\rm lead}(m,e)\geq 0$ for all $m,e$. A graph of ${\rm lead}(m,e)$ appears in Figure \ref{fig:lead}, which clearly hints at a $\Z_3$ ``rotation'' symmetry in the index as in \eqref{tettri}. We will describe a method of proving \eqref{tettri} in Section \ref{sec:tetdiff}, using difference equations for the index.

In order to acquire further intuition for the triality symmetry of $\CI_\Delta$, it is instructive to observe that \eqref{tettri} can be written as
\be \CI_\Delta(m,e;q) = \CI_{\sigma_eST\circ \Delta}(m,e;q)=\CI_{(\sigma_eST)^2\circ\Delta}(m,e;q)\,, \label{tetST} \ee
where $ST\in Sp(2,\Z)$ and $\sigma_e \equiv \sigma_e^R\sigma_e^F$ are affine symplectic transformations of the tetrahedron theory, as described in Section \ref{sec:actions}. The transformation $\sigma_eST$ has order three. Then we recall from \cite{DGG} that the tetrahedron SCFT $T_\Delta$ is invariant%
\footnote{The decoration of the symplectic transformation $ST$ by an R-charge and fermion number shift $\sigma_e=\sigma^R_e\sigma^F_e$ is correlated with the precise definition of the tetrahedron theory in the twisted $S^2\times S^1$ geometry, as discussed above \eqref{Iztet}. Had we chosen different $R$ and $F$ assignments for $T_\Delta$, the affine shift would look slightly different.} %
under the cyclic action of $\sigma_eST$, due to 3d $\CN=2$ mirror symmetry \cite{IS, AHISS, dBHOY, dBHO}. In other words, the following UV theories flow to the same IR fixed point:
\be
\begin{array}{r@{\quad}l}
T_\Delta\,: & \text{free chiral with global $U(1)$ symmetry at CS level $-\tfrac12$}\\[.2cm]
\sigma_eST\circ T_\Delta\,: & \text{gauged $U(1)$ theory at CS level $+\tfrac12$, coupled to a single chiral}\,.
\end{array}
\ee
Since these theories are mirror symmetric their indices must be the same, and that is precisely what we see in \eqref{tettri} and \eqref{tetST}.

\subsection{Difference equations}
\label{sec:tetdiff}

A final interesting property of the tetrahedron index is the fact that it obeys two difference equations. In terms of operators $\hat x_{\pm}=\exp(\hat X_\pm)$ and $\hat p_\pm =\exp(\hat P_\pm)$ as defined in Section \ref{sec:ops}, these are
\be \big(\hat p_++\hat x_+^{-1}-1\big)\CI_\Delta = 0\,,\qquad
  \big(\hat p_-+\hat x_-^{-1}-1\big)\CI_\Delta = 0\,.
\label{teteqs}
\ee
For example, writing these out in the charge basis we find
\begin{subequations} \label{teteqsem}
\begin{align}
 q^{\frac e2}\CI_\Delta(m+1,e)+q^{-\frac m2}\CI_\Delta(m,e+1)-\CI_\Delta(m,e) = 0\,, \\
 q^{\frac e2}\CI_\Delta(m-1,e)+q^{-\frac m2}\CI_\Delta(m,e-1)-\CI_\Delta(m,e) = 0\,,
\end{align}
\end{subequations}
for all $m,e$.
These equations are compatible with the $\rho$ symmetry of the index, which interchanges the two $(\pm)$ sets of operators.
The validity of \eqref{teteqs} can be checked easily by writing the index in a fugacity basis, and using the product formula \eqref{Iztet2}.
Also, in the semi-classical regime \eqref{tetasymp} each of the difference equations \eqref{teteqs} turns into
a differential equation, equivalent to the standard property of the dilogarithm function, $z\tfrac{d}{dz} \text{Li}_2 (z) =  \log (1-z^{-1})$.

Equations \eqref{teteqs} may look familiar from the study of ``quantum Lagrangians'' and $S^3_b$ partition functions in \cite{DG-Sdual,DGG}. As anticipated in Section \ref{sec:ops}, these are the \emph{same} difference equations obeyed by the $S^3_b$ partition function of $T_\Delta$, with an appropriate identification of the operators $\hat x_\pm,\,\hat p_\pm$. We will try to explain the reason behind this in Section \ref{sec:3d4d}. In essence, the same universal algebra of line operators is acting on both the index and the $S^3_b$ partition functions; Ward identities for the line operators then lead to \eqref{teteqs}.

\begin{wrapfigure}[15]{r}{2in}
\includegraphics[width=1.8in]{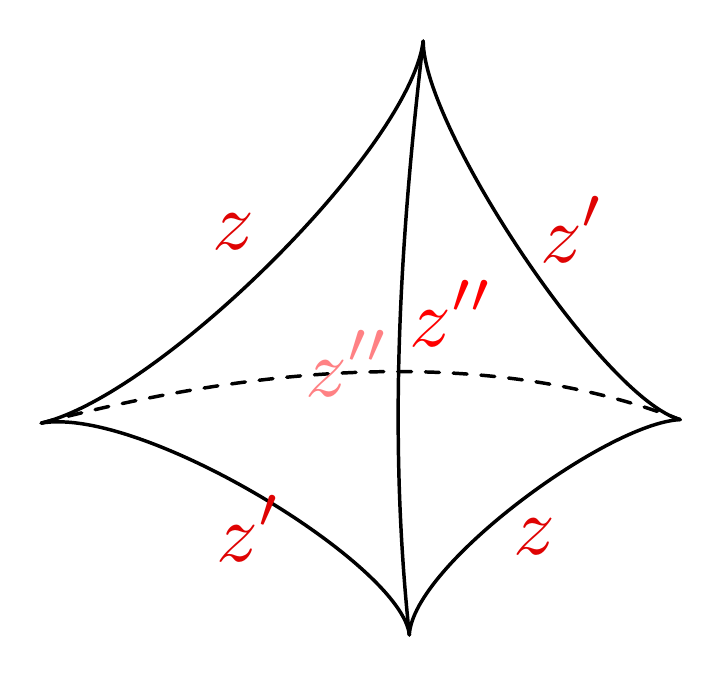}
\caption{An ideal tetrahedron, with edge parameters.}
\label{fig:tet}
\end{wrapfigure}

In terms of the tetrahedron $\Delta$ itself, Equations \eqref{teteqs} are two copies of the quantized Lagrangian $\hat \CL_\Delta$ that describes which flat $SL(2,\C)$ connections on the boundary $\pd \Delta$ can be extended as flat connections in the interior of $\Delta$. To see this, recall%
\footnote{Throughout this paper, we will be quite brief with details of flat connections and 3d geometry. We direct the reader to the summary of triangulations in Section 2 of \cite{DGG} and references therein (especially the classic \cite{Thurston-1980, NZ}) for some potentially useful background.}
from \cite{Dimofte-QRS, DGG} that the set of flat connections on $\pd\Delta$, \emph{a.k.a.} the phase space $\CP_{\pd \Delta}$, is described by three $\C^*$--valued edge coordinates $z,z',z''$, subject to the condition that $zz'z''=-1$. Upon quantization, these coordinates form an algebra of operators with $q$-commutation relations
\be \hat z\hat z'=q \hat z'\hat z\,,\qquad \hat z'\hat z''=q\hat z''\hat z'\,,\qquad \hat z''\hat z=q\hat z\hat z''\,,\ee
and a central constraint
\be \hat z''\hat z'\hat z= -1\,. \label{zconst} \ee
(For logarithms of the $\hat z$'s, written in uppercase letters, the constraint reads $\hat Z+\hat Z'+\hat Z''=i\pi+\hbar/2$.)
Similarly, the classical Lagrangian $\CL_\Delta = \{z''+z^{-1}-1=0\}$, describing flat connections in the interior of $\Delta$, becomes promoted to a quantum operator (\cf\ \cite{gukov-2003})
\be \hat\CL_\Delta = \hat z''-\hat z^{-1}-1\,, \label{tetLag} \ee
which must annihilate any putative wavefunction of the tetrahedron, in any representation. It should then be clear that Equations \eqref{teteqs} are just two copies of \eqref{tetLag}, with opposite quantization parameters $\hbar$ and $-\hbar$. Indeed, using the classical complex variables $z=q^{\frac m2}\zeta,\, \ol z=q^{\frac m2}\zeta^{-1}$, or logarithmically $Z = \tfrac m2\hbar+i\theta,\,\ol Z=\tfrac m2\hbar-i\theta$ (\cf\ \eqref{XPz}), we find that we can identify $(\hat P_+,\hat X_+) = (\hbar \pd_Z,Z)$ and $(\hat P_-,\hat X_-)=(-\hbar \pd_{\ol Z},\ol Z)$.

The Lagrangian \eqref{tetLag} is invariant under cyclic permutations $\hat z\to \hat z'\to\hat z''\to \hat z$ in the operator algebra for the tetrahedron. That is, the equation $\hat \CL_\Delta\,\psi=0$ for any putative wavefunction $\psi$ can be written using any pair of consecutive variables $(\hat z'',\hat z)$, $(\hat z,\hat z')$, or $(\hat z',\hat z'')$, thanks to the constraint \eqref{zconst}. Moreover, the generator of the $\Z_3$ cyclic symmetry is none other than our familiar affine $Sp(2,\Z)$ element $\sigma_eST$, \cf\ \eqref{tetST}, acting on logarithms of the $\hat z$ operators. For example, if we identify $(\hat P_+,\hat X_+) = (\hat Z'',\hat Z)$ and $(\hat P_+',\hat X_+')=(\hat Z,\hat Z')$, then
\be \label{XPST}
\begin{pmatrix} \hat X_+' \\ \hat P_+ '\end{pmatrix}
 = \sigma_eST \begin{pmatrix} \hat X_+ \\ \hat P_+ \end{pmatrix}
 = \begin{pmatrix} -1 & -1 \\ 1 & 0 \end{pmatrix} \begin{pmatrix} \hat X_+ \\ \hat P_+ \end{pmatrix}+\begin{pmatrix} i\pi+\tfrac\hbar2 \\ 0 \end{pmatrix}\,.
\ee
Together with the intertwining property \eqref{intertwine} for the index, the cyclic symmetry of the quantized Lagrangian guarantees that $\CI_\Delta$, $\CI_{\sigma_eST\circ \Delta}$, and $\CI_{(\sigma_eST)^2\circ\Delta}$ all satisfy the \emph{same} equations \eqref{teteqs}.
This constitutes the basis for a proof that $\CI_\Delta = \CI_{\sigma_eST\circ \Delta} = \CI_{(\sigma_eST)^2\circ\Delta}$, as in \eqref{tetST}, since the solutions to the difference equations are unique given appropriate boundary conditions. Details appear in Appendix \ref{app:ST}.

\section{The index of $T_M$}
\label{sec:TM}

Several copies of the tetrahedron theory $T_\Delta$ can be appropriately combined to construct an $\CN=2$ SCFT associated to any oriented 3-manifold $M$ that admits an ideal triangulation. The gluing rules for $T_\Delta$ theories, described in \cite{DGG}, immediately translate to a simple, combinatorial prescription for calculating the supersymmetric index $\CI_M \equiv \CI_{T_M}$ of $T_M$ on $S^2\times S^1$. Here we proceed to write down these rules explicitly, and to give several examples of the resulting 3-manifold indices $\CI_M$. Similar rules could be used to construct indices for more general theories in class $\CR$.

Just as the SCFT $T_M$ is independent of any chosen triangulation of $M$ by virtue of 3d mirror symmetry --- with different triangulations leading to equivalent UV Lagrangian descriptions --- the index $\CI_M$ must also be a topological invariant of~$M$. We will check this explicitly in Section \ref{sec:bip}, by calculating the index of a bipyramid and demonstrating its invariance under a ``2-3 move.'' This is sufficient (with a few technical caveats)
to guarantee triangulation invariance for a general 3-manifold.

Having obtained a new topological invariant, it is natural to ask how strong it is. We conjecture that the index $\CI_M$ is exactly as strong as the compact $SU(2)$ Chern-Simons partition function of $M$ (\emph{a.k.a.}\! the set of colored Jones polynomials, when $M$ is a knot complement). Equivalently, the index is just as strong as the ellipsoid partition function $\CZ_b[M]$ of $T_M$, which is a holomorphic $SL(2,\C)$ Chern-Simons invariant \cite{DGH, DGG}. The best way to see this is by identifying the index $\CI_M$ with a full, non-holomorphic $SL(2,\C)$ Chern-Simons partition function, as in Section~\ref{sec:CS}. For now, an excellent hint comes from the fact (demonstrated in Section \ref{sec:glueops}) that the same ``quantum Lagrangian'' operators that annihilate compact \cite{Gar-Le, garoufalidis-2004} and holomorphic \cite{gukov-2003, DGLZ, Dimofte-QRS} $SL(2,\C)$ Chern-Simons partition functions also annihilate the index. Then it is clear that (say) the compact $SU(2)$ CS partition function determines the difference operators, and these in turn determine the index, up to a finite number of $q$-dependent normalizations.

To test the conjectured strength (or weakness) of the index as a topological invariant, one could consider topologically distinct knot complements with the same colored Jones polynomials. A famous infinite family of such pairs is generated by the so-called ``mutation'' operation on knots \cite{Conway-mutation, MortonTraczyk}. In Section \ref{sec:mutant}, we will calculate the indices for the simplest pair of mutant knot complements, at charges $(m,e)=(0,0)$ and the first few orders in $q$, and show that they are identical. We then provide a new gauge-theoretic argument for mutation-invariance of the index (as well as $\CZ_b[M]$) using properties of 4d $\CN=2$ theories.

\subsection{Gluing rules}
\label{sec:rules}

Let's begin by recalling the gluing rules of \cite{DGG} for theories $T_M$. To construct $T_M$, we must choose an oriented 3-manifold $M$, an ideal triangulation $M = \bigcup_{i=1}^N\Delta_i$ of $M$ (which doesn't matter in the end), and a polarization $\Pi$ for the symplectic space $\CP_{\pd M}$ of flat connections on the boundary $\pd M$ (which does matter).

The choice of polarization was described carefully in Section 2 of \cite{DGG}, and we recall a few facts about it here. Physically, if we think of $T_M$ as a 3d boundary condition for a  4d $\CN=2$ SCFT $T[\pd M]$, the polarization specifies how to couple $T_M$ to bulk 4d degrees of freedom. In practice, for components of $\pd M$ that are triangulated by faces of tetrahedra $\Delta_i$, the polarization $\Pi$ involves a choice of independent external edges to which are associated canonically conjugate ``position'' and ``momentum'' coordinates on $\CP_{\pd M}$. For components of $\pd M$ that are torus cusps, coming from vertices of ideal tetrahedra, a polarization corresponds to a basis of canonically conjugate ``A and B cycles'' on the torus. We will see both of these cases appearing in examples later on.

Suppose then that we are given $M$, $\{\Delta_i\}_{i=1}^N$, and $\Pi$. To find $T_M$:
\begin{enumerate}

\item The semi-classical phase space $\CP_{\pd\Delta_i}$ of each tetrahedron $\Delta_i$ is described by three logarithmic edge parameters $Z_i,Z_i',Z_i''$ as in Figure \ref{fig:tet}, with a (quantum-corrected) constraint
\be Z_i + Z_i'+Z_i'' = i\pi + \hbar/2\, \label{Phbar} \ee
and a symplectic structure $\Omega_i = \frac{1}{\hbar} dZ_i\wedge dZ_i'$.
To each tetrahedron $\Delta_i$, in its ``canonical'' polarization $\Pi_i\equiv \Pi_{Z_i}$ with position and momentum $(X_i;P_i)=(Z_i;Z_i'')$, associate the tetrahedron theory $T_{\Delta_i,\Pi_i}$ as in \eqref{tetthy}. It has a $U(1)$ flavor symmetry, whose twisted mass parameter should be thought of as the position $Z_i$.

\item Form a product theory $T_{\{\Delta_i\},\{\Pi_i\}}=T_{\Delta_1,\Pi_1}\otimes\cdots\otimes T_{\Delta_N,\Pi_N}$. This corresponds to the collection of tetrahedra $\{\Delta_i\}$ with the natural product polarization $\Pi_1\times\cdots \times \Pi_N$ on the product phase space $\CP_{\pd \Delta_1}\times\cdots\times \CP_{\pd \Delta_N}$. The theory has $U(1)^N$ flavor symmetry, with each independent twisted mass corresponding to a position coordinate $Z_i$.

\item \label{step3} Choose a new polarization $\wt\Pi$ on the product phase space $\CP_{\pd \Delta_1}\times\cdots\times \CP_{\pd \Delta_N}$ such that
\begin{itemize}
\item it is compatible with the final desired polarization $\Pi$ for $\pd M$ (\ie\ the positions and momenta in $\Pi$ are also positions and momenta in $\wt\Pi$); and
\item the edge coordinates $C_I$ for all internal edges in the triangulation of $M$ are positions in $\wt\Pi$.%
\footnote{It may be useful to recall here that these internal edge coordinates are sums of edge coordinates $Z_i,Z_i',Z_i''$ of individual tetrahedra that come together to form an internal edge (the same holds for external edges); and all internal edge coordinates commute with all external coordinates on $\CP_{\pd M}$.}
\end{itemize}
This is possible by a classic result of \cite{NZ}.

\item Write $\wt \Pi = g\cdot \{\Pi_i\}$, where $g$ is an affine symplectic transformation,
\be g\;\in\; Sp(2N,\Z)\ltimes \big[(i\pi\Z)^{2N}\times (\tfrac\hbar2\Z)^{2N}\big]\,, \label{ggen} \ee
and act on the product theory $T_{{\Delta_i},\{\Pi_i\}}$ with $g$ as in Sections \ref{sec:Sp}--\ref{sec:aff} to obtain
\be T_{\{\Delta_i\},\tilde\Pi}\,=\,g\circ T_{\{\Delta_i\},\{\Pi_i\}}\,.\ee
This theory still has a $U(1)^N$ flavor symmetry, but the twisted mass parameters now correspond to position coordinates on $\CP_{\pd M}$ and to internal edge coordinates $C_I$.

\item \label{step5} Add a superpotential $\CW = \sum_I \CO_I$ to $T_{\{\Delta_i\},\tilde\Pi}$ that breaks the $U(1)$ symmetries associated to the internal edges $I$. This operation is the gauge-theory equivalent of symplectic reduction. We obtain a UV Lagrangian description of the theory $T_M$, which has a global symmetry group $U(1)^{\frac12\dim \CP_{\pd M}}$
left over. The twisted mass of each $U(1)$ is a position coordinate in the polarization $\Pi$ for $\CP_{\pd M}$.

\end{enumerate}

Two points here deserve further clarification. First, for defining a theory $T_M$ on $\R^3$, the affine shifts in \eqref{ggen} were irrelevant. In the context of the index, however, the theory is put on an $S^2\times S^1$ geometry, in the presence of magnetic flux. Then the affine shifts by $i\pi$ and $\hbar/2$ are related to $F$ and $R$ assignments, respectively, as discussed in Section \ref{sec:aff}. In order to see both shifts by $i\pi$ and $\hbar/2$ in the geometric description of phase spaces such as $\CP_{\pd\Delta_i}$ and $\CP_{\pd M}$, one must include a few $\hbar$ corrections in the relations among classical coordinates, as in \eqref{Phbar}. In the closely related context of analytically continued Chern-Simons theory, these semi-classical corrections were studied systematically in \cite{Dimofte-QRS}. The basic rule--of--thumb is that every $i\pi$ must be accompanied by an $\hbar/2$. Hence for gauge theory on $S^2\times S^1$ this means that every shift of R-charge $\sigma^R$ is coupled to a shift of fermion number $\sigma^F$.

Second, one might recall from \cite{DGG} that it was sometimes necessary to refine a given triangulation of $M$ in order to properly define the theory $T_M$. This is because the operators $\CO_I$ that one adds to the superpotential may not exist when triangulations are too coarse. (For example, the theory of the figure-eight knot complement built from two tetrahedra suffered from this problem.) For purposes of calculating the index $\CI_M$, such refinements of triangulation are not necessary. The index is insensitive to superpotential terms, aside from the simple fact that they break some flavor symmetry. Thus, when computing an index, we can often use unrefined, ``hard'' triangulations and just break flavor symmetries by hand, following the rules of Section \ref{sec:W}.

Now, translating the gluing rules for $T_M$ to gluing rules for the index, and taking into account the preceding remarks, we arrive at the following combinatorial construction of $\CI_M$. Let's again suppose that we have a manifold $M$, a triangulation $\{\Delta_i\}$, and a polarization $\Pi$ for $\CP_{\pd M}$. Then:
\begin{enumerate}

\item To each tetrahedron $\Delta_i$ with polarization $\Pi_i$, associate a tetrahedron index $\CI_{\Delta_i}(m_i,e_i)$ defined by \eqref{Imetet}. (We work in a charge basis, and also suppress the dependence on $q$.)

\item Form the product
\be \CI_{\{\Delta_i\},\{\Pi_i\}}(m,e) = \CI_{\Delta_1}(m_1,e_1)\times\cdots\times \CI_{\Delta_N}(m_N,e_N)\,. \ee
Now $m$ and $e$ are charge vectors of length $N$, and we can set $\gamma = {m \choose e}$.

\item Choose a polarization $\wt\Pi$ for the product phase space $\CP_{\pd\Delta_1}\times\cdots\times\CP_{\pd\Delta_N}$ as in Step \ref{step3} above. It is related to the obvious product polarization via an affine symplectic transformation $g$. Decompose $g$ as a product $g = \sigma(\alpha)\,g_{Sp}$, where $g_{Sp}\in Sp(2N,\Z)$ and $\sigma(\alpha) =\sigma^R(\alpha)\sigma^F(\alpha)$, $\alpha\in \Z^{2N}$, is an affine shift of position and momentum coordinates by $(i\pi+\tfrac\hbar2)\alpha$. (As noted above, shifts by $i\pi$ will always be coupled with shifts by $\hbar/2$.)

\item Apply the above affine symplectic transformation to the product index, following Sections \ref{sec:Sp}--\ref{sec:aff}. Using the notation $\gamma = {m\choose e}$, we obtain
\begin{align} \CI_{{\Delta_i},\tilde\Pi}(m,e) &= \big[\sigma^R(\alpha)\,\sigma^F(\alpha)\,g_{Sp}\big]\circ \CI_{\{\Delta_i\},\{\Pi_i\}}(m,e) \notag \\
 &= \big(-q^{\frac12}\big)^{\langle \alpha,\gamma\rangle} \CI_{\{\Delta_i\},\{\Pi_i\}}\big(g_{Sp}^{-1}\,\gamma\big)\,. \label{Iprod2}
 \end{align}

\item Finally, break the flavor symmetries corresponding to internal edges of the triangulation. In the polarization $\wt\Pi$, each independent internal edge coordinate $C_I$ corresponds to a distinct electric charge $e_I$. Then, according to Section \ref{sec:W}, we find
\be \CI_{M,\Pi}(m',e')\,=\, \sum_{e_I\in \Z} q^{\Sigma_Ie_I}\, \CI_{{\Delta_i},\tilde\Pi}(m,e)\Big|_{m_I\,=\,0}\,, \label{indsum} \ee
where the sum is over all internal-edge charges $e_I$, and we set all conjugate magnetic charges $m_I$ to zero. The extra factor of $q^{\Sigma_Ie_I}$ comes from the R-charge correction discussed in Section \ref{sec:W}, using the fact that our R-charge assignment for tetrahedron theories was $R_\phi = R_{\rm vac} = 0$.

\end{enumerate}
In the end, we obtain an index $\CI_M$ that depends on $\tfrac12\dim \CP_{\pd M}$ electric charges $e'$ and $\tfrac12\dim \CP_{\pd M}$ magnetic charges $m'$. They correspond directly to the $\tfrac12\dim\CP_{\pd M}$ $U(1)$ flavor symmetries of the theory $T_M$.

The sum \eqref{indsum} typically converges, in the sense that only finitely many terms are necessary for calculating $\CI_{M,\Pi}(m',e')$ to any desired order in $q$. Physically, the convergence of the sum is directly related to the existence of unconstrained chiral operators $\CO_I$ that can be added to the superpotential of $T_{\{\Delta_i\},\tilde\Pi}$ to obtain $T_M$. (This statement will become clearer in Section \ref{sec:tent}.) In particular, when using a refined, ``easy'' triangulation so that all operators $\CO_I$ exist, the sum should always converge. In practice, for an index computation, it actually appears that the only triangulations to be avoided are those with univalent internal edges --- \eg\ edges resulting from gluing two adjacent sides of a single tetrahedron together. Such triangulations are automatically ``hard''; but more seriously, they fail to describe the moduli space of flat connections on $M$ (\cf\ \cite{Dunfield-Mahler}, Sec. 10.3, and \cite{Segerman-def, DunGar}), and should never be expected to produce the correct theory $T_M$ or its index.

\subsection{The bipyramid}
\label{sec:bip}

\begin{figure}[htb]
\centering
\includegraphics[width=5.3in]{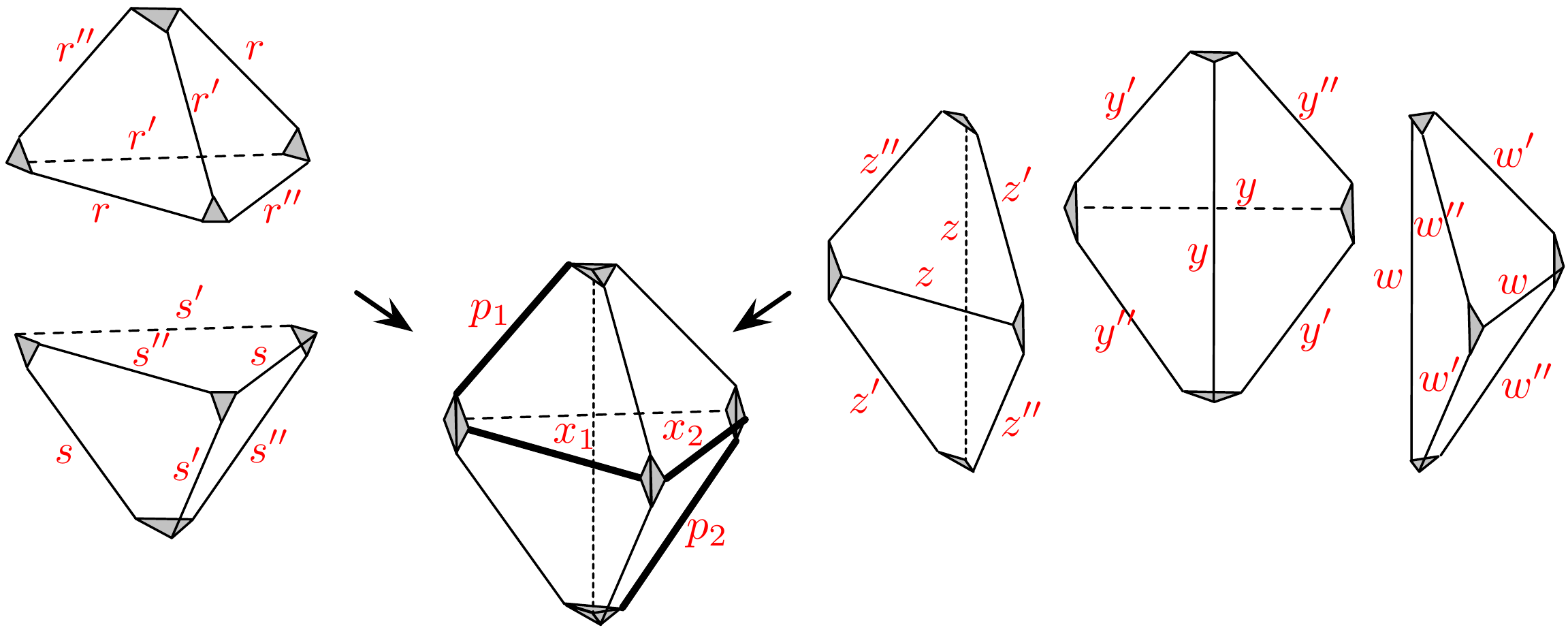}
\caption{The bipyramid, in its two triangulations. The equatorial polarization $\Pi = (X_1,X_2;P_1,P_2)$ has positions $X_1$ and $X_2$ corresponding to the two equatorial edges in the front.}
\label{fig:bip}
\end{figure}

As a simple but crucial example of a nontrivial 3-manifold, let us take $M$ to be the bipyramid, shown in the center of Figure \ref{fig:bip}. The bipyramid can be decomposed into either 2 or 3 tetrahedra, leading to two different UV descriptions of the theory $T_M$. It was shown in \cite{DGG} that, with an appropriate polarization, the gluing of two tetrahedra produces $N_f=1$ SQED, while the gluing of three tetrahedra produces the so-called XYZ model, a theory of three chiral multiplets coupled by a cubic superpotential. The fact that these theories are mirror symmetric \cite{AHISS} formed the basis of the argument that the construction of $T_M$ for general 3-manifolds is triangulation independent (see also \cite{CCV}).

Here we calculate the index $\CI_M$ for the bipyramid. The boundary phase space $\CP_{\pd M}$ is four-dimensional, with two position coordinates corresponding to the $U(1)^2$ flavor symmetry of either $N_f=1$ SQED or the XYZ model. Let us choose an equatorial polarization for $\CP_{\pd M}$, with position coordinates $x_{1,2}=\exp(X_{1,2})$ and momentum coordinates $p_{1,2}=\exp(P_{1,2})$ associated to the external edges shown in Figure \ref{fig:bip}.

For the decomposition into two tetrahedra $\Delta_R$ and $\Delta_S$, we first form the product index $\CI_{\{\Delta_R,\Delta_S\},\{\Pi_R,\Pi_S\}}(m_R,m_S,e_R,e_S) = \CI_\Delta(m_R,e_R)\,\CI_\Delta(m_S,e_S)$. (We continue to suppress the dependence of indices on the parameter $q$.) Then observe that
\be \begin{pmatrix} X_1\\ X_2\\ P_1\\ P_2 \end{pmatrix}
 = \begin{pmatrix} R+S'' \\ R''+S \\ R'' \\ S'' \end{pmatrix}
 = \begin{pmatrix} 1 & 0 & 0 & 1 \\ 0 & 1 & 1 & 0 \\ 0 & 0 & 1 & 0 \\ 0 & 0 & 0 & 1 \end{pmatrix}
 \begin{pmatrix} R \\ S\\ R'' \\ S'' \end{pmatrix}\,.
\label{pol2}
\ee
Therefore, $\Pi = \wt\Pi = g_{Sp}\cdot (\Pi_R\times\Pi_S)$, where $g_{Sp}$ is the symplectic matrix on the right side of \eqref{pol2}. There is no affine shift. Correspondingly, the index transforms as
\begin{align} \CI_{M,\Pi}(m_1,m_2,e_1,e_2) &=
 \big(-q^{\frac12}\big)^{\langle 0,(m,e)\rangle}\CI_{\{\Delta_R,\Delta_S\},\{\Pi_R,\Pi_S\}}\big(g_{Sp}^{-1}(m_1,m_2,e_1,e_2)\big) \notag \\
  &= \CI_\Delta(m_1-e_2,e_1)\,\CI_\Delta(m_2-e_1,e_2)\,.
 \label{ind2}
\end{align}
Since there are no internal edges, this is automatically the index of the bipyramid theory.

Physically, \eqref{ind2} counts states in $N_f=1$ SQED on $S^2\times S^1$. To be very explicit, this theory starts out with a flat-space Lagrangian%
\footnote{This particular Lagrangian is found after doing an $\sigma_eST$ rotation (a mirror symmetry) of the $\Delta_S$ tetrahedron; see Section 4.2 of \cite{DGG}.} %
\begin{align} &\CL_{M,\Pi} = \frac{1}{4\pi}\int d^4\theta\big( \Sigma_1\,V_2+(\Sigma_1+2 V_2)V\big) +\int d^4\theta \big(\phi_R^\dagger e^{V+\frac12 V_1}\phi_R + \phi_{S''}^\dagger e^{-V+\frac12 V_1}\phi_{S''}\big)\,,
\end{align}
where $V$ is dynamical, $V_1$ is a background vector multiplet for the axial $U(1)$ symmetry, and $V_2$ is a background vector multiplet for a slightly modified topological $U(1)$ symmetry. Some additional background Chern-Simons couplings are turned on, and one can also work out the appropriate R-charges and fermions numbers chosen for putting the theory on $S^2\times S^1$ (corresponding to our equatorial polarization $\Pi$). Then
\be \CI_{M,\Pi}(m_1,m_2;q,\zeta_1,\zeta_2) = \Tr_{\CH_{m_1,m_2}} (-1)^F q^{\frac{R}2+j_3}\zeta_1^{e_1}\zeta_2^{e_2} = \sum_{e_1,e_2\in \Z} \CI_{M,\Pi}(m_1,m_2,e_1,e_2;q)\zeta_1^{e_1}\zeta_2^{e_2}\,,\ee
where $e_1$ and $e_2$ count axial and topological charges, and $m_1$ and $m_2$ specify the amount of axial and topological flux, respectively, through $S^2$.

For the triangulation into three tetrahedra $\Delta_Z,\Delta_W,\Delta_Y$, we again start with a product index $\CI_{\{\Delta_i\},\{\Pi_i\}}(m,e) = \CI_\Delta(m_Z,e_Z)\,\CI_\Delta(m_W,e_W)\,\CI_\Delta(m_Y,e_Y)$. Now, however, the product phase space $\CP_{\pd \Delta_Z}\times \CP_{\pd \Delta_W}\times \CP_{\Delta_Y}$ is six-dimensional, and there is an internal edge. We choose an intermediate polarization $\wt\Pi = (X_1,X_2,C;P_1,P_2,\Gamma)$ on the product phase space that is compatible with $\Pi$ and also includes $C = Z+W+Y$, the internal edge parameter, as a position coordinate. Here $\Gamma = -Y'$ (say) is a conjugate momentum to $C$ that commutes with the external edges. Then
\be \begin{pmatrix} X_1\\X_2\\ C\\P_1\\P_2\\\Gamma \end{pmatrix}
= \begin{pmatrix} Z \\ W \\ Z+W+Y \\ Z''+Y' \\ W''+Y' \\ -Y' \end{pmatrix}=
\begin{pmatrix} 1 &0&0&0&0&0 \\ 0&1&0&0&0&0 \\1&1&1&0&0&0 \\ 0&0&-1&1&0&-1 \\ 0&0&-1&0&1&-1 \\ 0&0&1&0&0&1\end{pmatrix}
\begin{pmatrix} Z\\W\\Y\\Z''\\W''\\Y''\end{pmatrix} +
\begin{pmatrix} 0\\0\\0\\i\pi+\tfrac\hbar2\\i\pi+\tfrac\hbar2\\-i\pi-\tfrac\hbar2 \end{pmatrix}\,,
\label{pol3}
\ee
so the affine symplectic transformation from $\{\Pi_i\}$ to $\wt\Pi$ is $\sigma(\alpha)\,g_{Sp}$ with $g_{Sp}$ the matrix appearing in \eqref{pol3} and a shift vector $\alpha = (0,0,0,1,1,-1)$. Correspondingly,
\begin{align} \CI_{\{\Delta_i\},\tilde\Pi}(m,e) &= \big(-q^{\frac12}\big)^{\langle \alpha,(m,e)\rangle}\CI_{\{\Delta_i\},\{\Pi_i\}}\big(g_{Sp}^{-1}(m,e)\big) \notag \\
 &\hspace{-.8in}= \big(-q^{\frac12}\big)^{m_1+m_2-m_3}\CI_\Delta(m_1,e_1+e_3)\CI_\Delta(m_2,e_2+e_3)\CI_\Delta(m_3-m_1-m_2,e_3+m_1+m_2-m_3)\,. \notag
\end{align}
Finally, to obtain the index of the bipyramid theory, we sum over the electric charge $e_3$ corresponding to the internal edge, and set $m_3=0$\,:
\begin{align} \CI_{M,\Pi}(m,e) &= \sum_{e_3\in \Z} \big(-q^{\frac12}\big)^{2e_3+m_1+m_2}\CI_\Delta(m_1,e_1+e_3)\,\CI_\Delta(m_2,e_2+e_3)\,\CI_\Delta(-m_1-m_2,e_3+m_1+m_2) \notag \\
&= \sum_{e_3\in \Z} q^{e_3}\,\CI_\Delta(m_1,e_1+e_3)\,\CI_\Delta(m_2,e_2+e_3)\,\CI_\Delta(m_1+m_2,e_3)
\label{ind3}
\end{align}
(the last simplification follows by parity symmetry \eqref{tetPem}). It is not too hard to see that \eqref{ind3} is a reasonable index for the XYZ model, with the cubic superpotential breaking a diagonal $U(1)$ flavor symmetry and leading to a sum over its charge sectors.

The equivalence of \eqref{ind2} and \eqref{ind3} can be proven using difference equations, much in the same way that we demonstrate $\sigma_eST$--invariance of the tetrahedron index in Appendix \ref{app:ST}. Of course, these two expressions must be equal on physical grounds, because they are indices for mirror symmetric theories. Computationally, it is very easy to check equivalence at any fixed charge $\gamma=(m,e)$, order by order in $q$. For example, both expressions give
\begin{align*} \CI_{M,\Pi}(0,0,0,0) &= 1-2q-3q^2+4q^4+12q^5+14q^6+6q^8+\ldots\,, \\
\CI_{M,\Pi}(1,3,2,4) &= -q^6-2q^7-4q^8-6q^9-8q^{10}-9q^{11}-8q^{12}+\ldots\,,
\end{align*}
{\it etc.} Only a finite number of terms in the sum \eqref{ind3} is needed at any given order. In \cite{KSV-index, KW-index} the match was also proven at special values of $m=(m_1,m_2)$ using different methods.

The equivalence of \eqref{ind2} and \eqref{ind3} and the $\sigma_eST$-invariance of $\CI_\Delta$ are the basic nontrivial ingredients in a combinatorial argument that $\CI_{M,\Pi}$ is a topological invariant of $(M,\Pi)$ --- independent of triangulation or any other choices. Again, this must be the case physically as long as $T_M$ is well defined, but it us useful to have a more bottom-up understanding. The $\sigma_eST$-invariance shows that it does not matter how one labels edge parameters of individual tetrahedra in a triangulation, as long as their cyclic ordering (induced by the orientation of $M$) is preserved. The 2-3 invariance of the bipyramid index is then enough to show that the index of any triangulated 3-manifold is independent of triangulation. In particular, the 2-3 invariance must work for any boundary polarization of the bipyramid (it is trivial to show this), and seems to commute with the operation of gluing the bipyramid into a larger 3-manifold --- as long as the larger triangulation has no univalent edges.%
\footnote{\emph{Cf.} the end of Section \ref{sec:rules}. The potential issue here is that the sums \eqref{indsum} in the definition of the index may not converge uniformly --- so that order of summations can be interchanged. From all tested examples, it appears that avoiding univalent edges is sufficient for convergence. Otherwise, we should only use refined, ``easy'' triangulations. It is highly plausible that the set of ``easy'' triangulations is fully connected by 2--3 moves, but it is not yet known rigorously.} %
Then, conjecturally, the set of triangulations of $M$ with no univalent edges is fully connected by 2--3 moves, and triangulation invariance follows.

It would be useful to have a more rigorous understanding of convergence, its relation to combinatorics of triangulations, and how 2--3 moves act to connect restricted sets of triangulations --- such as those without univalent edges. Mathematically, this is still uncharted territory.

\subsection{Some knot complements}
\label{sec:knots}

A well-studied class of 3-manifolds are the complements of knots in the 3-sphere. One forms a knot complement $M_K$ by slightly thickening a knot $K\in S^3$ into a solid torus $N_K$, and then cutting it out,
\be M_K = S^3\bs N_K\,. \ee
With the single exception of the unknot complement, all knot complements seem to admit%
\footnote{All ideal triangulations of the unknot have a univalent (``peripherally homotopic'') edge. Otherwise, it has been proven that all hyperbolic knot complements have non-univalent triangulations \cite{Tillmann-def}, and the same seems to hold even for non-hyperbolic knots \cite{Segerman-def}.}
ideal triangulations that can be used to define the index $\CI_{M_K}$, a topological invariant of $M_K$.

\begin{wrapfigure}{r}{3in}
\includegraphics[width=2.8in]{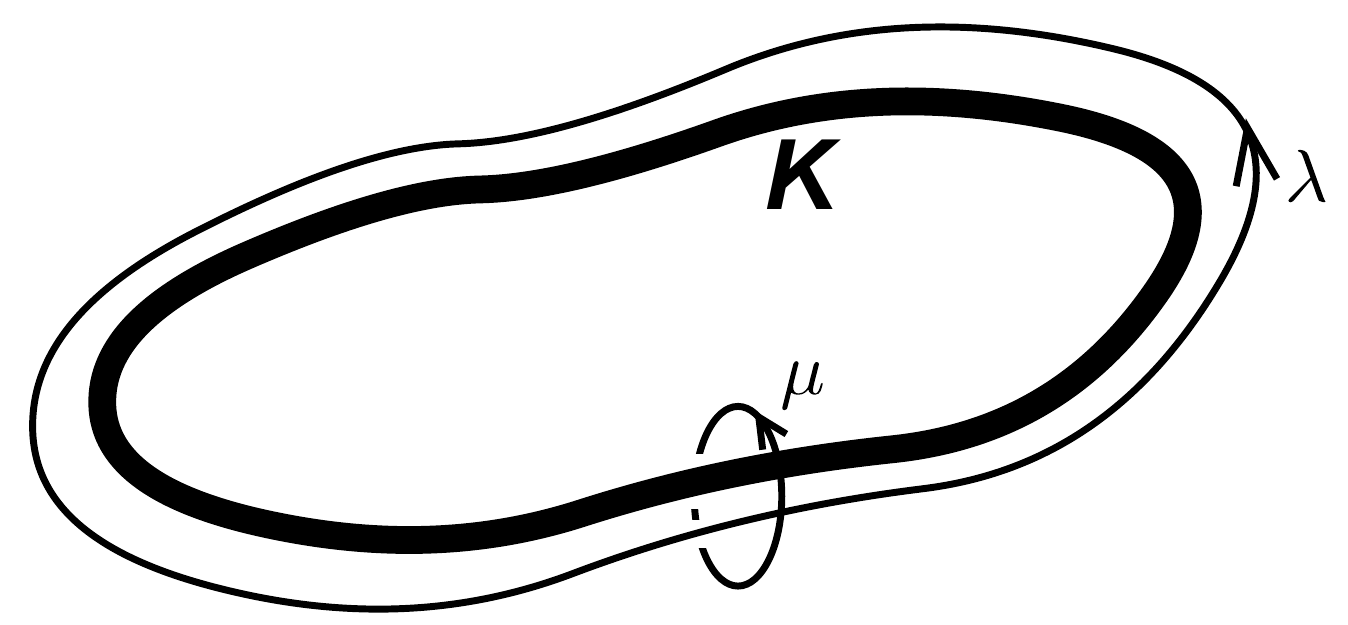}
\caption{Meridian and longitude cycles on the boundary of a knot complement.}
\label{fig:lm}
\end{wrapfigure}

The boundary of a knot complement is a torus $T^2$, and its phase space $\CP_{T^2}$ has a canonical polarization $\Pi$. To define it, one first identifies the so-called meridian and longitude cycles on the boundary: the meridian $\mu$ is a small loop linking the knot $K$, which would be contractible in the thickened knot neighborhood $N_K$; and the longitude $\lambda$ is a loop running parallel to $K$ that has zero linking number with $K$ and is null-homologous in $M_K$ (Figure \ref{fig:lm}). The orientation of $M_K$ induces a relative orientation on these cycles. The eigenvalues of the $SL(2,\C)$ holonomies along $\lambda$ and $\mu$, typically denoted $\ell$ and $m$, then provide $\C^*$ coordinates for the boundary phase space (see \eg\ \cite{gukov-2003}):
\be \CP_{\pd M} \;=\; \big(\C^*\times \C^*\big)\big/\raisebox{-.1cm}{$\Z_2$} \;=\; \{(\ell,m)\}\big/\raisebox{-.1cm}{$(\ell,m)\sim(\ell^{-1},m^{-1})$}\;\;. \label{PT2} \ee
Finally, we take the canonical position coordinate%
\footnote{The factor of $2$ in the definition of $U$ ensures that $U$ and $v$ are canonically conjugate. The minus sign in $v=\log(-\ell)$ is merely convenient when dealing with combinatorics of triangulations, to avoid unwanted factors of $i\pi$. Geometrically, this sign is correlated with the lift from $PSL(2)$ to $SL(2)$ holonomy on a knot complement.} %
to be $U\equiv 2\log(m)$ and its conjugate momentum to be $v\equiv \log(-\ell)$. These logarithmic coordinates are periodic, and have an identification $(U,v)\sim (-U,-v)$.

The coordinates $U$ and $v$ are easily expressed as sums and differences of edge parameters of tetrahedra in a triangulation of $M_K$ \cite{Thurston-1980, NZ}, and thus fit nicely into the general combinatorial framework of Section \ref{sec:rules} for constructing theories $T_{M_K}$ and indices $\CI_{M_K}$. In particular, any knot complement theory $T_{M_K}$ has a single $U(1)$ flavor symmetry, with twisted mass parameter $U$. Therefore, the index $\CI_{M_K}(m,e)$ depends on a single electric charge $e$ and magnetic flux $m$. We expect that the manifest $U(1)$ symmetry is actually enhanced to $SU(2)$, allowing knot complement theories to couple to the 4d theory $T[\pd M_K]=T[T^2]$, which is $\CN=4$ $SU(2)$ super-Yang-Mills. A classical indication of this enhancement appears in the $\Z_2$ Weyl symmetry of the phase space \eqref{PT2}. The enhancement would also imply that the index satisfies
\be \CI_{M_K}(m,e;q) = \CI_{M_K}(-m,-e;q)\,, \label{indZ2} \ee
which can be considered a quantum version of symmetry in the phase space.

We now give a few examples.

\subsubsection*{Figure-eight from two tetrahedra}

\begin{figure}[htb]
\centering
\includegraphics[width=6in]{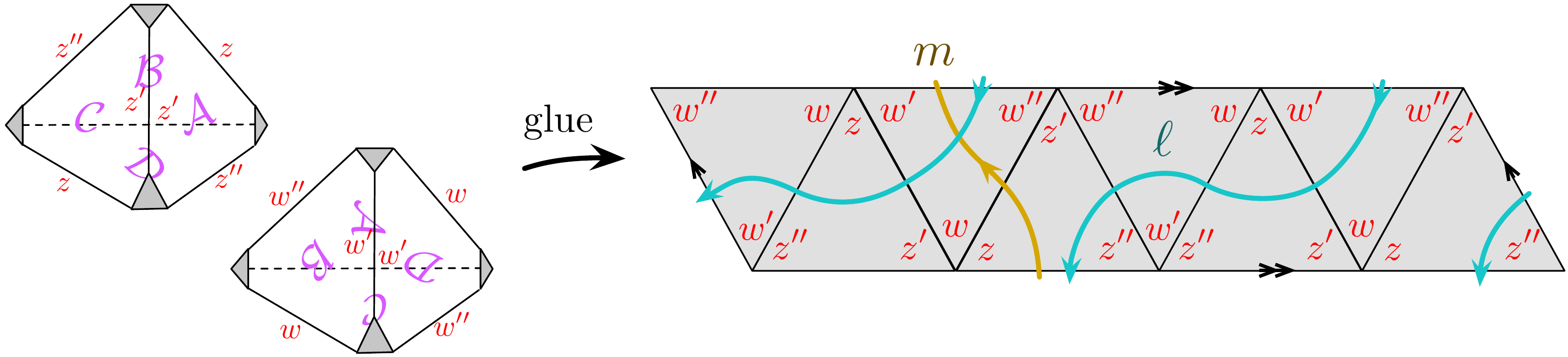}
\caption{The standard triangulation of the figure-eight knot complement}
\label{fig:fig8cusp}
\end{figure}

The standard triangulation of the figure-eight ($\mb{4_1}$) knot complement has two ideal tetrahedra, say $\Delta_Z$ and $\Delta_W$. All tetrahedron faces are glued together pairwise, and the $T^2$ boundary is made up from small, truncated ideal vertices of the tetrahedra (Figure \ref{fig:fig8cusp}). We find a meridian $U = Z-W''$ and a longitude $v= Z''-Z$. There are two internal edge coordinates, but only one of them is independent, and we can take it to be $C = 2Z''+Z'+2W''+W'$. Finally, the conjugate momentum to $C$ can be defined as (say) $\Gamma = -W''$. The change of polarization from $\Pi_Z\times \Pi_W$ to $\wt\Pi$ then becomes
\be \begin{pmatrix} U\\ C\\v\\\Gamma \end{pmatrix}
 = \begin{pmatrix} Z-W'' \\ 2Z''+Z'+2W''+W' \\ Z''-Z\\ -W'' \end{pmatrix}
 = \begin{pmatrix} 1&0&0&-1 \\ -1&-1&1&1  \\ -1&0&1&0 \\ 0&0&0&-1 \end{pmatrix}\begin{pmatrix} Z \\ W \\ Z'' \\ W''\end{pmatrix} + \begin{pmatrix} 0 \\ 2\pi i+\hbar \\ 0 \\ 0 \end{pmatrix}\,.
 \ee

Correspondingly, the product index $\CI_{\Delta_Z}(m_1,e_1)\CI_{\Delta_W}(m_2,e_2)$ gets transformed by the affine $Sp(4,\Z)$ action above to
\be \CI_{\{\Delta_i\},\tilde\Pi}(m_1,m_2,e_1,e_2) = \big(-q^{\frac12}\big)^{-2e_2}\CI_\Delta(m_1-e_2,m_1+e_1-e_2)\,\CI_\Delta(-m_2+e_1-e_2,-e_2)\,, \ee
and then after breaking the $U(1)$ symmetry associated to the internal edge $C$ we obtain
\begin{align} \CI_{\mb{4_1}}(m,e) &= \sum_{e_2\in \Z}\CI_\Delta(m-e_2,m+e-e_2)\,\CI_\Delta(e-e_2,-e_2)\,. \label{ind41}
\end{align}
It can be checked for any charges $(m,e)$ and to any order in $q$ that
\be \CI_{\mb{4_1}}(m,e) = \CI_{\mb{4_1}}(-m,e) = \CI_{\mb{4_1}}(m,-e) = \CI_{\mb{4_1}}(-m,-e)\,.\ee
For example,
\be \CI_{\mb{4_1}}(1,1)=\CI_{\mb{4_1}}(1,-1)= -q-q^2+2q^3+7q^4+11q^5+11q^6+3q^7+\dots\,, \ee
{\it etc.}
Thus, in addition to the Weyl symmetry \eqref{indZ2}, there is a parity-like symmetry that inverts the sign of a single charge. Geometrically, this is a result of the fact that the figure-eight knot complement is amphicheiral (\ie\ is equivalent to its mirror image).

\subsubsection*{Figure-eight from six tetrahedra}

In \cite{DGG}, we noted that the gauge theory arising from the simple, ``hard'' triangulation of the figure-eight knot complement above is a bit singular. Roughly, it is a $U(1)$ gauge theory with two chiral multiplets, both of charge $+1$. The $U(1)$ vector flavor symmetry (promoted to $SU(2)$) corresponds appropriately to the meridian coordinate $U$. However, the topological $U(1)$ symmetry (corresponding to the internal edge $C$) should be broken, and there appears to be no operator $\CO_I$ around that can break it.

To remedy this problem, one can use a refined, ``easy'' triangulation consisting of six tetrahedra, which leads to a perfectly good description of $T_{\mb{4_1}}$ with all desired operators present. We argued above that this refinement of triangulations should not be necessary in the calculation of the index, and we can now verify this. In Appendix \ref{app:6}, we use the six-tetrahedron decomposition to find the index. We have checked computationally that the more complicated expression \eqref{ind6} there agrees with \eqref{ind41} for $0\leq m \leq 3$ and $0\leq e\leq 3$, up to $7$th order in $q$. One should be able to prove the complete equivalence of these expressions using a sequence of 2--3 transformations on the index (or by using difference equations), but we do not do so here.

\subsubsection*{Trefoil}

The trefoil ($\mb{3_1}$) knot complement, like the figure-eight knot, has an ideal triangulation consisting of two tetrahedra $\{\Delta_Z,\Delta_W\}$. The triangulation is a bit asymmetric. The two internal edges have ``valency'' 2 and 10, with coordinates%
\footnote{The gluing data of this triangulation, and triangulations of any other knot or link complement, can be easily obtained from computer packages such as \texttt{snap} \cite{snap} or \texttt{SnapPy} \cite{SnapPy}.}
\begin{align} C_2 &= Z+W  \hspace{2.12in}\text{(2 dihedral angles)}\,, \notag \\
 C_{10} &= Z+2Z'+2Z''+W+2W'+2W''  \qquad\text{(10 dihedral angles)}\,.
\end{align}
Note that $C_2+C_{10}=4\pi i+2\hbar = (2\pi i+\hbar)\times$(\#\! tetrahedra). The meridian and longitude holonomies can be described as $U = W''-Z''$ and $v = -2Z''+2W''-W+i\pi+\tfrac\hbar2$. Then, using $C\equiv C_2$ as the independent internal gluing constraint, and taking $\Gamma = Z$ as its canonical conjugate, the change of polarization $\wt\Pi=g\circ(\Pi_Z\times\Pi_W)$ becomes
\be \begin{pmatrix} U\\ C\\v\\\Gamma \end{pmatrix}
 = \begin{pmatrix} -Z''+W''\\ Z+W \\ -2Z''+2W''-W+i\pi+\tfrac\hbar2\\ Z \end{pmatrix}
 = \begin{pmatrix} 0 & 0 & -1 & 1 \\
 1 & 1 & 0 & 0 \\
 0 & -1 & -2 & 2 \\
 0 & 0 & 1 & 0\end{pmatrix}\begin{pmatrix} Z \\ W \\ Z'' \\ W''\end{pmatrix} +\begin{pmatrix}0\\0\\ i\pi+\tfrac\hbar2\\0\end{pmatrix}\,.
\label{pol31}
\ee
From this, we compute the index and find a small surprise:
\begin{align} \CI_{\mb{3_1}}(m,e) &= \sum_{e_2\in\Z}\big(-q^{\frac12}\big)^{2e_2+m}\,\CI_\Delta(e-2m,e_2)\,\CI_\Delta(2m-e,e_2+m)\,.\notag \\
&= \delta_{e,3m}\,. \label{ind31} \end{align}
Alternatively, in a fugacity basis, we could write $\CI_{\mb{3_1}}(m;q,\zeta) = \zeta^{3m}$.

Such a simple index indicates that $T_{\mb{3_1}}$ could, for example, be a pure $\CN=2$ Chern-Simons theory in the IR. Properly verifying this guess would require refining the above triangulation of the knot complement, because it contains a ``hard'' internal edge $C_{10}$, and thus cannot be used to define $T_{\mb{3_1}}$. Nevertheless, the triangulation is perfectly reasonable for calculating the index. Due to the anticipated relation between the index and $SL(2,\C)$ Chern-Simons theory (Section \ref{sec:CS}), we actually expect that any torus knot complement has a delta-function index. For example, an $(a,b)$ torus knot with $ab$ even should have $\CI(m,e) = \delta_{e,\frac{ab}{2}m}$\,.

\subsection{Mapping tori and $T[SU(2)]$}
\label{sec:mtori}

One could also try constructing indices for some 3-manifolds without explicit reference to ideal triangulations.
Another popular construction of 3-manifolds starts with a Riemann surface $C$ and builds a 3-manifold
by identifying the ``top'' and ``bottom'' boundaries of a mapping cylinder $M_{\varphi} = C \times_{\varphi} I$,
\be
M \; = \; C \times_{\varphi} S^1 \equiv C \times_{\varphi} I \big/ (x,0) \sim (\varphi (x), 1) \,,
\label{maptorus}
\ee
with an element of the mapping class group $\varphi : C \to C$.
For example, if $C = T^2 \setminus \{ \text{pt} \}$ is a punctured torus,
the mapping class group is $PSL(2,\Z)$ generated by the $S$ and $T$ elements.
Many interesting 3-manifolds can be represented as punctured torus bundles over $S^1$,
{\it e.g.} the figure-eight knot complement is a punctured torus bundle over $S^1$ with monodromy
\be
\varphi \; = \; TS T^{-1} S^{-1} \,.
\ee

\begin{wrapfigure}{r}{2.2in}
\includegraphics[width=2in]{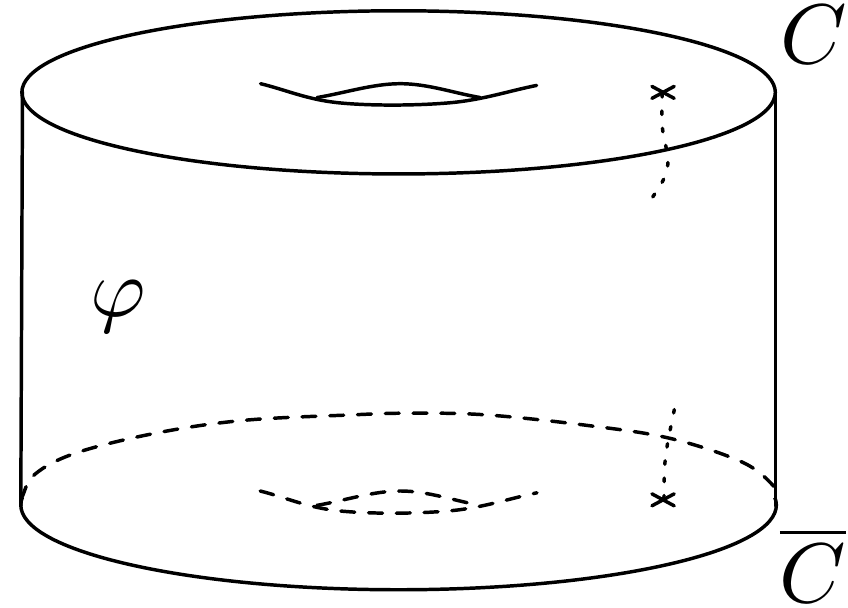}
\caption{A mapping cylinder for an element $\varphi$ acting on the punctured torus $C=T^2\bs\{p\}$.}
\label{fig:MCgeneric}
\end{wrapfigure}

This construction of 3-manifolds has a natural interpretation in combined 3d/4d system
as a periodic array of duality walls (determined by $\varphi$) in the 4d $\CN=2$ gauge theory $T[C]$ \cite{DGG-defects, Yamazaki-3d, DG-Sdual}.
For instance, if $C$ is a punctured torus,
then $T[C]$ is the so-called $\CN=2^*$ theory in four dimensions.
Moreover, every element $\varphi \in PSL(2,\Z)$ in this example
can be represented as a word (a sequence) of $S$ and $T$ generators,
each associated to a basic duality wall in the four-dimensional gauge theory.

Relegating further details of the combined 3d/4d system to Section \ref{sec:3d4d},
we can briefly summarize here the rules for calculating the index of 3d theories $T_M$,
at least in the large class of examples \eqref{maptorus} where $C$ is a punctured torus.
Roughly speaking, for every word $\varphi = \gamma_1 \cdot \gamma_2 \cdot \ldots$
in the basic duality generators $\gamma_i$ one can associate a periodic array of 3d theories $T_{\gamma_i}$
on the corresponding duality walls, such that
\be
\CI_{M_{\varphi}} \; = \; \text{``Tr"} \left( \CI_{\gamma_1} \CI_{\gamma_2} \ldots \right)\,,
\label{imtrace}
\ee
reflecting the geometry \eqref{maptorus} of the mapping torus $M_{\varphi}$.
For punctured torus bundles, one needs to describe only two duality walls that correspond to
the $S$ and $T$ generators of the mapping class group $SL(2,\Z)$.

The theory on the duality wall associated with $T^k$ transformation is very simple:
it simply carries a Chern-Simons action for the flavor symmetry at level $k$.
Hence, it contributes to the integrand of \eqref{imtrace} a factor
\be
\CI_{T^k} \; = \; \zeta^{k m}\,,
\ee
written in terms of the fugacity $\zeta$ and magnetic flux $m$.

Similarly, the $S$ transformation corresponds to a duality wall
described by a three-dimensional $\CN=2$ SQED with $N_f =2$ flavors,
also known as the mass-deformed theory $T[SU(2)]$:
\be
\begin{array}{l@{\;}|@{\;}ccccc@{\;}|@{\;}c@{\;}|@{\;}c}
\multicolumn{8}{c}{\text{theory}~T[SU(2)]} \\[.1cm]
& q_1 & q_2 & q_3 & q_4 & \phi_0 & \text{fugacity} & \text{flux} \\\hline
U(1)_{\rm gauge} & 1 & 1 & -1 & -1 & 0 & z & s \\
U(1)_{\rm bottom} & 1 & -1 & 1 & -1 & 0 & u & v \\
U(1)_{\rm puncture} & 1 & 1 & 1 & 1 & -2 & \alpha & m \\
U(1)_{\rm top} & 0 & 0 & 0 & 0 & 0 & w & n
\end{array}
\label{TSU2charges} \ee
Here, the global symmetry $U(1)_{\rm bottom}$ associated to the ``bottom'' boundary
of the mapping cylinder is actually a Cartan subgroup of the $SU(2)$ flavor symmetry group,
as suggested by the charge assignments \eqref{TSU2charges}.
This symmetry is gauged when one glues the bottom boundary to something else.
Similarly, the top boundary of the mapping cylinder shown on Figure \ref{fig:MCgeneric} corresponds to the topological symmetry $U(1)_{\rm top}$.
Furthermore, we denote the axial symmetry by $U(1)_{\rm puncture}$ since it corresponds to the puncture of $C=T^2\bs\{p\}$.

The index of this theory is
\begin{align}
\CI_{T[SU(2)]} \; = \; \chi(q \alpha^{-2}, -2m)
\sum_{s \in \Z} \int \frac{dz}{ 2 \pi i z} z^{n} w^{s}
& \; \chi (z \alpha u, m + s + v)
\; \chi (z^{-1} \alpha u^{-1}, m-s - v) \notag \\
& \cdot \chi (z \alpha u^{-1}, m + s - v)
\; \chi (z^{-1} \alpha u, m - s + v)
\label{tsu2index}
\end{align}
where $(w,n)$ are the parameters (fugacity and flux) for the topological symmetry $U(1)_{\rm top}$, and
\be
\chi (\zeta, m) \; = \; (q^{1/2} \zeta^{-1})^{-m/2} \CI_\Delta(\zeta,m)  \; = \;  (-1)^{\frac{m + |m|}{2}} (q^{1/2} \zeta^{-1})^{|m|/2}  \prod_{r=0}^\infty \frac{1- \q^{r+\frac12|m|+1}\zeta^{-1}}{1-\q^{r+\frac12|m|}\zeta}\,
\label{chiralchi}
\ee
is the contribution of a single chiral multiplet of R-charge $0$, which agrees up to a sign with {\it cf.} \cite{KW-index}.

The index of this theory is written in the charge basis as
\begin{align}
\CI_{T[SU(2)]} \; = \; \sum_{e_1,e_2} & q^{e_1 + e_2} \chi(e_1 + e_2,-2 m) \cr \cdot & \chi(\frac{e_\alpha - n+ e_u}{4} + e_1,m+s+v) \chi(\frac{e_\alpha + n -e_u}{4} + e_1,m-s-v)\cr \cdot & \chi(\frac{e_\alpha -n -e_u}{4} +e_2,m+s-v)\chi(\frac{e_\alpha + n+ e_u}{4} + e_2,m-s+v)
\label{tsu2indexem}
\end{align}
where
\be
\chi (e, m) \; = \; q^{-m/4} \CI_\Delta(e-m/2,m) \ee
is the index of a free chiral of R-charge $0$ in the charge basis.

Alternatively, adopting the results of \cite{KW-index} one can be easily evaluate the integral in \eqref{tsu2index}.
There are four sets of poles at
\be
z^{-1} =
\begin{cases}
\alpha u q^{\tfrac{|s + m + v|}{2} + j + \tfrac{1}{4}} & \\
\alpha u^{-1} q^{\tfrac{|s + m - v|}{2} + j + \tfrac{1}{4}} &
\end{cases}
\qquad \text{and} \qquad
z =
\begin{cases}
\alpha u^{-1} q^{\tfrac{|s - m + v|}{2} + j + \tfrac{1}{4}} & \\
\alpha u q^{\tfrac{|s - m - v|}{2} + j + \tfrac{1}{4}} &
\end{cases}
\ee
with $j \in \Z_{\ge 0}$.
Moreover, we can assume\footnote{Besides $|q|<1$, which is required for $q$-expansions to make sense,
this assumption also involves $|\alpha u q^{1/4}|<1$ and $|\alpha u^{-1} q^{1/4}|<1$.
The answer with parameters outside of this range can be obtained by analytic continuation.}
that these two groups correspond to poles outside and inside of the unit circle, respectively.
Therefore, taking the residues of the first group of poles for $n<0$ and the residues of the second
group of poles for $n>0$, one easily finds a closed form of expression \eqref{tsu2index}
that depends on three sets of parameters (fugacities and fluxes), namely $(\alpha,m)$, $(u,v)$ and $(w,n)$.

We might emphasize that although the index of $T[SU(2)]$ here was not defined with respect to a triangulation, there \emph{does} exist a triangulation of the appropriate mapping cylinder $M$ that can be used to construct both $T[SU(2)]=T_M$ and its index \eqref{tsu2indexem}. Describing the triangulation is a focus of \cite{DGGV-hybrid}.

\subsection{Quantum Lagrangian operators}
\label{sec:glueops}

Just as the index of the tetrahedron theory $T_\Delta$ is annihilated by two difference operators \eqref{teteqs}
\be \big(\hat p_{\pm}+\hat x_{\pm}{}^{-1}-1\big) \CI_\Delta(m,e;q) \,=\, 0\,,\ee
we find that the index of any 3-manifold theory $T_M$ satisfies pairs of difference equations
\be \hat \CL_M^{(i)}(\hat p_\pm,\hat x_\pm;q^{\pm1})\cdot\CI_M(m,e;q) \,=\, 0\,, \label{Meqs} \ee
with $\hat x,\hat p$ as in Section \ref{sec:ops}.
Generally, there are just as many pairs of equations as pairs of electric and magnetic flavor charges $(m_i,e_i)$ for $T_M$. Thus, once one knows $\CI_M(m,e;q)$ at finitely many values of $(m,e)$, \eqref{Meqs} completely determine the index everywhere. Moreover, the difference equations \eqref{Meqs} govern the asymptotics of the index in a fairly simple way --- for example, the behavior of $\CI_M(m,e)$ at large charges $(m,e)$, or as $q\to 1$. The equations always come in mutually commuting $\pm$ pairs due to the $\rho$ symmetry of the index.

Physically, difference equations for the index arise from identities in the algebra of line operators acting on $T_M$; these line operators will be the focus of Section \ref{sec:3d4d}. For now, we can understand the difference operators geometrically and combinatorially. The notation $\hat \CL_M$ in \eqref{Meqs} is meant to be suggestive. Indeed, geometrically, the operators $\hat \CL_M$ are just quantizations of the classical Lagrangians $\CL_M$ that describe the subset of flat $SL(2,\C)$ connections on the boundary $\pd M$ that can be extended as flat connections in the bulk:
\be \CL_M = \{\text{flat conn$^{s}$ on $\pd M$ that extend to $M$}\}\quad \subset \; \CP_{\pd M} \label{classLM} \ee
(\cf\ \eqref{tetLag}). Such a Lagrangian is generically cut out by $\tfrac12 \dim_\C \CP_{\pd M}$ polynomial equations $\CL_M^{(i)}(x,p)=0$ in the complex coordinates $(x_i,p_i)$ on $\CP_{\pd M}$, and each of these equations leads to a pair of operators $\hat\CL_M^{(i)}(\hat p_\pm,\hat x_\pm;q)$.

To explicitly construct the operators $\hat \CL_M^{(i)}$, we can translate the gluing rules for the index $\CI_M(m,e)$ from Section \ref{sec:rules} into gluing rules for operators. We find the following:
\begin{enumerate}

\item For a triangulation $M=\{\Delta_i\}_{i=1}^N$, begin with $N$ pairs of operators
\be \hat \CL_{\Delta_i}(\hat x_\pm,\hat p_\pm;q^{\pm 1}) = \hat p_{i\,\pm}+\hat x_{i\,\pm}^{-1}-1\,. \label{tetiops} \ee
Each pair annihilates a tetrahedron index $\CI_{\Delta_i}(m_i,e_i)$.

\item The collection of all $N$ pairs \eqref{tetiops} annihilates the product index $\CI_{\{\Delta_i\},\{\Pi_i\}}(m,e)$. We can say that the $\hat \CL_{\Delta_i}$'s define a left ideal in the algebra of operators generated by $\{\hat x_{i\,\pm},\hat p_{i\,\pm}\}_{i=1}^N$, with nontrivial commutation relations
\be \hat p_{i\,+}\hat x_{i\,+}=q\,\hat x_{i\,+}\hat p_{i\,+}\,,\qquad
p_{i\,-}\hat x_{i\,-}=q^{-1}\hat x_{i\,-}\hat p_{i\,-}\,.\ee
All elements of this left ideal annihilate the product index.

\item Change variables in the algebra of operators according to the change of polarization $\wt\Pi = g\circ\{\Pi_i\}$. That is, if $g=\sigma^R(\alpha)\,\sigma^F(\alpha)\,g_{Sp}$, define a new basis of logarithmic operators via the affine linear transformation
\be \begin{pmatrix} \hat X_{i\,\pm}'\\\hat P_{i\,\pm}' \end{pmatrix} = g_{Sp}\begin{pmatrix} \hat X_{i\,\pm}\\\hat P_{i\,\pm} \end{pmatrix} \pm \alpha\big(i\pi+\tfrac\hbar2\big)\,,\ee
as in Section \ref{sec:ops}.
We can then exponentiate to obtain the new basis of $q$-commuting operators $\hat x_{i\,\pm}'=\exp{\hat X_{i\,\pm}'},\, \hat p_{i\,\pm}'=\exp{\hat P_{i\,\pm}'}$\,.

\item Rewrite the tetrahedron Lagrangians \eqref{tetiops} in terms of the new $(\hat x_\pm',\hat p_\pm')$, obtaining $N$ pairs of operators
\be \hat \CL^{(i)}(\hat x_\pm',\hat p_\pm';q^{\pm 1})\,. \label{tetiops2} \ee
Due to the crucial intertwining property \eqref{intertwine} from Section \ref{sec:ops}, these $N$ pairs all annihilate the transformed product index $\CI_{\{\Delta_i\},\tilde\Pi}(m',e') = \big(-q^{\frac12}\big)^{\langle\alpha,\gamma'\rangle}\CI_{\{\Delta_i\},\{\Pi_i\}}(g_{Sp}^{-1}\gamma')$ of \eqref{Iprod2}. Note that the generators $(\hat x_\pm',\hat p_\pm')$ now act on $(m',e')$ as $\hat x_{i\,\pm}'=\exp\big(\tfrac\hbar2m_i'\mp\pd_{e_i'}\big)$ and $\hat p_{i\,\pm}'=\exp\big(\tfrac\hbar2e_i'\pm\pd_{m_i'}\big)$.

\item Finally, suppose that the addition of a superpotential $\sum_I \CO_I$ to $T_M$ breaks the $U(1)$ symmetries with electric charges $e_i'$ for $i=\tfrac12\dim\CP_{\pd M}+1,...,N$. Then, working in the left ideal defined by the $N$ pairs \eqref{tetiops2}, eliminate the corresponding $\hat p_{i\,\pm}'$, and set $\hat x_{i\,\pm}' = q^{\pm 1}$. (Working in a left ideal means that we are allowed to add and subtract operators \eqref{tetiops2}, and to multiply only on the \emph{left} --- since some index should be sitting on the right.) What remains are $\tfrac 12\dim \CP_{\pd M}$ pairs%
\footnote{We slightly oversimplify the counting for purpose of exposition: in general the result of elimination may be $\geq \tfrac 12\dim \CP_{\pd M}$ pairs of operators, even in the classical limit (\eg\ if the equations are not a complete intersection).} %
of operators that only involve the fundamental generators $(\hat x_{i\,\pm}',\hat p_{i\,\pm}')$ for $i=1,...,\tfrac12\dim\CP_{\pd M}$, corresponding to the charges $(e_i',m_i')$, $i=1,...,\tfrac12\dim\CP_{\pd M}$ for the unbroken $U(1)$ symmetries of $T_M$. Call these remaining pairs of operators
\be \hat \CL_M^{(i)}(\hat x_\pm',\hat p_\pm ';q^{\pm 1})\,,\qquad i=1,...,\tfrac12\dim \CP_{\pd M}\,. \label{rempairs} \ee
By construction, they will annihilate the final index $\CI_M(m',e')$

\end{enumerate}
The last step here --- elimination in an operator algebra --- may seem a little complicated. However, it follows directly from the final sum \eqref{indsum} defining the index $\CI_M$. For every broken $U(1)$ symmetry, we set some $m_i'=0$. Then it no longer makes sense to shift this charge $m_i'$, so $\hat p_{i\,\pm}'$ must be eliminated from \eqref{tetiops2}. Moreover, we multiply the index by $q^{e_i'}$ and sum over electric charge sectors $e_i'$ for the broken $U(1)$. Acting on this sum, $\exp(\pd_{e_i'})$ is equivalent to multiplication by $q^{-1}$. Therefore,
\be \hat x_{i\,\pm}' \;\to\; q^{\pm1} \,,\ee
just as dictated above. From now on, we will remove the ``primes'' from $(\hat x_\pm',\hat p_\pm')$ as well as from charges $(m',e')$ when discussing the final index $\CI_M$.

The rules found here for constructing the ``quantum Lagrangian operators'' \eqref{Meqs} are identical to the construction of quantized Lagrangians discussed in \cite{Dimofte-QRS}. More precisely, we find here two independent copies of the quantized Lagrangians of \cite{Dimofte-QRS}: one involving fundamental generators $(\hat x_+,\hat p_+)$ and a quantization parameter $q=e^\hbar$, and another involving generators $(\hat x_-,\hat p_-)$ and a quantization parameter $q^{-1}=e^{-\hbar}$. This correspondence forms the basis of our argument in Section \ref{sec:CS} that the index is an $SL(2,\C)$ Chern-Simons wavefunction.

To get a feeling for how quantized Lagrangians actually look, we can consider a few examples. First, let's take the bipyramid. By triangulating it into either 2 or 3 tetrahedra and applying the gluing rules above, we find two pairs of operators
\be \hat \CL_{\rm bip}^{(1)} = \hat p_{1\,\pm}+\hat x_{1\,\pm}^{-1}\hat p_{2\,\pm}-1 \,,\qquad \hat \CL_{\rm bip}^{(2)} = \hat p_{2\,\pm} + \hat x_{2\,\pm}^{-1}\hat p_{1\,\pm}-1\,.
\ee
that both annihilate the index $\CI_{\rm bip}(m_1,m_2,e_1,e_2) = \CI_\Delta(m_1-e_2,e_1)\,\CI_\Delta(m_2-e_1,e_2)$ from \eqref{ind2}. A complete, detailed derivation of these Lagrangian operators appears in Appendix \ref{app:L}.

In the case of a knot complement $M_K$, we saw that the theory $T_K$ has a single $U(1)$ symmetry. In knot theory, the single equation that cuts out the classical Lagrangian $\CL_K$ is usually called the A-polynomial of $K$ \cite{cooper-1994}, and correspondingly $\hat \CL_K$ is the ``quantum A-polynomial'' \cite{garoufalidis-2004, gukov-2003}. Conforming to the knot theory literature, let us denote the exponentiated operators acting on the index $\CI_K(m,e)$ as
\be \hat M_\pm = e^{\hat U_\pm} \equiv \hat x_\pm = \exp\big(\tfrac\hbar2m\mp\pd_e\big)\,,\qquad  \hat\ell_\pm=-e^{\hat v_\pm} \equiv -\hat p_\pm =-\exp\big(\tfrac\hbar2e\pm\pd_m\big) \,. \ee
Then, for example, it is easy to see that the index for the trefoil $\CI_{\mb{3_1}}(m,e) = \delta_{e,3m}$, is annihilated by
\be \hat \CL_{\mb{3_1}\,+} = \hat \ell_++q^{\frac32}\hat M_+^3\,,\qquad \CL_{\mb{3_1}\,-} = \hat \ell_-+q^{-\frac32}\hat M_-^3\,,
\ee
which are both quantizations of the classical A-polynomial $A =\ell+M^3$.

The figure-eight knot is a little less trivial. Following the above gluing rules (detailed in Appendix \ref{app:L}) leads to an operator
\be \hat \CL_{\mb 4_1} = \big(q^{\frac12}\hat M-q^{-\frac12}\hat M^{-1}\big)\ell^{-1}-\big(\hat M-\hat M^{-1}\big)\big(\hat M^{-2}-\hat M^{-1}-q-q^{-1}-\hat M+\hat M^2\big)+\big(q^{-\frac12}\hat M-q^{\frac12}\hat M^{-1}\big)\hat\ell \label{A41M} \ee
in its `$+$' version (with `$+$' subscripts suppressed). This is the well known quantum A-polynomial of the figure-eight knot \cite{garoufalidis-2004}, in the normalization of \cite{DGLZ, Dimofte-QRS}. It can be checked computationally that both \eqref{A41M} and its `$-$' version, obtained by sending $\hat M\to \hat M_-$, $\hat \ell\to\hat \ell_-$, $q\to q^{-1}$, annihilate the index $\CI_{\mb{4_1}}(m,e)$ in \eqref{ind41}. We emphasize that the above gluing rules actually \emph{prove} algebraically that this must be the case.

\subsection{Tentacles and vacua}
\label{sec:tent}

Some interesting physical consequences of the difference equations \eqref{Meqs} result from the fact that they control the behavior of the index at large charges $(m,e)$, in a fairly simple manner. There are actually two ways to send $m$ and/or $e$ to infinity: we can either keep $|q|<1$ fixed, or simultaneously send $m\to \infty$ and $q\to 1$ ($\hbar\to 0$) so that $q^m$ stays fixed. In the latter case, the index $\CI_M$ diverges, with leading asymptotics governed by the volume of $M$. Mathematically, this is a familiar phenomenon, closely related to the ``Volume Conjecture'' for Chern-Simons partition function. Physically, it seems closely related to $Z$-extremization \cite{Jafferis-Zmin}, though the precise connection is still not well understood.

We will mention some aspects of the $q\to 1$ limit at the end of this section. For now, let's instead consider the less familiar limit $(m,e)\to \infty$ with $|q|<1$ fixed. The leading behavior of the index%
\footnote{The leading behavior is all we will look at here. It would be very interesting to also consider subleading corrections and their physical implications.} %
in this limit is simply governed by the \emph{classical} Lagrangian $\CL_M(x,p)$ \eqref{classLM}, and turns out to detect the presence of unconstrained chiral operators in the theory $T_M$. Equivalently, the leading behavior can detect when $T_M$ has a moduli space of vacua.

In order to make a more precise statement, note that superconformal theories $T_M$ (and more generally theories in class $\CR$) should have an index $\CI_M(m,e;q)$ such that the powers of $q$ appearing at fixed $(m,e)$ are bounded from below (\cf\ Section \ref{sec:actions}). Then we can define $\lead_M(m,e)$ to be the lowest power of $q$ the appears in $\CI_M(m,e;q)$ --- this is the first nontrivial $R+\tfrac{j_3}{2}$ contribution of an operator with charge $e$ in the presence of flux $m$. We claim that $\lead_M(m,e)$ generically grows quadratically as $m,e\to \infty$, but that restricted to special rays in charge space it may instead grow linearly. These are exactly the rays along which the ``amoeba'' of $\CL_M$ has a ``tentacle.'' Moreover, when we are in an $Sp(2N,\Z)$ duality frame such that a ray/tentacle lies in a purely electric direction, the theory $T_M$ should have an unconstrained chiral operator $\CO$ that prodces the leading contribution to the index, and parametrizes a 3d moduli space of vacua.

\begin{figure}[htb]
\centering
\hspace{-.4in}\includegraphics[width=6.5in]{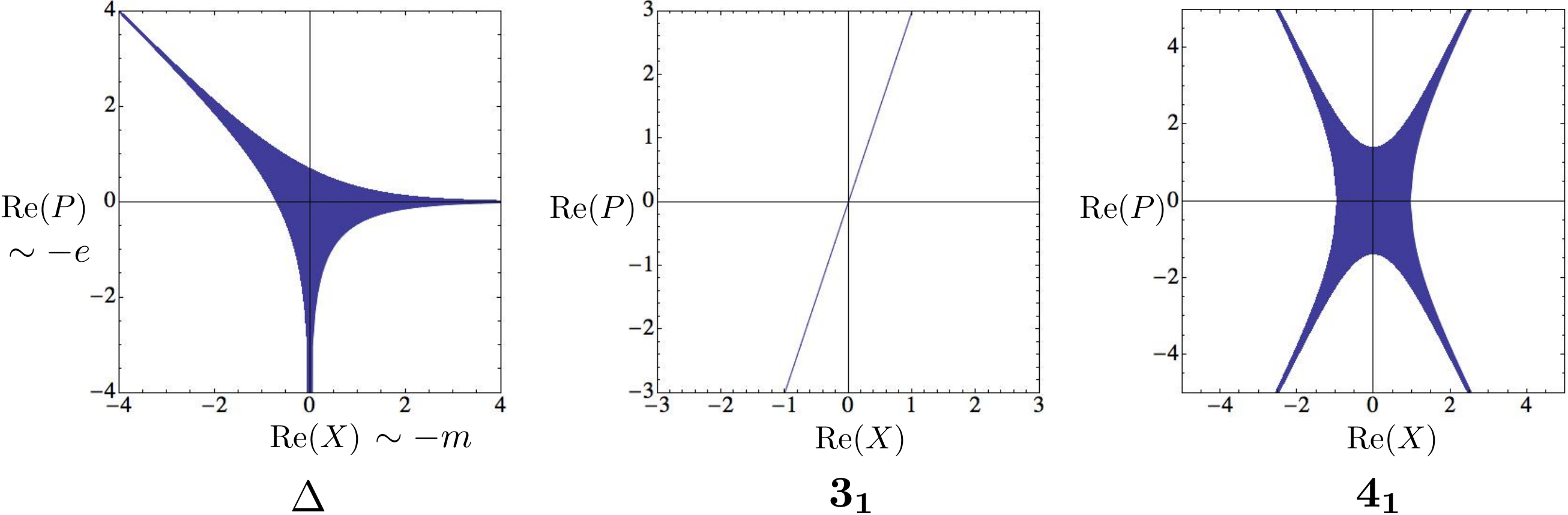}
\caption{Amoebas for the tetrahedron, trefoil, and figure-eight knot Lagrangians.}
\label{fig:amoebas}
\end{figure}

We will try to motivate these statements physically and mathematically. To begin, let us recall the formal definition of an amoeba \cite{GKZ-amoeba}. Given an algebraic variety in $(\C^*)^{2N}$, say with coordinates $(x_i,p_i)$, the amoeba is its projection to the real subspace spanned by the magnitudes
\be \Re X_i = \log|x_i|\,,\qquad \Re P_i = \log|p_i|\,.\ee
For example, the classical Lagrangians $\CL_M$ for the tetrahedron, trefoil, and figure-eight knot,%
\footnote{It is easy to see that these are classical $q\to 1$ limits of the corresponding operators in Section \ref{sec:glueops}, with $x=M$ and $p=-\ell$ in the case of knot complements. For the figure-eight knot, one must throw out an extra factor of $(x^2-1)$ that only arises in \eqref{A41M} as a quantum correction. Evidently, this factor is not relevant for analyzing vacua or flat directions in the index.}%
\begin{subequations}
\begin{align} &\CL_\Delta\,=\, p+x^{-1}-1 = 0\,, \\
 &\CL_{\mb{3_1}} \,=\,  p-x^3 = 0\,,\\
 &\CL_{\mb{4_1}} \,=\,  p^{-1}+(x^{-2}-x^{-1}-2-x+x^2)+p=0\,,
\end{align}
\end{subequations}
have the amoebas shown in Figure \ref{fig:amoebas}. A salient feature of the amoebas is that they extend asymptotically along a finite collection of semi-infinite rays, or tentacles. These tentacles occur whenever there is a solution to the defining equations of $\CL_M$ in the limit $(|x|,|p|)=(e^{nr},e^{ns})$, $n\to \infty$, for some nontrivial integer slope vector $(r,s)$.

The tentacles of amoebas in $\CL_M$ actually arise physically as parameter spaces of vacua for the 3d theories $T_M$ in flat $\R^3$. One indirect but instructive way to see this is to recall from \cite{DG-Sdual} or \cite{DGG} that the entire complex Lagrangian $\CL_M$ is the SUSY parameter space for $T_M$ compactified on (untwisted) $\R^2\times S^1_R$. This effective 2d $\CN=(2,2)$ theory has a complex, periodic twisted mass associated to each $U(1)$ global symmetry, obtained by complexifying the 3d real mass with the Wilson line of the background $U(1)$ gauge field. We can multiply this mass by the radius $R$ to make it dimensionless, and express it as
\be X = R\,m_{3d} +i\oint_{S^1_R} A\,. \ee
The 2d theory also has a complexified, periodic FI parameter (or moment map), naturally obtained by combining the 3d FI parameter $\xi_{3d}$ and the 2d background $\theta$-angle:
\be P = R\,\xi_{3d} +i\theta\,. \ee
The 2d twisted superpotential preserves supersymmetry when
\be \exp\left( \frac{\pd \wt \CW}{\pd X}\right) = \exp(P) \label{WXP} \ee
(put differently, the FI parameter $P$ contributes to $\CW$ as $XP$), and it was argued in \cite{DG-Sdual,DGG} that \eqref{WXP} are just the defining equations for $\CL_M$.

To lift back up to a SUSY parameter space in 3d, we can first restrict to the real parts of $X$ and $P$, and then send $R\to\infty$. The first operation projects $\CL_M$ onto its amoeba, and the second scales the amoeba so that only the tentacles --- a collection of semi-infinite rays --- are left. Sufficiently far along these rays, $\Re\,X$ and $\Re\,P$ (or rather $m_{3d}$ and $\xi_{3d}$) are the 3d masses and FI parameters consistent with SUSY. Near the origin, quantum corrections may still play an interesting role, which will not affect the current story.%
\footnote{In 2d, quantum corrections are responsible for smoothing the asymptotic regions of the parameter $\CL_M$ into a connected algebraic variety. Otherwise, $\CL_M$ would be a collection of ``cigars'' centered around each tentacle.} %
We emphasize that the tentacles map onto a \emph{parameter} space of vacua. However, if any of the tentacles happen to align with the $P$ axis (or plane) at $X=0$, an actual SUSY moduli space in $T_M$ also opens up. This is because the effective background FI parameter $\Re\,P$ ($\xi_{3d}$) is a real moment map for a $U(1)$ symmetry. Then, for example, SUSY requires that $\xi_{3d}$ sets the vev for a sum of chiral fields $\sum_j Q_j|\phi_j|^2$, where $Q_j$ are the $U(1)$ charges. Preserving SUSY at any value of $\Re\,P\sim \xi_{3d}$ while simultaneously having zero mass $\Re\,X\sim m_{3d}$ means there must be an infinite flat direction in dynamical field space.

This phenomenon is simple to illustrate in the tetrahedron theory. Its amoeba does have a tentacle on the negative $\Re\,P$ axis (or $|p|\to 0$). Correspondingly, $T_\Delta$ has a free chiral operator $\phi_Z$ charged under the single global $U(1)$, whose vev parametrized a moduli space. We have $|\phi_Z|^2+\xi_{3d}=0$, or $|\phi_Z|^2\sim -\xi_{3d}\sim -\Re\,P$. The other tentacles of the amoeba lie along $|x|\to\infty$ and $|p|\to \infty,\,|x|\to 0$. In either case, we can happily preserve SUSY, though the field $\phi_Z$ becomes massive. If we apply a $T^k\in Sp(2,\Z)$ transformation to $T_\Delta$ to shift the Chern-Simons level by $k$, the story remains essentially the same; we clearly still get a moduli space, while the amoeba gets ``skewed'' horizontally without lifting its tentacle from the $P$-axis. (Recall that $Sp(2,\Z)$ simply acts by multiplication on the symplectic vector ${X\choose P}$, so $T^k:{X \choose P}\mapsto {X \choose {P+kX}}$.) On the other hand, generic $Sp(2,\Z)$ images of $T_\Delta$ have no moduli space at all.

Now, let us return to the index. Suppose that $T_M$ has a moduli space corresponding to a tentacle of $\CL_M$ at $\Re\,X=0$, along a ray in the $\Re\,P$ plane pointing in direction $s$ (with $s$ an $N$-dimensional vector of coprime integers). The direction $s$ selects a $U(1)$ symmetry inside $U(1)^N$ whose moment map can be nonzero. We would then expect that the moduli space is parametrized by the vev of a chiral operator $\CO$ with electric charge $e=-s$. This operator and its powers $\CO^n$ can potentially contribute to the index. It is not completely clear that the contribution will be the dominant one (at leading order in $q$), but if it is, then
\be \lead(0,-ns) = \frac{R_\CO}{2}n\,,\qquad n\gg 0\,,\ee
where $\CR_\CO$ is the R-charge of $\CO$. This behavior should at east hold true for sufficiently large $n$. Thus the index (potentially) develops an asymptotically ``flat'' direction along the tentacles of the amoeba.

\begin{wrapfigure}{r}{2.7in}
\includegraphics[width=2.5in]{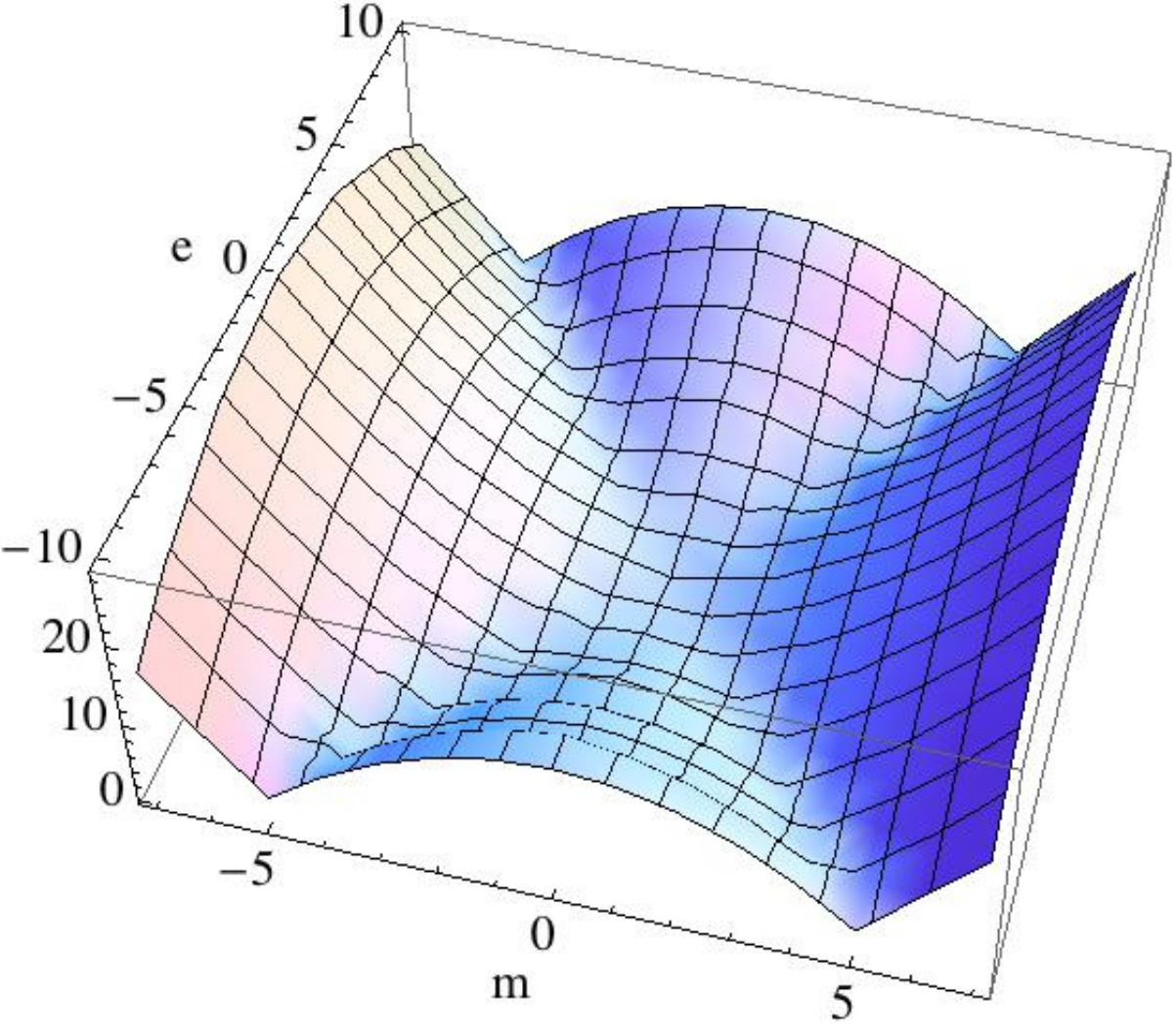}
\caption{Graph of the leading exponent ${\rm lead}(m,e)$ for the figure-eight index.}
\label{fig:lead41}
\end{wrapfigure}

Here we used the presence of a moduli space to argue for linear growth of $\lead(m,e)$ in a distinguished electric direction --- corresponding to a tentacle in the $P$ plane. However, an $Sp(2N,\Z)$ transformation could be used to align \emph{any} tentacle with the $P$ plane. Since $Sp(2N,\Z)$ acts covariantly on the index, preserving the growth of $\lead(m,e)$, this implies that we could actually expect linear growth along every tentacle, in any direction. Explicitly, if a tentacle extends in direction $(r,s)$, so that $(|x|,|p|)\sim (e^{nr},e^{ns})$, $n\to\infty$, lies in $\CL_M$, then
\be \lead(-rn,-sn)\,\sim\,n\,,\qquad n\to \infty\,. \label{leadrs} \ee
It is easy to see from Figure \ref{fig:lead} that this holds for $T_\Delta$: flat directions (anti)align with tentacles. In Figure \ref{fig:lead41}, we similarly plot $\lead_{\mb{4_1}}(m,e)$ for the figure-eight knot.%
\footnote{The tentacles and flat directions of knot complements always occur back-to-back, due to the $\Z_2$ Weyl symmetry \eqref{indZ2}. This is nicely illustrated in Figures \ref{fig:amoebas} and \ref{fig:lead41}.} %
For the trefoil, \eqref{leadrs} holds in a trivial way: the index $\CI_{\mb{3_1}}=\delta_{e,3m}$ vanishes except right on the tentacles, where it is constant.

The behavior \eqref{leadrs} is (roughly) predicted mathematically by the general structure of the difference equations $\hat \CL_M$, and the fact that they are quantizations of the classical Lagrangian $\CL_M$. The simple property that the operators $\hat\CL_M$ are polynomials in $\hat x_\pm,\,\hat p_\pm,$ and $q$, along with the assumption that indices $\CI_M(m,e;q)$ are $q$-series bounded from below, implies that the generic growth of $\lead(m,e)$ is quadratic (\cf\ \cite{Garoufalidis-slopes}). This is essentially because the operators $\hat x_\pm, \hat p_\pm$ only involve linear factors $q^{\frac m2}$ and $q^{\frac e2}$, and can also only generate other linear factors when shifting $m$ and $e$ in quadratic parts of the index. For example, $\hat x_+\cdot q^{e^2} = (q^{\frac m2-2e-1})q^{e^2}$. All these linear factors then enable cancellations between different terms in the equations $\hat\CL_M\CI_M=0$.

Nevertheless, linear growth of the index is permitted along directions (anti)parallel to tentacles of the amoeba of $\CL_M$. The tentacles are normal vectors to boundary surfaces of the Newton polygon of $\CL_M$, and also (roughly) the Newton polygon of $\hat \CL_M$. This allows extra cancellations to happen in the difference equations, along these distinguished directions. A more detailed, albeit heuristic, explanation is given in Appendix \ref{app:tent}.

One potential application of the observed growth rates of the index is to shed some light on the convergence of sums \eqref{indsum} that define $\CI_M$. Specifically, suppose that we start with a triangulation of $M$ and first construct a product index $\CI_{\{\Delta_i\},\{\Pi_i\}}=\CI_{\Delta_1}\times\cdots\times \CI_{\Delta_N}$. We then transform this to a polarization $\wt\Pi$, in which all internal edges of $M$ are electric. If we chose a refined, ``easy'' triangulation --- so that it properly defines a theory $T_M$ --- there must be a chiral operator $\CO_I$ associated to every internal edge, and charged under the edge coordinate's $U(1)$ symmetry. Moreover, before doing the final index sum \eqref{indsum}, we shift R-charge so that every $\CO_I$ has $R_{\CO_I}=2$. In this affine $Sp(2N,\Z)$ frame, we know that the leading exponent function $\lead_{\{\Delta_i\},\tilde\Pi}(m,e)$ of the product index $\CI_{\{\Delta_i\},\tilde\Pi}(m,e)$ should grow linearly in any electric direction $e_I$ corresponding to the charge of $\CO_I$, with slope $R_{\CO_I}/2$. In the negative direction $-e_I$, we generically expect quadratic growth instead. To compute the final index $\CI_M$, we set magnetic charges $m_I \to 0$ (just like in the analysis above), and sum along both positive and negative $e_I$ directions. We would then expect the sum not only to converge, but to do so ``uniformly'': in order to compute $\CI_M(m,e)$ at fixed external charges $(m,e)$ to $n^{\rm th}$ order in $q$, we should need to sum only from about $-ne_I$ to $ne_I$ in each direction. This has certainly been observed in examples. \\

To close this discussion, let us also return to the other, 't Hooft-like, asymptotic limit of the index. Rather than taking $(m,e)\to\infty$ along a ray, with $|q|<1$, we send both $m\to\infty$ and $q\to 1$ ($\hbar\to 0$) with $q^m$ held fixed. Working in an electric fugacity basis, we can use $\zeta$ to complexify the naturally real number $q^m$, and set
\be x = q^m\zeta \ee
as in \eqref{tetasymp}. Then, just as we found in Section \ref{sec:T1} that the tetrahedron index is dominated approximately by the hyperbolic volume of a tetrahedron,
\be \CI_\Delta(m;q,\zeta)\;\sim\; \exp\left(\frac{2}{i\hbar}V_\Delta(z)+\ldots\right)\,,\qquad V_\Delta(z) = -\Im\,\Li_2(z^{-1})\,,\ee
we now find that the index $\CI_M$ is dominated approximately by the hyperbolic volume%
\footnote{If $M$ does not admit a deformed hyperbolic structure with specified boundary conditions, the formula still holds, but ``volume'' means volume of a flat $SL(2,\C)$ connection.} %
of $M$,
\be \CI_M(m;q,\zeta)\;\sim\;\exp\left(\frac{2}{i\hbar}V_M(x)+\ldots\right)\,. \label{volM}\ee
Note that this volume depends on boundary conditions $x$. For example, the volume depends on the external dihedral angles of a geodesic boundary, or on the metric (or $SL(2,\C)$) holonomy around a cusp boundary of $M$. It differs very slightly from the actual hyperbolic volume, but the difference is easy to correct: the actual volume is
\be {\rm Vol}_M(x) = V_M(x) + (\Im\,P)\cdot(\Re\,X)\,, \ee
where $(x,p)=(e^X,e^P)$ is a point on the Lagrangian $\CL_M$ (a solution to the defining equations for $\CL_M$ at fixed $x$). Note that while $V_M(x)$ and $\Im\, P$ have branch cuts, ${\rm Vol}_M(x)$ should be well defined.

One way to check \eqref{volM}, up to a constant, is to note that the leading asymptotics of the index must be governed by the classical $\hbar\to 0$ ($q\to 1$) limit of the difference operators $\hat \CL_M$. In particular, the point $(X,\,P = \frac{\pd V_M}{\pd X})$ must lie on the classical $\CL_M$. This property is known to characterize volumes of 3-manifolds \cite{NZ}. Physically, the asymptotics \eqref{volM} are extremely familiar from the study of ellipsoid partition functions $\CZ_b[T_M]$ and twisted $\R^2\times S^1$ partition functions of $T_M$, both of which are very closely related to the index. We will discuss the relations further in Section \ref{sec:6d}.

\subsection{Mutation invariance from gauge theory}
\label{sec:mutant}

\begin{figure}
\centering
\includegraphics[width=4.8in]{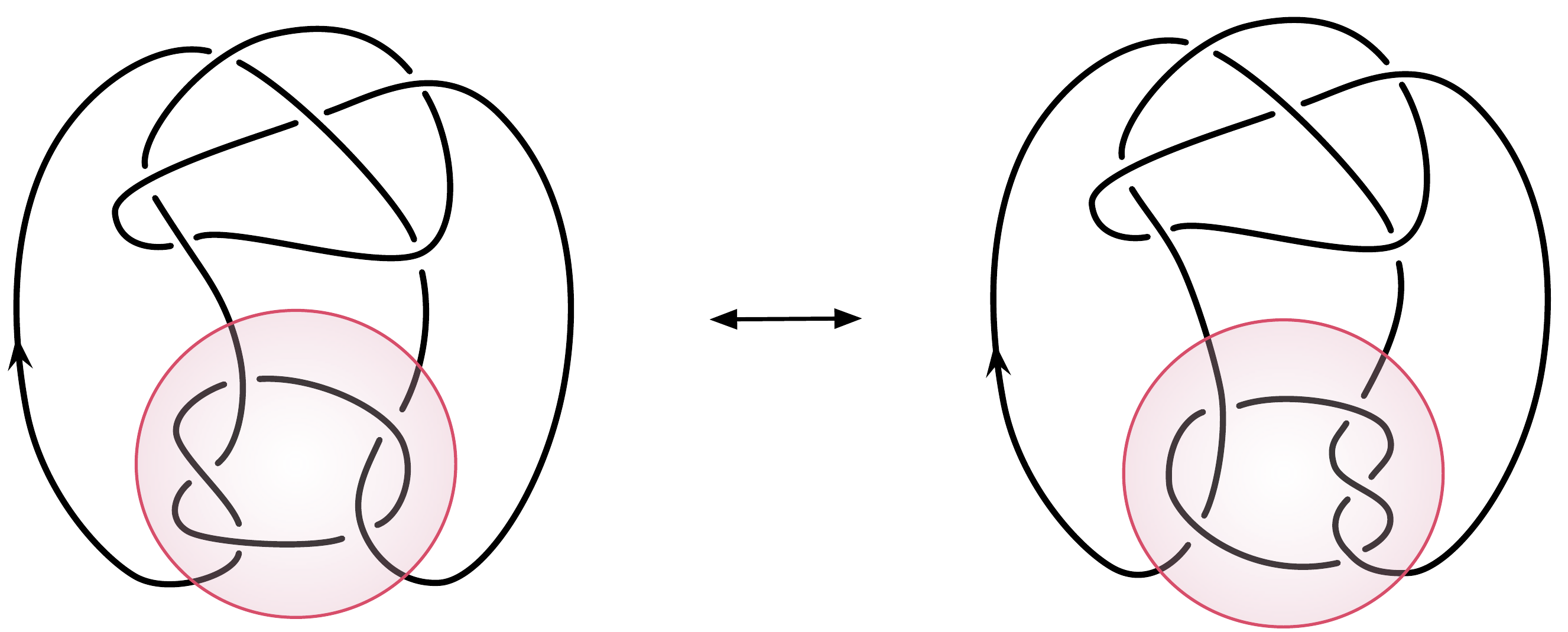}
\caption{The Kinoshita-Terasaka (left) and Conway (right) knots, related by a nontrivial mutation operation.}
\label{fig:mutation}
\end{figure}

In this section, we begin to test the strength of the index as a topological invariant of 3-manifolds --- \ie\ its ability to distinguish 3-manifolds from one another. We will specialize to the case of knot complements $M$, either in $S^3$ or in some closed 3-manifold $\ol M$, and consider the operation of mutation \cite{Conway-mutation}. To perform a mutation, the knot complement $M$ is cut along an $S^2$ that is punctured exactly four times by the knot, separating $M$ into two halves%
\footnote{It is perfectly possible that $S^2$ does not split $M$ in two, but rather into a manifold $\tilde M$ with two $S^2$ boundaries.
Everything we say in this section will go through with very minor modifications in that case.
} %
$M_1$ and $M_2$; then the halves are glued back together with a ``$180^\circ$ rotation'' along some axis of $S^2$, forming a new manifold $M_\mu$. More precisely, the rotation of $S^2$ interchanges two pairs of punctures. This is illustrated in Figure \ref{fig:mutation} for the simplest topologically distinct mutant pair of knot complements.

It is famously known that mutation acts trivially on colored Jones polynomials, the $SU(2)$ Chern-Simons wavefunctions of knot complements \cite{MortonTraczyk} (see also \cite{StoimenowTanaka}). We have also mentioned several times that the index should correspond to a complex $SL(2,\C)$ Chern-Simons wavefunction. Thus a natural expectation might be that the index is mutation invariant. We will prove this physically below, by using the $6d$ realization of theories $T_M$, and also by arguing that mutation is implemented by a 3d superpotential deformation that leaves the index untouched. First, however, we can do a simple check. The triangulations of the KT and Conway knot complements in Figure \ref{fig:mutation} are easily found with the program \texttt{SnapPy} \cite{SnapPy}. In their simplest (non-univalent) form, they have 12 tetrahedra each, leading to an expression for the corresponding indices as 11-dimensional sums. Then we find, for example, that%
\footnote{As per the linear/quadratic behavior of the index explained in Section \ref{sec:tent}, evaluating these indices to order $q^n$ requires summing over an 11-dimensional cube with sides approximately of length $2n$. This is feasible for low order. The computation in \eqref{ConKT}, done naively, required $11^7$ terms and took 28 hours per knot.}
\be \CI_{\rm KT}(0,0;q) = \CI_{\rm Con}(0,0;q) = 1 + 6 q - 32 q^2 + O(q^3)\,. \label{ConKT} \ee
If the indices $\CI_{\rm KT}$ and $\CI_{\rm Con}$ were to differ, they could only do so by a finite number of $q$-dependent normalizations --- since the colored Jones polynomial determines operators $\hat \CL_M$, and these in turn fix the index up to $q$-dependent boundary conditions. Thus, evaluating the index even at $(m,e)=(0,0)$ as in \eqref{ConKT} is quite a nontrivial check of mutation invariance.

\begin{wrapfigure}{r}{2.6in}
\includegraphics[width=2.7in]{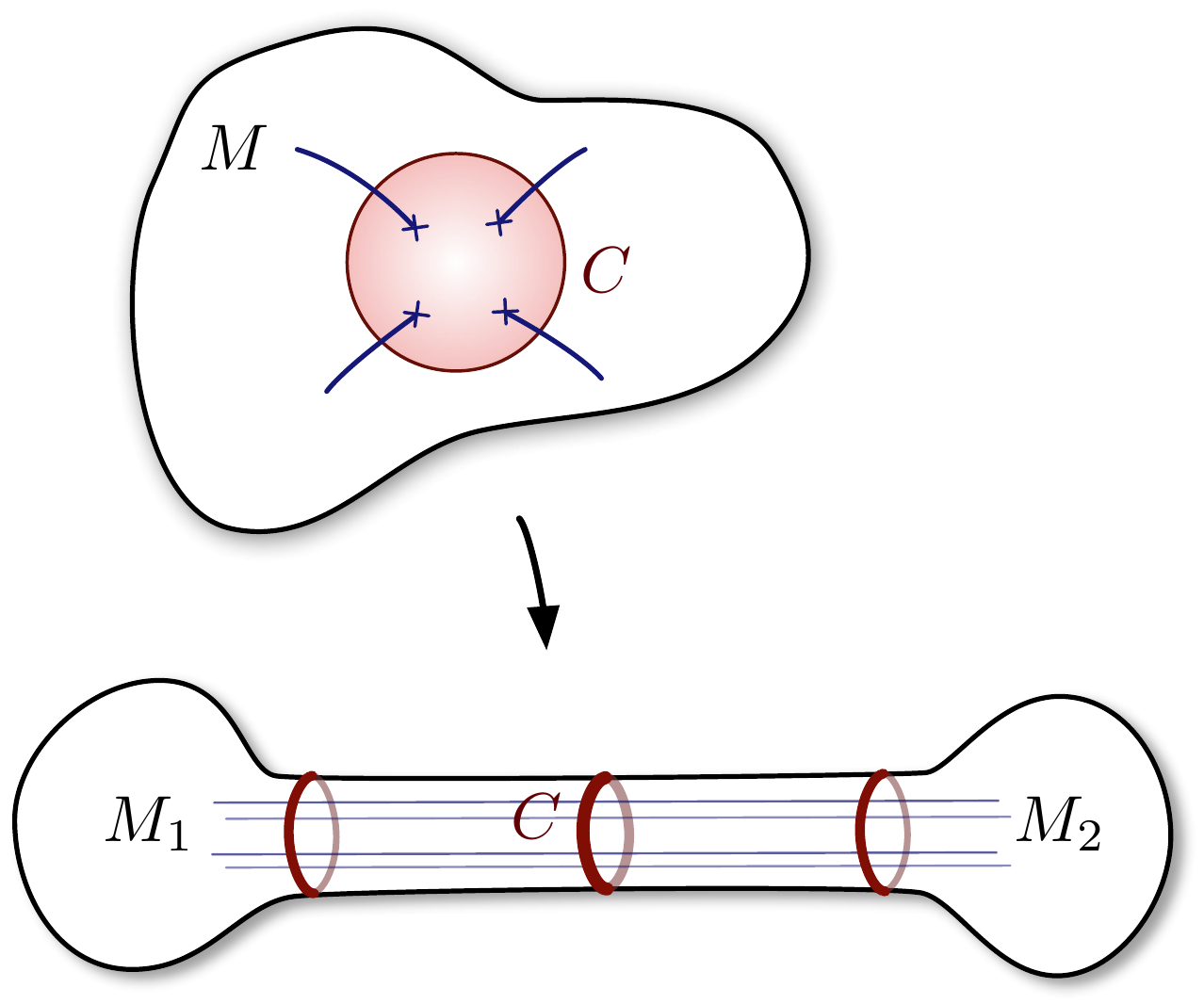}
\caption{Stretching $M$ along $C$. Compactification on $C$ gives a 4d theory $T[C]$ on $\R^3\times I$.}
\label{fig:Mstretch}
\end{wrapfigure}

Now, let us proceed to the six-dimensional argument. Consider the six-dimensional $A_1$ theory on a three-manifold $M$, and identify
a two-sphere $C=S^2$ in $M$ that intersects exactly four strands of codimension 2 regular defects.
We will denote as $M_1$ and $M_2$ the two halves of $M$ joined along $C$.
We can put a metric on $M$ that elongates a neighborhood of $C$ into a long cylinder of cross-section $C$ (Figure \ref{fig:Mstretch}).
At low energy, the six-dimensional setup admits a simple four-dimensional description, as
a four-dimensional $\CN=2$ gauge theory on a segment, with boundary conditions $\CB_1$ and $\CB_2$ at the endpoints that are determined by the geometry of $M_1$ and $M_2$. The four-dimensional theory is $SU(2)$ gauge theory with $N_f=4$ fundamental hypermultiplets.
In the far IR, the four-dimensional setup will flow back to $T_M$.

This cutting construction, of course, can be done for any surface $C$, but the four-punctured sphere is special. While for generic $C$
the four-dimensional theory has an $SU(2)$ flavor symmetry for each puncture, inherited from the flavor symmetry of codimension-$2$ defects in the six-dimensional theory, the $SU(2)$ $N_f=4$ theory has an $SO(8)$ flavor symmetry.
The flavor symmetries with a six-dimensional origin compose an $SO(4) \times SO(4)$ block-diagonal subgroup of $SO(8)$.
The remaining $16$ generators of $SO(8)$ are accidental IR symmetries that appear in the four-dimensional limit.
As $T_M$ can be defined by the four-dimensional setup, the enhanced $SO(8)$ flavor symmetry has important consequences for~$T_M$.

One important consequence is that any boundary condition $\CB$ for $SU(2)$ $N_f=4$ theory
sits in an $SO(8)$ orbit of boundary conditions $g\circ \CB$, $g \in SO(8)$.
In principle it is possible to imagine an $SO(8)$-invariant boundary condition. For example:
set Dirichlet boundary conditions for the gauge fields,
split the $16$ chiral fields in the hypermultiplets into sets $X^i$, $Y^i$, each in a vector representation of $SO(8)$
($X^i$ and $Y^i$ are the top and bottom components respectively of the $8$  $SU(2)$ doublets),
and then give Dirichlet b.c. to $X^i$ and Neumann b.c. to $Y^i$. However, we are not aware of a geometric realization for such a boundary condition.
Boundary conditions $\CB$ obtained from a generic six-dimensional configuration will typically break $SO(8)$ down to $SU(2)^4$,
or even a smaller subgroup $H_\CB$, and hence will sit in a non-trivial continuous family $SO(8)/H_{\CB}$ of boundary conditions.

Although it will not be too important for us, it is simple to show that these exactly marginal deformations of an $\CN=2$ supersymmetric boundary condition
$\CB$ are actually superpotential deformations. Indeed, they correspond to deformations of the boundary conditions for the hypermultiplets only, and
can be implemented by adding superpotential terms at the boundary. The true space of exactly marginal deformations of $\CB$ will be some complex manifold
that locally is the complexification of  $SO(8)/H_{\CB}$. We do not have a good handle over the full space of exactly marginal deformations,
but for our purpose $SO(8)/H_{\CB}$ is sufficient.

Notice that even if $\CB$ has a six-dimensional realization as some cobordism with a single non-empty boundary $C\,(\times \R^3)$, generically the boundary condition $g \circ \CB$ may not have such a realization. There is an exception, though: if $g$ is an $SO(8)$ rotation that permutes the four $SU(2)$'s among themselves, then $g \circ \CB$ can be also realized by the same cobordism as $\CB$, with the four punctures suitably permuted.
There is an obvious example: the diagonal matrix $(1,1,1,-1,1,1,1,-1)$ acts as a reflection in each $SO(4)$,
and therefore permutes the two $SU(2)$'s in each $SO(4)$.
Thus we can find an $SO(8)$ rotation $\mu$ that simultaneously permutes any two distinct pairs of punctures.
Then $\CB$ and $\mu \circ \CB$ are defined by two possibly inequivalent six-dimensional cobordisms, but are related by a continuous family of deformations that
do not admit a simple six-dimensional definition.

Now, we can define a deformation $T^\mu_M$ of the 3d theory $T_M$ by taking four-dimensional $SU(2)$ $N_f=4$ gauge theory on the segment,
with modified boundary conditions $\CB_1$ and $\mu\circ \CB_2$ at the endpoints.
Clearly $T^\mu_M$ is the theory associated to a new three-manifold $M_\mu$,
obtained by permuting the two pairs of strands across $C$. This is the definition of a mutation of $M$.
But the theories $T_M$ and $T^\mu_M$ are related by a continuous family of exactly marginal deformations, hence must have the same index!

This is the main conclusion of this section. Next, we would like to recast it in a language that makes use only of the definition of $T_M$
through the gluing of tetrahedra, with no reference to the six-dimensional setup.
We are after a continuous family of exactly marginal deformations of $T_M$, possibly including a subset $H_{\CB_1} \backslash SO(8)/H_{\CB_2}$ and
modeled locally on the complexification of that locus. If $H_{\CB_1}$ and $H_{\CB_2}$ are $SU(2)^4$, the exactly marginal locus should be four-dimensional.
Notice that in concrete examples, such exactly marginal deformations may be directly visible in the index, as chiral operators of $R$-charge $2$ and specific flavor charges under the flavor symmetries associated to the codimension two defects.

\begin{figure}[htb]
\centering
\includegraphics[width=5.7in]{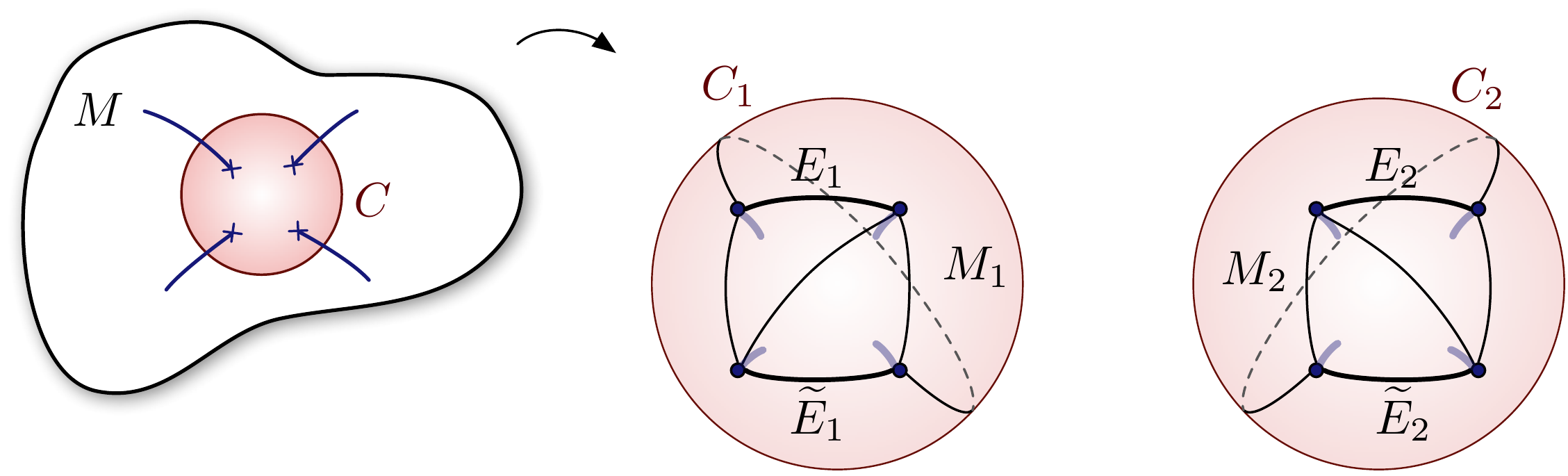}
\caption{Splitting $M = M_1\cup_C M_2$ along a 2-sphere $C$ to form two new geodesic boundaries, triangulated by faces of tetrahedra.}
\label{fig:mutMC}
\end{figure}

In order to proceed, we will assume that $M$ admits a triangulation that can be split along $C$ without cutting any tetrahedra.
If we split along $C$,  we will get triangulated versions of $M_1$ and $M_2$, each with a triangulated $C$ boundary component, $C_1$ and $C_2$; see Figure \ref{fig:mutMC}.%
\footnote{The triangulations of $M_1$ and $M_2$ contain annular cusps, and are of the ``hybrid'' type discussed in Section 2 of \cite{DGG}. Such triangulations will be a major focus of \cite{DGGV-hybrid}.} %
Clearly, $C_1$ and $C_2$ are triangulated in the same way, and gluing $M_1$ and $M_2$ back along $C$ can be done by imposing a
standard gluing constraint for each edge in the triangulation of $C$.
So we can start from the  theories $T_{M_1}$ and $T_{M_2}$ and implement the gluing constraints by adding superpotential terms $\CO_E$, one for each edge $E$ of $C$.
In a polarization where the images $E_1$ and $E_2$ of $E$ in $C_1$ and $C_2$ carry position edge variables,
we can write
\begin{equation}
\CO_E =  \CO_{E_1} \CO_{E_2}
\end{equation}
for operators $ \CO_{E_1}$ and $ \CO_{E_2}$ in the theories $T_{M_1}$ and $T_{M_2}$ respectively.

Again, this is true for a generic $C$, but if $C$ is a four-punctured sphere something special happens.
Consider any of the many triangulations with the topology of a tetrahedron (not to be confused with the tetrahedra we use for gluing!).
Then any pair of opposite edges $E$, $\tilde E$ have the same gauge charges, and only differ by the flavor charges associates with the punctures. This means that in any polarization where the $E$ edge coordinate and the meridians of defects are positions, the edge coordinate of $\tilde E$ is also a position.

Thus we can simultaneously decompose
\begin{equation} \label{mutO}
\CO_E =  \CO_{E_1} \CO_{E_2}\,,  \qquad   \CO_{\tilde E} =  \CO_{\tilde E_1} \CO_{\tilde E_2}\,.
\end{equation}
Now, $\CO_{E_1}$ and $\CO_{\tilde E_1}$ have the same gauge charges, and different flavor charges.
The same is true for $\CO_{E_2}$ and $\CO_{\tilde E_2}$. Thus we can also build two new gauge-invariant operators
\begin{equation} \label{mutOmixed}
\CO_{12} =  \CO_{E_1} \CO_{\tilde E_2}\,,  \qquad   \CO_{21} =  \CO_{\tilde E_1} \CO_{E_2}\,.
\end{equation}

We can repeat the analysis for all the pairs of edges in the triangulation of $C$. This gives us a total of six new
gauge-invariant marginal operators of ``mixed'' type \eqref{mutOmixed} in addition to the six operators of type \eqref{mutO}. It should be clear that a mutation $M \to M_\mu$ simply modifies the geometric gluing by permuting two opposite pairs of edges in, say, $C_2$. In field theory, this corresponds to a different choice of
gluing superpotential, replacing four of the standard operators with four of the new operators.
This has no effect on the index. In particular, either configuration breaks exactly the same set of global $U(1)$ symmetries.

We can even try to see part of the space of exactly marginal deformations of $T_M$:
consider the most general superpotential, a linear combination of the six standard operators and the six new operators.
There is a choice of R-charges that makes all the twelve operators marginal. The general superpotential breaks
many flavor symmetries of the product theory $T_{M_1} \times T_{M_2}$. The standard superpotential forces the
meridians (puncture eigenvalues) of $C_1$ to be equal to the meridians of $C_2$. The general superpotential forces all meridians to be equal .
The number of flavor symmetries broken by the general superpotential is strictly smaller than $12$, but there are $12$ marginal operators.
By the results of \cite{GKSTW}, this indicates that at least one linear combination of the $12$ operators is exactly marginal.

We can try to be more precise. Assume that $T_{M_1} \times T_{M_2}$ has $SU(2)^8$ flavor symmetry associated to the defects, broken to $SU(2)^4$ by the standard gluing superpotential.
Then the $6$ new operators have charges
$\pm (\frac12,\frac12,-\frac12,-\frac12)$, $\pm (\frac12,\frac12,-\frac12,-\frac12)$, $\pm (\frac12,-\frac12,-\frac12,\frac12)$, $\pm (\frac12,-\frac12,\frac12,-\frac12)$ respectively.
Because of the $SU(2)^4$ flavor symmetry, the operators must really be part of a set of $16$ marginal operators in the tensor product of
the four doublet representations of the four $SU(2)$ flavor groups. By the results of \cite{GKSTW}, the space of exactly marginal deformations
is locally the K\"ahler quotient of ${\mathbb C}^{16}$ by $SU(2)^4$, which is indeed a local complexification of the expected
$SU(2)^4 \backslash SO(8)/SU(2)^4$. Thus we recovered the full space of exactly marginal deformations predicted in the UV.

\section{The index from four dimensions and line operators}
\label{sec:3d4d}

Our discussion until now has been purely three dimensional. When the three-manifold $M$ has boundary components,
the three-dimensional theory $T_{M,\Pi}$ we defined by the gluing construction depends on a choice of polarization,
and different polarizations are related by the action of $Sp(2N,\Z)$.
However, there is a simple way to erase the polarization dependence: consider a combined 3d/4d system,
where $T_{M,\Pi}$ lives at the boundary of a half-space,
and a free $U(1)^N$ gauge theory lives in the half space,
with Neumann-type boundary conditions that gauge the $U(1)^N$ flavor symmetry of $T_{M,\Pi}$.%
\footnote{In this section we assume that the boundary of $M$ is a triangulated, geodesic boundary, in the language of Section 2 of \cite{DGG}. Then the theory $T_M$ naturally couples to the IR degrees of freedom of a 4d $\CN=2$ theory on its Coulomb branch, with abelian gauge group $U(1)^N$. Here $N = 3g-3+s$, where $g$ is the genus of $\pd M$ and $s$ is the number of punctures (codimension-two defects) on $\pd M$.} %
Then  $Sp(2N,\Z)$ is identified with the group of electric-magnetic dualities of the four-dimensional theory.

In the previous work concerning the moduli space of
$T_{M,\Pi}$ on $S^1 \times \R^2$ \cite{DGH, DGG}, this 3d/4d setup is rather useful. Upon compactification on a circle,
the four-dimensional gauge theory reduces to a sigma-model on $(\C^*)^{2N}$ parameterized by
the vevs of 't Hooft-Wilson supersymmetric line operators wrapping the circle.
The coordinates $x=e^X$ and $p=e^P$ are literally the vevs of the basic Wilson loop and the basic 't Hooft loop supported on the $S^1$.
The Lagrangian submanifold $\CL \in (\C^*)^{2N}$ that represents the parameter space of $T_{M,\Pi}$ on $S^1 \times \R^2$
defines a boundary condition for the bulk sigma model.

It is natural to wonder if the 3d index of $T_{M,\Pi}$ can be also given a 3d/4d interpretation.
At least for a conformal field theory, the index at zero magnetic flux counts some protected operators in the 3d theory in flat space.
If we couple the 3d theory to 4d degrees of freedom, operators in $T_M$ that carry a flavor charge
will not be gauge-invariant anymore. Rather, they give bulk Wilson loops a way to end.
By electric-magnetic duality, the 't Hooft-Wilson loops of the bulk theory should also be able to end on the boundary.
Because of how the 3d index transforms under $Sp(2N,\Z)$, it is fairly clear that the index at magnetic charge $m$
and electric charge $e$ must ``count'' the possible ways a bulk 't Hooft-Wilson loop of the same charge can end (supersymmetrically) at the boundary.

In order to make this notion more precise, we should identify a setup involving
the half-BPS bundary condition and half-BPS straight line defects ending at a point (the origin) at the boundary,
which preserves at least one supercharge $Q$, and one superconformal charge $S$. Actually, as long as the setup has inversion symmetry,
the superconformal generator $S$ will come for free as long as we identify $Q$. Then we can organize the operators at the origin
in representations of the $SU(1|1)$ algebra generated by $Q$ and $S$, and count the short representations
graded by other conserved charges that commute with $Q$ and $S$ (besides $\{Q,S\}$, which is zero on short representations).
We can also use the standard state-operator map at the origin, to map the counting to an index of the Hilbert space of the
theory on the ``half 3-sphere'' (= a 3-dimensional ball $B^3$) with the half-BPS boundary condition at the equator $S^2 = \partial B^3$,
and in the presence of half-BPS line defects.

A flat half-BPS boundary in an $\CN=2$ superconformal four-dimensional gauge theory
preserves a copy of the 3d $\CN=2$ superconformal group, $OSp(2|4)$,
embedded the four-dimensional superconformal group, $SU(2,2|2)$. In particular, it breaks the $U(1)_r$
R-symmetry, and breaks the $SU(2)_R$ R-symmetry down to the three-dimensional R-symmetry group $SO(2)_R$.
For a fixed geometry of the boundary, there is a whole one-parameter family of choices of embedding,
rotated among each other by the broken $U(1)_r$ symmetry.
The choice of embedding controls which linear combination of the two real scalars in the vector multiplet is a superpartner of the gauge field parallel to the boundary
under the preserved SUSY, and which is a superpartner of the gauge field perpendicular to the boundary.

In contrast, a half-BPS line operator in the 4d theory preserves an appropriate real form of $OSp(4|2)$, including the full $SU(2)_R$ symmetry of the bulk, and $SO(3)$ rotations around the line operator.
Again, each line operator comes labeled by a $U(1)_r$ phase $\vartheta$.
Among other things, this choice determines which linear combination of the two real scalars in the vector multiplet goes into the definition of the Maldacena-Wilson loop,
and which in the definition of the 't Hooft loop \cite{Kapustin-WtH, GMNIII}. Several line operators lying in a common plane can preserve two out
of eight supercharges (a $SU(2)_R$ doublet), as long as the respective phases $\vartheta_i$ are aligned with the slope in the common plane.

It is easy to see that a half-BPS line operator and a half-BPS boundary can form
a common $1/4$-BPS configuration in two natural ways: the line operator can lie in the boundary, or can be orthogonal to the boundary.
We can use the supersymmetric Wilson lines as an example. A Wilson line parallel to the boundary only involves the component of the gauge field
parallel to the boundary. Thus we can define the Maldacena-Wilson loop using the real scalar which is in the same supermultiplet as the gauge field parallel to the boundary.
A line operator that lies in the boundary will simply preserve the same supersymmetries as a half-BPS line defect in a $\CN=2$
three-dimensional theory. On the other hand, a supersymmetric Maldacena-Wilson loop perpendicular to the boundary can be defined using
the real scalar field in the same supermultiplet as the perpendicular component of the gauge field. From the point of view
of the $\CN=2$ three-dimensional superconformal algebra, this Wilson loop behaves as a chiral operator,
and can be used to dress non-gauge-invariant boundary chiral operators to give gauge-invariant chiral operators in the 3d-4d setup.

Notice that the $U(1)_r$ phase $\vartheta$ of the parallel line defect and of the orthogonal line defect differ by $\pi/2$
as they use orthogonal real components of the vectormultiplet scalar field. The two operators are also orthogonal in space-time,
and thus they preserve a common set of supercharges, which is also preserved by the boundary condition. More precisely, the two line operators preserve
$1/4$ of the original SUSY, an $SU(2)_R$ doublet. Then, the boundary condition will select a single supercharge $Q$ in the doublet.
It should be clear that the same supersymmetry is also preserved by all line defects that lie in the common plane
of the parallel and orthogonal line defects, which is a generic plane orthogonal to the boundary, as long as the $\vartheta_i$ parameters are properly chosen.
This single supercharge (together with the corresponding $S$) is exactly what we need in order to define an index.

\begin{wrapfigure}{r}{2.6in}
\includegraphics[width=3.0in]{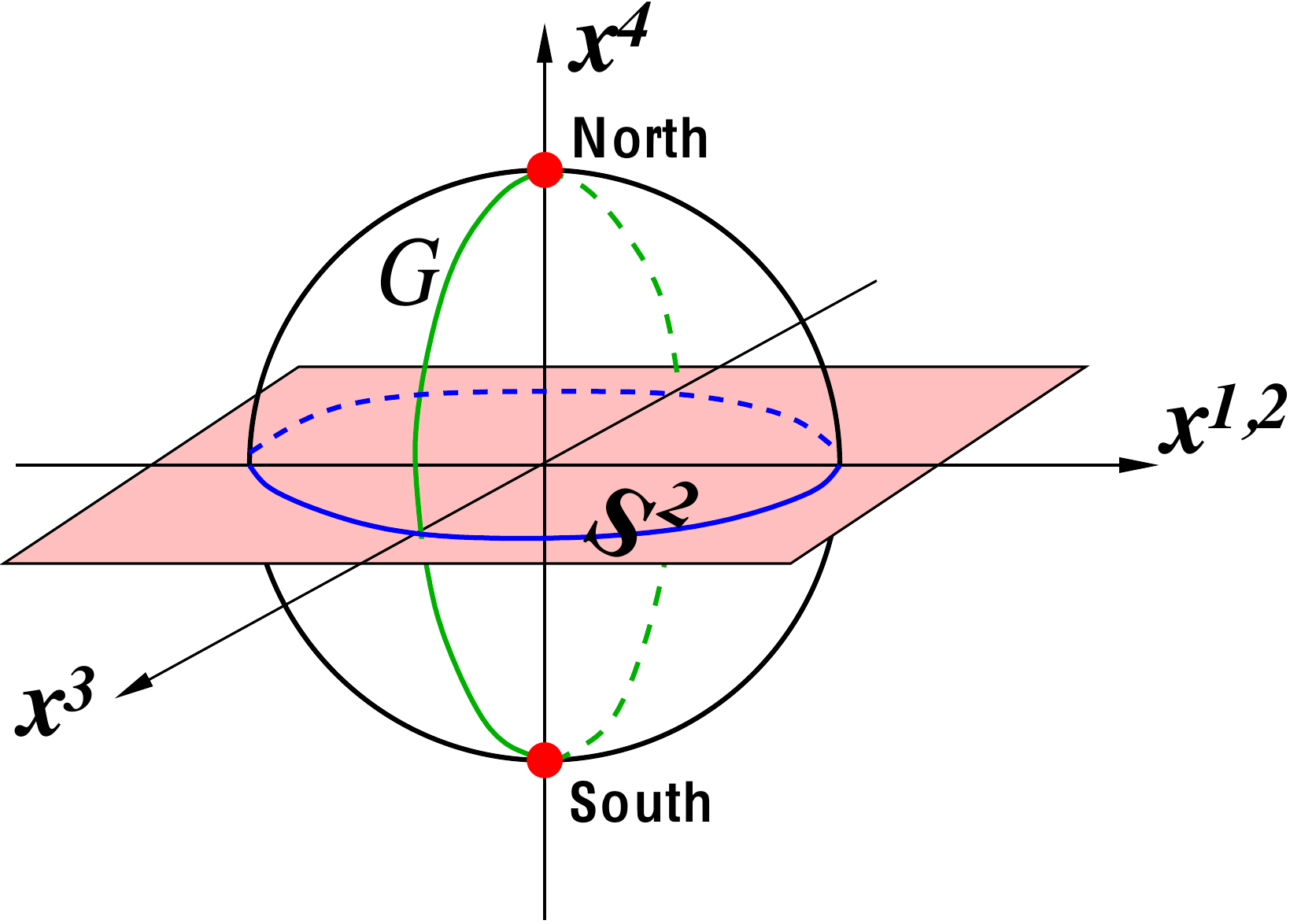}
\caption{3d theory $T_M$ can be put on the equator $S^2$ of a 3-dimensional (hemi)sphere.
Line operators can be placed at any point on the (semi)circle $\CG$, shown in green.}
\label{S3}
\end{wrapfigure}

After the state-operator map, we get a simple setup (see Figure \ref{S3}): the four-dimensional theory on $S^1 \times B^3$,
with the boundary condition at the ``equator'' $S^1 \times S^2 = \partial (S^1 \times B^3)$,
and with line operators supported on $S^1 \times p_*$, where the point $p_* \in \CG$
is located on the ``Greenwich meridian" (= the semi-circle in $B^3$ fixed by a rotation symmetry $U(1)_E$ associated with angular momentum $j_3$).

A closely related setup --- that one can consider in parallel --- comprises the four-dimensional theory on $S^1 \times S^3$,
with a duality wall at the ``equator'' $S^1 \times S^2$, and with line operators supported on $S^1 \times p_*$, where $p_* \in \CG$
is a point on the circle $\CG \cong S^1$ that is fixed by~$U(1)_E$.
In Figure \ref{S3}, this corresponds to considering the full figure, not just the upper half of it.

To make this a little bit more explicit, in either case, we can realize $S^3$ as a unit sphere in
a 4-dimensional Euclidean space $\R^4$ parametrized by the coordinates $x_i$, $i=1, \ldots, 4$,
\be
S^3 ~: \qquad x_1^2 + x_2^2 + x_3^2 + x_4^2 = 1 \,.
\ee
The rotation symmetry $SO(4) \cong SU(2)_1 \times SU(2)_2$ has various subgroups
that leave invariant different submanifolds in $S^3$.
For example, the diagonal subgroup $SU(2)_E \subset SU(1)_1 \times SU(2)_2$
acts as a rotation symmetry of the plane $\R^3 \subset \R^4$ parametrized by $(x_1,x_2,x_3)$
or, equivalently, as a rotation symmetry of the ``equator" $S^2 \subset S^3$.
Therefore, this symmetry is relevant for describing the angular momentum in the 3d theory on $S^1 \times S^2$.
Note that the only fixed points of the rotation symmetry $SU(2)_E$ are the North pole ($x_1 = x_2 = x_3 = 0$, $x_4 = 1$)
and the South pole ($x_1 = x_2 = x_3 = 0$, $x_4 = -1$) of the $S^3$.

Furthermore, the subgroup $U(1)_E \subset SU(2)_E$ corresponds to rotations of the $(x_1,x_2)$ plane $\R^2 \subset \R^3$
around the $x_3$-axis. The corresponding quantum number is what we call $j_3$ throughout this paper.
The fixed points of this rotation symmetry have $x_1 = x_2 = 0$ and, therefore, form a semi-circle (resp. a circle)
in half 3-sphere $B^3$ (resp. in $S^3$):
\be
\CG  ~: \qquad x_1 = x_2 = 0 \,, \qquad x_3^2 + x_4^2 = 1 \,.
\ee
Since the rotation symmetry $U(1)_E$ suffices for defining the index,
in this setup we can consider line operators supported on $S^1 \times p_*$, for any $p_* \in \CG$, without breaking this symmetry.
Note that the curve $\CG$ meets the equator $S^2 = \{ x_1^2 + x_2^2 + x_3^2 = 1 \}$ at two points,
\be
\CG \cap S^2 ~: \qquad x_1 = x_2 = x_4 = 0 \,, \qquad x_3 = \pm 1 \,.
\ee

To summarize,
this setup preserves three isometries: scale invariance, rotations $U(1)_E$ around a chosen plane, and the $SO(2)_R$ symmetry with the generator $R$.
The anticommutator $\{Q,S\}$ is a combination of the dilatation and two other charges, commuting with $Q$ and $S$.
Then, a second combination, say $R/2 + j_3$, will commute with $Q$ and $S$, and can be used to grade the short representations.
Thus we recover the index
\begin{equation}
{\rm Tr}_{\CH_{3d/4d}} (-1)^F \q^{\tfrac{R}{2} + j_3}\,.
\label{3d4dindex}
\end{equation}
{}From the point of view of the 4d theory, this is a specialization of the standard superconformal index \cite{KMMR-index},
which is called the ``Schur index'' in \cite{GRRY}.
The most general index for an $\CN=2$ four-dimensional theory has three fugacities $(p,q,u)$
coupled to the corresponding combinations of R-symmetries and $SO(4) \cong SU(2)_1 \times SU(2)_2$ rotations:
\be
\CI_{4d} \; = \; {\rm Tr}_{\CH_{4d}} (-1)^F \; p^{j_2 + j_1 + \tfrac{R - r}{2}} \; q^{j_2 - j_1 + \tfrac{R - r}{2}} \; u^{-(r+R)}\,.
\label{4dindex}
\ee
This index has a curious property: upon specialization to $p=qu^2$ naively one would expect that
the result should depend on two variables $q$ and $u$. However, after this specialization
$u$ becomes a fugacity for the quantum number $j_1 + j_2 - r$ that commutes not only with $Q$
but also with another supercharge $Q'$.
As a result, the $u$-dependence disappears and the index becomes a function of $q$ only.

Moreover, after the specialization to $p=qu^2$ the variable $q$ in the 4d index \eqref{4dindex}
becomes a fugacity for the combination of the $SU(2)_E$ angular momentum and the R-charge,
precisely as in \eqref{3d4dindex}.
Therefore, this specialization of the 4d index can be used for the combined 4d/3d system,
where $j_3$ is the angular momentum for the $SU(2)_E$ rotation symmetry and $R$ is the R-charge
for the unbroken $SO(2)_R$ symmetry in the presence of a boundary or a duality wall.
When line operators are included at the generic points on $\CG$ the $SU(2)_E$ rotation symmetry
is broken further to $U(1)_E$ which, however, suffices for defining the index \eqref{3d4dindex}.

We can readily compute the index for our setup in the $S^1\times B^3$ geometry, simply by making the gauge coupling very weak. For the moment, let us assume that the 4d theory is a pure $U(1)^N$ gauge theory --- \ie\ with $N$ $U(1)$ vector multiplets but no hypermultiplets.
The index does not depend on the continuous parameter that labels the position of the line operator along $\CG$.
For simplicity, we can place the line operator at the North pole.
The 3d theory lives in the background of the line operator's 't Hooft charge $m$, under the $U(1)^N$ flavor symmetry.
Moreover, the restriction to gauge-invariant states forces us to look at states of the 3d theory that have charge equal to the Wilson loop charge.
The 4d gauge fields contribute in a simple way: the usual Schur index receives contributions only from modes of two
gauginos with various angular momenta. The boundary condition at the equator sets half of the gauginos to zero.
Hence, in the presence of a 't Hooft-Wilson loop of charge $(m,e)$, the index of the 3d/4d system becomes
\begin{equation}
\CI_\CT(m,e;\q) \prod_{n>0} (1-q^n)^N
\end{equation}
Of course, this answer is consistent with the interpretation of $Sp(2N,\Z)$ as the electric-magnetic duality group of
the bulk theory.

\subsection{Line operator algebra}

One real payoff of the 3d/4d construction is an explanation of why the index of the 3d theory satisfies the difference equations
built from the operators $(\hat{x}_{\pm}, \hat{p}_{\pm})$, which we are about to identify with line operators.
First, we need a simple observation about the OPE of line operators.
Consider a setup with two line operators, wrapping $S^1$ and lying at different points $p_1$ and $p_2$ in $\CG$, as defined above. Although the positions $p_1$, $p_2$ locally do not matter,
the relative order along $\CG$ is a topological invariant, as we cannot bring one operator around the other without
breaking the supersymmetry preserved by the index. On the other hand, we can consider the OPE of two line operators.
It is known that the OPE of two supersymmetric  line operators in an abelian $\CN=2$ gauge theory
is rather simple \cite{GMNIII}%
\footnote{More generally, one can consider line operators localized on the 2-dimensional world-sheet $D \times p \times \{ 0 \}$
of a surface operator (= codimension-4 defect) in the six-dimensional $(2,0)$ theory on $D \times C \times R^2_{\hbar}$.
It was argued in \cite{Ramified} that such line operators generate an affine Hecke algebra with parameter $q = e^{\hbar}$.
Note that this affine Hecke algebra is ``local on $C$.''
In other words, it does not depend on the details of the Riemann surface $C$ away from the point $p$.
For application to the problem in hand, one needs to consider a trivial surface operator and take $D = S^1 \times \CG$.}%
:
\begin{equation}
L_\gamma L_{\gamma'} \; = \; V_{\langle \gamma, \gamma' \rangle} \otimes L_{\gamma + \gamma'} \,.
\end{equation}
Here $\gamma$, $\gamma'$ are the charges of the line operators and $\langle \gamma, \gamma' \rangle$
is the usual symplectic pairing. The ``coefficient'' $V_{\langle \gamma, \gamma' \rangle}$ of the OPE
is a one-dimensional vector space, whose only role is to carry a charge $\langle \gamma, \gamma' \rangle/2$ under the action of $SO(2)$ rotations
around the axis that goes through the pair of line operators (and the appropriate fermion number).

In the present context, the curve $\CG$ plays the role of such axis and the $SO(2)$ rotation is generated by $j_3$.
Therefore, we find that the OPE of line operators inside the index becomes
\begin{equation}
L_\gamma L_{\gamma'} \; = \; (q^{1/2})^{\langle \gamma, \gamma' \rangle} L_{\gamma + \gamma'} \,.
\label{qLLOPE}
\end{equation}

Now, consider the setup with a line operator $L_{\gamma}$ of charge $\gamma = (e,m)$ at the North pole
and then introduce another Wilson loop of charge $1$ at some point on the curve $\CG$ away from the North pole.
We can compute the index by using the OPE of line operators, under which the electric charge $e$
of the original line operator is shifted by one unit and, according to \eqref{qLLOPE},
the result is multiplied by $(q^{1/2})^{\pm m}$, depending on which side of the North pole of $B^3$ (resp. $S^3$)
the extra Wilson operator was added.
Similarly, if we add a 't Hooft operator of charge $1$, we will shift $m$ by one unit, and multiply  everything by $(q^{1/2})^{\mp e}$.
This is exactly how our operators $(\hat{x}_{\pm}, \hat{p}_{\pm})$ act on the index!

Therefore, we can identify Wilson and 't Hooft operators with the following operators acting
on the 3d index ({\it cf.} an analogous identification \cite{DGG} with operators acting on the $S^3_b$ partition function):
\begin{align}
\text{Wilson}_{\pm} \quad \longleftrightarrow \quad & (\hat{x}_{\pm})^{\mp 1} = e^{\partial_e \mp \tfrac{\hbar}{2} m} \\
\text{'t Hooft}_{\pm} \quad \longleftrightarrow \quad & (\hat{p}_{\pm})^{\pm 1} = e^{\partial_m \pm \tfrac{\hbar}{2} e}
\end{align}

Now we see why the index satisfies recursion relations modeled on $\CL$: line operators $L_{\gamma_i}$ supported on $S^1 \times p_i$
can be brought all the way to the boundary, where they satisfy Ward identities dictated by the boundary condition, as in \cite{DGG}.

\subsection{Boundary conditions for hypermultiplets}

So far, we have considered $U(1)^N$ vector multiplets coupled to a boundary theory $T_M$. Generically, the 4d theories $T[\pd M]$ arising from a geodesic boundary $M$ will also involve hypermultiplet matter, which leads to further interesting observations in our 3d/4d setup.

A theory of a single hypermultiplet has two natural classes of boundary conditions. In one class,
one of the chiral fields $Y$ in the hypermultiplet  has Dirichlet-type boundary conditions,
\begin{equation}
Y = \CO
\end{equation}
for some chiral operator $\CO$ in the boundary theory,
and the other chiral field $\tilde Y$ has Neumann-type boundary conditions.
In the other class, the roles of $Y$ and $\tilde Y$ are exchanged. These boundary conditions were defined, \eg, in Section 3 of \cite{DGG}.
It turns out that each boundary condition in one class is mirror to a boundary condition in the other class.
The corresponding boundary theories are related by an ``F'' transformation,
which maps the theory with chiral operator $\CO$ to a new theory with a new 3d chiral
field $\phi$ coupled by the superpotential $\phi \CO$. In the new theory, $\phi$ plays the role of the
special chiral operator $\tilde \CO$.

This mirror symmetry is manifest in the 3d/4d index calculations.
The contribution of one hypermultiplet to the Schur index
looks like
\begin{equation}
\CI_{hyper}= \frac{1}{\prod_{n>0} ( 1 - \zeta q^{n+1/2})( 1 - \zeta^{-1} q^{n+1/2})}\,.
\end{equation}
This simply counts angular momentum modes of the two chiral fields
$Y$ and $\tilde Y$, of charge $\pm 1$ under the flavor symmetry with fugacity $\zeta$.

If we impose Dirichlet boundary conditions on $\tilde Y$,
we are left with the contribution from $Y$ only,
\begin{equation}
\CI_Y \; = \; \frac{1}{\prod_{n>0} ( 1 - \zeta q^{n+1/2})} \,.
\end{equation}
Similarly, in the case of Dirichlet boundary conditions for $Y$ we get
\begin{equation}
\CI_{\tilde Y} \; = \; \frac{1}{\prod_{n>0} ( 1 - \zeta^{-1} q^{n+1/2})} \,.
\end{equation}
And, of course, the product of $\CI_{\tilde Y}$ with the 3d index of a free 3d chiral field of charge $1$, in the absence of magnetic flux,
gives $\CI_Y$, as expected from the discussion of the F move.%
\footnote{Furthermore, $\CI_{hyper} = \CI_Y \CI_{\tilde Y}$, while obvious, can be given a neat interpretation:
in order to describe a hypermultiplet on the whole space, one can start with Dirichlet boundary conditions on $\tilde Y$ for a half-space, and
Dirichlet boundary conditions on $Y$ for the second half-space, and ``glue'' them together by a $Y_- \tilde Y_+$ 3d superpotential
integrated over the equator. Here $Y_-$ and $\tilde Y_+$ denote the boundary values of $Y$ and $\tilde Y$ from the two sides.
This ``gluing'' prescription is akin to the idea that a gauge theory on the full space can be reconstructed from
the gauge theory on two half-spaces with Dirichler boundary conditions by gauging the diagonal flavor symmetry
at the boundary}

Crucially, we expect the mirror relation to hold even in the presence of 't Hooft operators at the North pole.
Thus we can predict, up to a prefactor, how the Schur index for a hypermultiplet should look in the presence of such a
magnetic charge:
\begin{equation}
\CI_{hyper}(m; q,\zeta) \;= \; \frac{1}{\prod_{n>0} ( 1 - \zeta q^{n+1/2+|m|/2})( 1 - \zeta^{-1} q^{n+1/2+|m|/2})} \,. \label{eq:maghyper}
\end{equation}

\subsection{Non-abelian line defects}

There are natural ways to use the three-dimensional theories $T_M$ to define domain walls in the UV for four-dimensional ${\cal N}=2$ $SU(2)$ gauge theories.
A full account of that construction will be the subject of a future publication \cite{DGGV-hybrid}. For now, we will content ourselves with sketching some interesting directions of inquiry.
For simplicity, we will focus on the example of $SU(2)$ ${\cal N}=2^*$ SYM.

We would like to establish a very concrete parallel between the general structure of Pestun's $S^4$ partition function and
the refined index. In the presence of line defects or domain walls at the equator, localization reduces Pestun's partition function to a matrix element of a self-adjoint
operator \cite{DGOT,AGGTV,DGG-defects}
\begin{equation}
\langle Z_S| \hat O | Z_N \rangle \equiv \int d\nu(a') d\nu(a) \bar Z_{\mathrm{inst}}(a') O(a',a)  Z_{\mathrm{inst}}(a)
\end{equation}
Here $Z_{\mathrm{inst}}(a)$ is the instanton partition function (together with the tree-level contribution) for the gauge theory,
which is naturally thought of as a wavefunction, $| Z_N\rangle=Z_{\mathrm{inst}}(a)$, and $\langle Z_S |$ is the complex conjugate;
$d\nu(a)$ is the one-loop integration measure, and $ O(a',a)$ is an integration kernel
that encodes the effect of the domain wall or line defect. For domain walls, $O(a',a)$ is basically the $S^3$
partition function of the domain wall 3d degrees of freedom. For 'tHooft-Wilson line defects $L$,
$O(a,a')$ is a sum over delta functions, so that $\hat O$ takes the form of a sum over shift operators
\begin{equation}
\hat O_L = \sum_n R_n(a) e^{2 \pi n \partial_a}
\end{equation}
The functions $R_n(a)$ can in principle be computed from the geometry of monopole moduli spaces, as they account for
``bubbling monopole'' configurations near the line defect \cite{GOP-tHooft} that interpolate between the ``bare'' 'tHooft monopole magnetic flux
and an Abelian magnetic flux $n/2$.

A simple way to understand the physical meaning of the matrix element is to realize the configuration starting from
two hemispheres with Dirichlet boundary conditions for the gauge fields, and then glue together the hemispheres, along with any
extra degrees of freedom at the equator, by gauging appropriate 3d flavor symmetries. In particular,
it is natural to see the wavefunction $|Z_N\rangle$ as the partition function on a hemisphere with Dirichlet boundary conditions.
Notice that the notion of Dirichlet boundary condition is not S-duality invariant, but depends on a choice of weakly coupled duality frame.
Hence $| Z_N\rangle$ transforms in interesting ways under S-duality, while the $S^4$ partition function
$\langle Z_S| Z_N \rangle$ is S-duality invariant. Incidentally, in the gluing procedure one also has to
pick adequate boundary conditions for the hypers, as discussed before. Different choices split the one-loop measure in different way
among North and South hemisphere, and change the normalization of the wavefunctions.

In the context of AGT \cite{AGT}, the wavefunctions are identified with BPZ conformal blocks, and the matrix elements
with 2d CFT correlation functions. The operators $O_L$ coincide with the so-called Verlinde line operators, or with
topological defects in the 2d CFT. Crucially, they can be computed explicitly through a laborious algorithm \cite{AGGTV,DGOT},
sidestepping the difficult localization calculations of gauge theory. The whole structure described above
can be deformed to generic $b$, though the corresponding deformation $S^4_b$ of the $S^4$ geometry has not been described yet.
We will need the generalized formulae in the following.
The simplest 'tHooft-Wilson loop for ${\cal N}=2^*$ SYM is the operator of magnetic charge $1$ and electric charge~$s$,
\begin{equation}
\hat O_{1,s} = \frac{\sin \pi b(2 a - Q -\mu)}{\sin \pi b(2a-Q)}e^{i \pi b (2a-Q) s - \frac{b}{2}\partial_a} + \frac{\sin \pi b(2 a - Q + \mu)}{\sin \pi b(2a-Q)}e^{-i \pi b (2a-Q) s + \frac{b}{2}\partial_a}\,.
\end{equation}
It is self-adjoint against the standard Liouville measure $d\nu(a) = \sin \pi b (2 a-Q) \sin \pi/b (2a-Q) da$, as long as the mass parameter $\mu$ lies in the physical line $Q/2+ i \R$.
Notice that the first factor in the measure cancels against the denominators in $\hat O_{1,s}$ and the second factor in the measure commutes with the shift operators.
The opposite is true for a second family of operators, obtained from $\hat O_{1,s}$ by $b \to b^{-1}$. The operator is explicitly symmetric under $a \to Q-a$ but not under
$\mu \to Q-\mu$, due to an asymmetric split of the one-loop factors between North and South hemispheres. The symmetry $\mu \to Q-\mu$
can be implemented by conjugation by an appropriate rescaling factor.

In order to set up our analogy, we can break down the calculation of the index of a 4d theory, possibly in the presence of a domain wall
on the equator of the $S^3$ or of line defects, by first defining a ``half-index'' $I\!I_m(q,\zeta,\Gamma)$ as the index of the 4d gauge theory on half of $S^3$ with Dirichlet boundary conditions.
This will be a function of the fugacities and magnetic fluxes $\zeta,m$ associated to the
3d flavor symmetries and of the choice of S-duality frame $\Gamma$.
The full index of the 4d theory can be recovered by combining the half-indices for the two hemispheres with the measure for
a three-dimensional non-abelian gauge theory:
\begin{equation}
\left( I\!I_S | I\!I_N \right) \equiv \sum_m \oint \frac{d\zeta}{2\pi i\zeta} \Delta_m(\zeta) I\!I_m(q,\zeta,N) I\!I_m(q,\zeta,S)\,.
\end{equation}
In the absence of 'tHooft operators in the bulk, we expect $I\!I_m(q,\zeta,\Gamma)$ to be non-zero only at $m=0$.
The measure $\Delta_0(\zeta)$ is the usual Vandermonde measure.
In the presence of 'tHooft loops, we will have contributions at
 non-zero (possibly half-integral) $m$ and  we will find a natural generalization of the measure,
\begin{equation}
\Delta_m(\zeta)= \frac{1}{2} (q^{m/2} \zeta-q^{-m/2} \zeta^{-1})(q^{-m/2} \zeta-q^{m/2} \zeta^{-1})\,.
\end{equation}

How can we add 'tHooft operators to the half-index? The hard way would be to do a careful localization computation, taking into account bubbling monopole contributions
that screen the magnetic charge at the pole, giving a variety of possible Abelian fluxes $m$ at the equator. But there is a shortcut. We can try to borrow the calculation of $R_n(a)$
from the Verlinde loop operators. Notice that if we assume the existence of a dictionary, these are very constrained. Namely, the OPE of line defects and the Ward identities satisfied when they are brought to
collide with domain walls must take the same form for the  $S^4_b$ partition function and for the index, upon the dictionary  established in the rest of this paper between vector-multiplet vevs and fugacities, and
 $2 \pi i b^2 \to \hbar$. In particular, we will map
 \begin{equation} \label{opdict}
 e^{i \pi b (2a-Q)} \to \hat x_+ = q^{m/2} \zeta  \qquad  e^{b \partial_a} \to \hat p_+ = e^{\partial_m + \frac{\hbar}{2} \partial_{\log \zeta}} \qquad  e^{i \pi b (2m-Q)} \to \eta
 \end{equation}
 We can replace $(\hat x_+,\hat p_+)$ with $(\hat x_-,\hat p_-)$ for operators acting from the opposite side of the equator.

We are led to conjecture that the half-index in the presence of a 'tHooft-Wilson loop $L$ can be computed by acting on the half-index in the absence of the loop
with the image of $\hat O_L$ under the above dictionary. Let's consider a concrete example.
The 4d index of $SU(2)$ ${\cal N}=2^*$ is
\begin{align}
&I(\eta,q) = \oint \frac{d\zeta}{4\pi i\zeta} (1-\zeta^2)(1-\zeta^{-2}) \\\notag &\times \prod_{n=0}^{\infty} \frac{\left(1-q^{n+1}\right)^2 \left(1-\frac{q^{n+1}}{\zeta^2}\right)^2 \left(1-\zeta^2 q^{n+1}\right)^2}{\left(1-\eta^{-1}q^{n+\frac{1}{2}}\right) \left(1-\eta q^{n+\frac{1}{2}}\right)
   \left(1-\frac{q^{n+\frac{1}{2}}}{\eta \zeta^2}\right) \left(1-\frac{\eta q^{n+\frac{1}{2}}}{\zeta^2}\right) \left(1-\frac{\zeta^2 q^{n+\frac{1}{2}}}{\eta}\right) \left(1-\eta \zeta^2 q^{n+\frac{1}{2}}\right)}\,,
\end{align}
and the half-index is
\begin{align}
&I\!I_m(\zeta,\eta,q) = \delta_{m,0} \prod_{n=0}^{\infty} \frac{\left(1-q^{n+1}\right)\left(1-\frac{q^{n+1}}{\zeta^2}\right) \left(1-\zeta^2 q^{n+1}\right)}{ \left(1-\eta^{-1} q^{n+\frac{1}{2}}\right)
   \left(1-\frac{\eta^{-1} q^{n+\frac{1}{2}}}{\zeta^2}\right) \left(1-\eta^{-1} \zeta^2 q^{n+\frac{1}{2}}\right)}\,,
\end{align}
so that
\begin{equation}
I(\eta,q) =  \sum_m \oint \frac{d\zeta}{2\pi i\zeta} \Delta_m(\zeta) I\!I_m(q,\zeta,\eta) I\!I_m(q,\zeta,\eta^{-1})\,.
\end{equation}

Now we can act with a Wilson loop operator, $2 \cos \pi b (2a-Q) \to q^{m/2} \zeta+q^{-m/2} \zeta^{-1}$, following \eqref{opdict}.
We get that the half-index in the presence of a Wilson loop is simply multiplied by the appropriate character, thanks to the $\delta_{m,0}$ constraint:
\begin{align}
\hat O_{0,1} I\!I_m(\zeta,\eta,q) = \left(\zeta + \zeta^{-1} \right) I\!I_m(\zeta,\eta,q)\,.
\end{align}

On the other hand, if we act with the translated 'tHooft-Wilson operator
\begin{align}
\hat O_{1,s} &\,=\, \frac{q^{m/2-1/4} \zeta \eta^{-1/2} - q^{-m/2+1/4} \zeta^{-1} \eta^{1/2}}{q^{m/2} \zeta - q^{-m/2} \zeta^{-1}}q^{- s/4} q^{s m/2} \zeta^s p_+^{-1/2} \notag \\ &\qquad+\, \frac{q^{m/2+1/4} \zeta \eta^{1/2} - q^{-m/2-1/4} \zeta^{-1} \eta^{-1/2}}{q^{m/2} \zeta - q^{-m/2} \zeta^{-1}}q^{- s/4} q^{-s m/2} \zeta^{-s} p_+^{1/2}\,,
\end{align}
we get
\begin{align}
& \hspace{-.4cm}\hat O_{1,s}  I\!I_m(\zeta,\eta,q) = q^{1/4} \eta^{1/2} \left( \zeta^s \delta_{m,1/2} + \zeta^{-s} \delta_{m,-1/2}  \right) \prod_{n=0}^{\infty} \frac{\left(1-q^{n+1}\right)\left(1-\frac{q^{n+3/2}}{\zeta^2}\right) \left(1-\zeta^2 q^{n+3/2}\right)}{ \left(1-\eta^{-1} q^{n+\frac{1}{2}}\right)
   \left(1-\frac{\eta^{-1} q^{n+1}}{\zeta^2}\right) \left(1-\eta^{-1} \zeta^2 q^{n+1}\right)}\,.
\end{align}
Several useful cancellations led to a simple result. We would have derived the same result if we had used $(\hat x_-,\hat p_-)$ in the dictionary,
as it should be: the half-index for a bare hemisphere should intertwine the line defects acting on opposite sides of the equator.
Notice that we chose a measure $\Delta_m(\zeta) = \frac{1}{2} (x_+ - x_+^{-1})(x_- - x_-^{-1})$, which mimics
the properties of the Liouville measure.

We can now compare the index in the presence of two basic Wilson loops, one for each hemisphere,
 \begin{align}
&I_{el}(\eta,q) = \oint \frac{d\zeta}{4\pi i\zeta} (1-\zeta^2)(1-\zeta^{-2}) \\\notag &\times\prod_{n=0}^{\infty} \frac{(\zeta + \zeta^{-1})^2 \left(1-q^{n+1}\right)^2 \left(1-\frac{q^{n+1}}{\zeta^2}\right)^2 \left(1-\zeta^2 q^{n+1}\right)^2}{\left(1-\eta^{-1}q^{n+\frac{1}{2}}\right) \left(1-\eta q^{n+\frac{1}{2}}\right)
   \left(1-\frac{q^{n+\frac{1}{2}}}{\eta \zeta^2}\right) \left(1-\frac{\eta q^{n+\frac{1}{2}}}{\zeta^2}\right) \left(1-\frac{\zeta^2 q^{n+\frac{1}{2}}}{\eta}\right) \left(1-\eta \zeta^2 q^{n+\frac{1}{2}}\right)}\,,
\end{align}
with the index in the presence of two 'tHooft loops (of zero electric charge)\,,
\begin{align}
&I(\eta,q) =2 \oint \frac{d\zeta}{4\pi i\zeta} (1-q^{1/2} \zeta^2)(1-q^{1/2} \zeta^{-2}) \\\notag &\quad\times \prod_{n=0}^{\infty} \frac{\left(1-q^{n+1}\right)^2 \left(1-\frac{q^{n+3/2}}{\zeta^2}\right)^2 \left(1-\zeta^2 q^{n+3/2}\right)^2}{\left(1-\eta^{-1}q^{n+\frac{1}{2}}\right) \left(1-\eta q^{n+\frac{1}{2}}\right)
   \left(1-\frac{q^{n+1}}{\eta \zeta^2}\right) \left(1-\frac{\eta q^{n+1}}{\zeta^2}\right) \left(1-\frac{\zeta^2 q^{n+1}}{\eta}\right) \left(1-\eta \zeta^2 q^{n+1}\right)}\,.
\end{align}
The factor of $2$ arises from the sum of the abelian contributions of magnetic charge $\pm 1/2$. We absorbed the factors $q^{1/4} \eta^{1/2}q^{1/4} \eta^{-1/2}$ in the measure.
Amazingly, the two results agree, computationally, to as high an order as we could check in the $q$ expansion.
This is a striking verification that our prescription of the line defects is compatible with S-duality.


\section{The index as $SL(2,\C)$ Chern-Simons theory}
\label{sec:CS}

Throughout the preceding sections, there have been various hints that the index $\CI_M(m,e;q)$ of a theory $T_M$ may have some relation to complex Chern-Simons theory on $M$ itself, \ie\ Chern-Simons theory with complex gauge group $SL(2,\C)$. For example, the constructions of $T_M$ and $\CI_M$ involve choosing a polarization on the phase space $\CP_{\pd M}$, and this is precisely the phase space of complex Chern-Simons theory on a 3-manifold $M$ with boundary \cite{Wit-Jones, Witten-cx}. In addition, the flat $SL(2,\C)$ connections in the bulk of $M$, which project to the Lagrangian submanifold $\CL_M$,
\begin{align} \CL_M &\;=\;\{\text{flat $SL(2,\C)$ connections on $\pd M$ that extend to $M$}\} \notag \\
\cap\;\; & \notag \\
\CP_{\pd M} &\;=\;\{\text{flat $SL(2,\C)$ connections on $\pd M$}\}\,, \notag
\end{align}
correspond to spaces of SUSY vacua for $T_M$ \cite{DGH, DGG}; and these flat connections on $M$ are the classical solutions to Chern-Simons theory. Moreover, we saw in Sections \ref{sec:TM} and \ref{sec:3d4d} that a quantization of the Lagrangian's defining equations, $\hat\CL_M$, provides difference operators that annihilate the index.
This again is an expected property of Chern-Simons wavefunctions: the constraints describing classical solutions $(\CL_M)$ become promoted to quantum operators $(\hat\CL_M)$ that annihilate the quantum wavefunctions \cite{gukov-2003}.

To be even more suggestive, consider the tetrahedron index in the electric fugacity basis \eqref{Iztet2}. Note that in the definition of the index as a partition function on $S^2\times S^1$, the parameter $q=e^\hbar$ is real, while $\zeta = e^{i\theta}$ is a pure phase. We can then recombine $m$ and $\theta$ into complex variables, as in Section \ref{sec:T1}\,:
\be Z = \frac{m}{2}\hbar+i\theta\,,\qquad \ol Z = \frac{m}{2}\hbar-i\theta\,, \label{tmZ}
\ee
or
\be z = e^Z = q^{\frac m2}\zeta\,,\qquad \ol z = e^{\ol Z}=q^{\frac m2}\zeta^{-1}\,; \label{tmz} \ee
so that the tetrahedron index takes the form
\be \CI_\Delta(m;q,\zeta) = \CI_\Delta(z,\ol z;q) = \prod_{r=0}^\infty \frac{1-q^{r+1}z^{-1}}{1-q^r\,\ol{z}^{-1}}\,.\ee
If we define $\CZ_\Delta(z;q)\equiv\prod_{r\geq 1} (1-q^r z^{-1})$ for $|q|<1$, and allow ourselves the ``Fock space reorganization'' $\prod_{r\geq 1}(1-q^{-r}z)=\prod_{r\geq 0}(1-q^{r}z)^{-1}$, we would then have an index
\be \CI_\Delta(z,\ol z;q) \;\text{``}\!=\!\text{''}\; \CZ_\Delta(z;q)\,\CZ_\Delta(\ol z;q^{-1})\,, \label{Dfact} \ee
factorized into identical holomorphic and antiholomorphic pieces.

Such a factorization is familiar from the studies of $SL(2,\C)$ Chern-Simons partition functions in \cite{gukov-2003, DGLZ, Wit-anal}. More generally, the physical Chern-Simons partition function on a 3-manifold $M$ would be a sum over classical flat connections $\alpha$ with fixed boundary conditions~$x$,
\be \CZ^{\rm CS}_M(x,\ol x;q)\; = \; \sum_{\alpha}n_{\alpha\bar\alpha}\,\CZ_M^\alpha(x;q)\,\CZ_M^{\bar \alpha}(\ol x;q^{-1})\,, \label{CSfact} \ee
and with $n_{\alpha\bar\alpha}$ real and diagonal.%
\footnote{Explicit expansions of this type, not for indices but for the closely related ellipsoid partition functions of some theories in class $\CR$, have recently appeared in \cite{Pasquetti-fact}.}
In the case of the tetrahedron, there is a unique flat connection at fixed boundary condition $x=z$, and so a single term in \eqref{CSfact}.
Note that \eqref{CSfact} has a symmetry under complex conjugation and exchange of $q$ and $q^{-1}$, which looks like $\rho$ symmetry \eqref{kind} for the index.

Our goal in this section is to make the relation between $SL(2,\C)$ Chern-Simons theory and the index as precise as possible. There actually exist several inequivalent quantizations of complex Chern-Simons theory --- corresponding to different real symplectic structures on the complex phase space $\CP_{\pd M}$ --- and we will see that one special choice is related to the generalized index $\CI_M(m;q,\zeta)$. An interesting open question is to understand what the other quantizations might mean. In Section \ref{sec:6d}, we will give a ``top-down,'' qualitative view of the equivalence between Chern-Simons theory and the index by considering the theory of M5-branes in the geometry $S^1\times S^2\times M$ and applying various known dualities.

\subsection{Quantization at $k=0$}
\label{sec:k0}

Let's begin by reviewing a few facts about Chern-Simons theory with gauge group $G_\C=SL(2,\C)$ \cite{Wit-CSgrav, Witten-cx, gukov-2003}. The most general complex bulk action on a 3-manifold $M$ is
\be S_{CS}(M) \,=\, \frac{t}{8\pi}\int_M \Tr\Big(\CA\, d\CA+\frac{2}{3}\CA^3\Big)\,+\,\frac{\tilde t}{8\pi}\int_M \Tr\Big(\ol\CA\, d\ol\CA+\frac{2}{3}\ol\CA^3\Big)\,,
\label{CSaction} \ee
where $\CA$ is a local $\mathfrak{g}_\C=\mathfrak{sl}(2,\C)$-valued one-form, defined modulo the standard gauge transformations $\CA\to g^{-1}\CA\, g+ g^{-1}dg$. Unlike the case of real Chern-Simons theory, the complex action here contains both holomorphic and antiholomorphic terms, and each comes with its own coupling constant, here $t$ and $\tilde t$. Let us write
\be t=k+is\,,\qquad \tilde t=k-is\,.\ee
A priori, $k$ and $s$ are independent complex parameters. However, independence of the path integral measure $\exp\big[ i S_{CS}(M) \big]$ under large gauge transformations forces $k$ to be an integer. Moreover, unitarity --- the statement that the path integral be conjugated under a change of orientation on $M$ --- requires $s\in \R$.%
\footnote{An alternative unitarity structure with $s\in i \R$ was also discussed in \cite{Witten-cx}. However, it does not appear to have direct relevance for 3d indices.}

If the boundary of $M$ is nonempty, then the Chern-Simons path integral on $M$ is a wavefunction in a boundary Hilbert space $\CH_{\pd M}$. This space is the quantization of the classical phase space associated to the boundary,
\be \CP_{\pd M} \;\simeq\; \big\{\, \text{flat connections $\CA$ on $\pd M$}\,\big\}\,. \ee
This space is endowed with a real symplectic structure induced from the Chern-Simons action:
\be \omega_{k,s} = \frac{t}{8\pi}\Omega+\frac{\tilde t}{8\pi}\ol\Omega\,,
\label{wtt} \ee
where $\Omega = \int_{\pd M}\Tr\big(\delta\CA\wedge \delta \CA)$ is a  holomorphic symplectic form. By setting $\Omega  = \omega_I-i\omega_K$, we could also write this as
\be \omega_{k,s} = \frac{k}{4\pi}\omega_I+\frac{s}{4\pi}\omega_K\,.\ee
Here $\omega_I$ and $\omega_K$ coincide with two of the standard real symplectic forms on $\CP_{\pd M}$, viewed as a hyperkahler manifold \cite{Hitchin-SD}, hence the notation. It turns out that, as cohomology classes, $[\omega_K]=0$ while $[\omega_I/(4\pi)] \in H^2(\CP_{\pd M};\Z)$ for any 2d boundary $\pd M$. Thus, as long as $k\in \Z$, we have $[\omega_{k,s}/(2\pi)]\in H^2(\pd M;\Z)$, and the pair $(\CP_{\pd M},\omega_{k,s})$ is quantizable (\cf\ \cite{woodhouse-1992}).

The Hilbert space $\CH_{\pd M}$ is the quantization of $(\CP_{\pd M},\omega_{k,s})$. However, this quantization depends critically on the relative values of $k$ and $s$. To illustrate this dependence more concretely, let us consider a ``model'' phase space $\CP\simeq \C^*\times\C^*$, with coordinates $x$ and $p$, and a holomorphic symplectic form
\be \Omega = 2\,\frac{dp}{p}\wedge \frac{dx}{x}\,. \label{Omega}\ee
It is also convenient to take logarithms
\be X = \log x\,,\qquad P=\log p\,, \ee
defined modulo $2\pi i\Z$\,. $\CP$ is the actual phase space for the boundary $\pd\Delta$ of a tetrahedron --- thinking of $\pd \Delta$ as a four-punctured sphere, with unit $SL(2,\C)$ holonomy eigenvalues at each puncture \cite{DGG} --- and it can be used as a building block to approximate the phase space of any more complicated boundary \cite{DGGV-hybrid}. For example, the phase space of a torus is $\CP_{T^2}\simeq \CP/\Z_2$, while for a genus $g\geq 2$ surface with $s$ punctures an open patch of the phase space looks like a quotient of $\CP^{3g-3+s}$. The coordinates $z$ and $p$ can be thought of as models for eigenvalues of general $SL(2,\C)$ holonomies.

Now, to demonstrate the sensitivity of quantization to the values of $k,s$, suppose first that we quantize $(\CP,\omega_{k,s})$ for a choice of coupling constants $s=0$ and $k\neq 0$. In other words, we choose the real symplectic form
\be \omega_{k,0} = \frac{k}{4\pi}\omega_I = \frac{k}{2\pi}\big[d\,\Re(P)\wedge d\,\Re(X)- d\,\Im(P)\wedge d\,\Im(X)\big]\,. \label{wI}\ee
Thus, writing $\CP = \C^*_1\times \C^*_2\simeq (\R_1\times \R_2)\times(S^1_1\times S^1_2)$, \eqref{wI} indicates that the two noncompact real directions are canonically conjugate to each other, as are the two $S^1$ directions. Moreover, it is easy to see that $[\omega_{k,0}/(2\pi)]\in H^2(\CP;\Z)$, since $\int_{S^1_1\times S^1_2}\omega_{k,0}=-2\pi k$. If we choose a polarization such that (say) $\Re(Z)$ and $\Im(Z)$ are both ``positions'' while $\Re(P)$ and $\Im(P)$ serve as ``momenta,'' the Hilbert space becomes
\be \CH_{k,0} \;\simeq\; L^2(\R)\otimes L^2(\Z_k)\,, \ee
with an infinite-dimensional factor coming from quantization of $\R_1\times \R_2$, and a $|k|$-dimensional factor coming from quantization of the compact $S^1_1\times S^1_2$. The latter factor is much like the Hilbert space of Chern-Simons theory with compact gauge group $SU(2)$.%

The choice $s=0$ and $k\neq 0$ turns out not to be the relevant one for the 3d index. Rather, we conjecture that the index is related to Chern-Simons theory quantized in the opposite limit $k=0$ and $s\neq 0$. The Hilbert space in this case looks very different, and we will proceed to analyze it for the remainder of this section. The general situation $k,s\neq 0$ is a slightly twisted variation of $k=0,\,s\neq 0$; it would be interesting to see if it were related to a generalization of the index.

The real symplectic structure at $k=0$ is
\begin{align} \omega_{0,s} &\,=\,\frac{s}{4\pi}\omega_K \,=\, \frac{i}{\hbar} dP\wedge dX - \frac{i}\hbar d\ol P\wedge d\ol X \label{ws1}  \\
 &\,=\, -\frac{2}\hbar\,\big[ d\,\Re P\wedge d\,\Im X+d\,\Im P\wedge d\,\Re X\big]  \,, \label{ws2}
\end{align}
with
\be \boxed{\hbar \equiv \frac{4\pi}{s}}\,. \ee
Then, we see that the coordinate on each noncompact $\R$ factor in $\CP$ is canonically conjugate to the coordinate on a compact $S^1$. In fact, we could more naturally write the phase space $(\CP,\omega_{0,s})$ as $T^*(S^1_1\times S_2^1)$, with the standard symplectic form of a cotangent bundle. In order to quantize $(\CP,\omega_{0,s})$ we must choose a polarization, and there are several nice choices, summarized in Table \ref{tab:pol}. The polarizations differ in the combinations of compact and noncompact coordinates that are designated as positions and momenta.%
\footnote{In the terminology of geometric quantization, all the polarizations in Table \ref{tab:pol} are \emph{real} polarizations. One advantage of this is that quantum wavefunctions can then be expressed as standard functions, rather than sections of a line bundle, \cf\ \cite{Witten-cx}.} %
This choice is somewhat distinct from the coarser distinction of $Z$ vs. $P$ as position and momentum, and the related $Sp(2,\Z)$ action that featured prominently in the construction of theories $T_M$ and indices $\CI_M$. We will return to the $Sp(2,\Z)$ action in Section \ref{sec:CSsymp}.

\begin{table}[htb]
\centering
$\begin{array}{|@{\;\;}l@{\quad}l@{\quad}l@{\qquad}l@{\;\;}|} \hline
&\text{\underline{Positions}} & \text{\underline{Momenta}} & \text{\underline{Hilbert space $\CH\equiv \CH_{0,s}$}} \\[.1cm]
\text{a)} & \Im\, X\,,\;\Im\, P\;\;(S^1_1\times S^1_2) & -2\,\Re\, P\,,\;2\,\Re\, X\;\; (\R_2\times \R_1) & L^2(S^1)\times L^2(S^1) \\[.1cm]
\text{b)} & \Im\, X\,,\;\Re\, X\;\; (S^1_1\times \R_1) & -2\,\Re\, P\,,\;-2\,\Im\, P \;\;(\R_2\times S^1_2) & L^2(S^1)\times L^2(\Z) \\[.1cm]
\text{c)} & \Re\, P\,,\;\Im\, P\;\; (\R_2\times S^1_2) & 2\,\Im\, X\,,\;2\,\Re\, X \;\;(S^1_1\times \R_1) & L^2(\Z)\times L^2(S^1) \\[.1cm]
\text{d)} & \Re\, X\,,\;\Re\, P \;\;(\R_1\times \R_2) & -2\,\Im\, P\,,\; 2\,\Im\, X \;\;(S^1_2\times S^1_1) & L^2(\Z)\times L^2(\Z) \\\hline \end{array}$
\caption{Real polarizations and Hilbert spaces at $k=0$. All Hilbert spaces are isomorphic, and related by compact Fourier transform.}
\label{tab:pol}
\end{table}

Two of the polarizations in Table \ref{tab:pol} may begin to look familiar from our discussion of indices. In particular, in polarization (d), the compactness of the momenta quantizes the position coordinates $\Re\,X$ and $\Re\,P$. Specifically, we find
\be \boxed{\Re\,X = m\tfrac{\hbar}2\,,\qquad \Re\,P = e\tfrac{\hbar}{2}\,,\qquad m,\,e\,\in\,\Z}\,, \ee
so that wavefunctions in $\CH$ are just complex-valued functions of the two integers $(m,e)$ and the real parameter $\hbar$. This appears identical to the form of the index in a ``charge basis'' \eqref{chargeindex}. Alternatively, if we treat both real and imaginary parts of $Z$ as positions, as in (b), the wavefunctions become non-holomorphic functions of $x = \exp(\Re\,X+i\,\Im\,X)$. The compact momentum $\Im\, P$ still quantizes $\Re\,Z\in \tfrac\hbar2\Z$, so
\be \hspace{.3in} \boxed{x = q^{\frac m2}\zeta}\,,\qquad\text{with}\quad q\equiv e^\hbar\,,\; m\in \Z\,;\;\;\; \zeta =e^{i\theta} \equiv e^{i\,\Im\,X}\,.
\label{zzeta} \ee
Therefore, wavefunctions in polarization (b) look like indices in an electric fugacity basis \eqref{Izeta}. To go from one polarization to another, one should use Fourier transform,
$f^{(b)}(x,\bar x;\hbar)= f^{(b)}(m,\zeta;\hbar) = \sum_{e\in \Z} f^{(d)}(m,e;\hbar)\,\zeta^e\,,$
just as described below \eqref{chargeindex}.

Further confirmation of the relation between Chern-Simons wavefunctions and indices comes from analyzing the algebra of operators acting on the Hilbert spaces in Table \ref{tab:pol}. In \emph{any} polarization, the coordinates $X,\ol X,P,\ol P$ become promoted to operators with commutation relations
\be \big[\hat P,\hat X\,\big]=\hbar\,,\qquad \big[\,\hat{\ol P},\hat{\ol X}\,\big]=-\hbar\,,\qquad \big[\hat P,\hat{\ol X}\,\big]=\big[\,\hat{\ol P},\hat X\,\big]=0\,. \label{CSops} \ee
These relations can be read off directly from the form \eqref{ws1} of $\omega_{0,s}$. One can then work out the action of the operators on wavefunctions, in various polarizations. In particular, acting on $f^{(d)}(m,e;\hbar)$ we find
\be \hat X=m\tfrac\hbar 2-\pd_e\,,\qquad \hat P=e\tfrac\hbar2+\pd_m\,,\qquad \hat{\ol X}=m\tfrac\hbar 2+\pd_e\,,\qquad \hat {\ol P}=e\tfrac\hbar2-\pd_m \ee
whereas acting on $f^{(b)}(m,\zeta;\hbar)$ we have
\be \hat X=X\,,\qquad \hat P=\hbar \,\pd_X\,,\qquad \hat{\ol X}=\ol X\,,\qquad \hat {\ol P}=-\hbar\,\pd_{\ol X}\,. \ee
Even in the $(b)$ polarization, it is nice to note that the compactness of the position $\Im\,Z$ necessarily quantizes the spectrum of the conjugate momentum operator $\widehat{\Re P}=\tfrac12\big(\hat P+\hat {\ol P}\big) = e\tfrac\hbar2\,,$ $e\in \Z$.

It should be evident that the operator algebra just described is identical to the algebra of index operators from Sections \ref{sec:ops} and \ref{sec:tetdiff}, with the identification $(\hat X,\hat{\ol X},\hat P,\hat {\ol P})=(\hat X_+,\hat X_-,\hat P_+,\hat P_-)$. With this information in hand, we can move beyond Hilbert spaces and compare the actual wavefunctions of nontrivial 3-manifolds. As reviewed at the beginning of this section (and in many other places), the classical solutions to complex Chern-Simons theory on $M$ cut out a Lagrangian submanifold $\CL_M$ in the boundary phase space $\CP_{\pd M}$; and the quantum wavefunction of $M$ is annihilated by a quantization of the defining equations for the Lagrangian, $\hat \CL_M\cdot \CZ_M=0$. The quantization $\CL_M\to \hat \CL_M$ is universal in Chern-Simons theory, in the sense that it does not depend on a certain real or complex form of the gauge group, or on choices of polarization, \cf\ \cite{gukov-2003, GW-branes}. Therefore, we can borrow results of \cite{Dimofte-QRS}, where Lagrangians were quantized in analytically continued Chern-Simons theory, and simply apply them to find quantized Lagrangians in the present case of physical $SL(2,\C)$ Chern-Simons at $k=0$.

The simplest example of such quantization is for a tetrahedron itself. The phase space is just our model phase space, $\CP_{\pd \Delta}=\CP$ with standard coordinates $x=z$, $p=z''$, and the classical Lagrangian $\CL_\Delta$ is cut out by the equation $z''+z^{-1}-1=0$. Since the real symplectic form $\omega_{0,s}$ manifestly breaks the holomorphic structure of $\CP$, it is perhaps better to write the Lagrangian as
\be \CL_\Delta\,:\quad z''+z^{-1}-1 = 0\,,\qquad  \ol z''+\ol z^{-1}-1 = 0\,.\ee
Then quantization with respect to $\omega_{0,s}$ produces operators $\hat \CL_{\Delta+}=\hat z''+\hat z^{-1}-1$, $\hat \CL_{\Delta-}=\hat{\ol z}''+\hat{\ol z}^{-1}-1$ that must annihilate the tetrahedron wavefunction. (As usual, we define $\hat z=\exp \hat Z$, etc.) In polarization (b), the equations determine
\be \CZ^{CS}_\Delta(z,\ol z;\hbar) = \prod_{r=0}^\infty \frac{1-q^{r+1}z^{-1}}{1-q^r \,\ol z^{-1}} \label{CStet} \ee
up to an overall function of $q$. Given the dictionary $z=q^{\frac m2}\zeta$ of \eqref{zzeta}, this establishes the equivalence of the Chern-Simons wavefunction with the index in an electric fugacity basis, up to normalization.

In order to understand $SL(2,\C)$ Chern-Simons theory on a more general 3-manifold $M$ with boundary, we can decompose $M$ into ideal tetrahedra, and then try to apply the symplectic gluing framework of \cite{Dimofte-QRS}. This requires defining an affine symplectic $Sp(*,\Z)$ action on the tetrahedron Hilbert spaces $\CH = \CH_{0,s}$ of Table \ref{tab:pol} and their products. It also requires an appropriate quantum version of symplectic reduction, in order to implement gluing. We will describe these operations in Section \ref{sec:CSsymp}. Of course, we will find that they coincide with the operations on the index $\CI_M$ found in Section \ref{sec:actions}, so that Chern-Simons wavefunctions can be constructed using the index-gluing rules of Section \ref{sec:rules}. Moreover, asking for $SL(2,\C)$ Chern-Simons wavefunctions at $k=0$ to be independent of a triangulation (invariant under 2-3 moves) even fixes the normalization of the single-tetrahedron wavefunction \eqref{CStet} to agree with that of the tetrahedron index. Taken together, these facts constitute an axiomatic proof%
\footnote{The same technicalities involving convergence and univalent edges from Section \ref{sec:TM} appear in Chern-Simons theory too. We mean a non-rigorous, physical ``proof.''} %
that the index of $T_M$ is equivalent to an $SL(2,\C)$ Chern-Simons wavefunction, $\CI_M = \CZ_M^{CS}$.

We may finally note that $SL(2,\C)$ Chern-Simons theory has an obvious conjugation symmetry, stemming from the symmetry of the action \eqref{CSaction}. It acts on a semi-classical phase space $(\CP,\omega_{k,s})$ as complex conjugation and simultaneous exchange of $t$ and $\tilde t$, to preserve the symplectic form \eqref{wtt}. In the case $k=0$, this means $\hbar\to-\hbar$, or $q\to q^{-1}$. Following the dictionary \eqref{zzeta} or \eqref{tmZ}, we see that complex conjugation is precisely $\rho$ symmetry for the index, from Section \ref{sec:kappa}. Interestingly, the action of conjugation on wavefunctions suffers from the same ``Fock space reorganization'' subtleties as the index --- discussed briefly, for example, in Section 6.5 of \cite{Wfiveknots}.

\subsection{Symplectic action}
\label{sec:CSsymp}

We would like to define an affine symplectic action on the Hilbert space $\CH\equiv \CH_{0,s}$ (or products thereof) that intertwines an obvious symplectic action in the algebra of operators \eqref{CSops}. This was  already done in Section \ref{sec:actions} for the index, but we want to rediscover it here from the point of view of Chern-Simons theory and geometric quantization.

Fortunately, a holomorphic version of the desired symplectic action was  defined in \cite{Dimofte-QRS}. There, the relevant tetrahedron (say) phase space was viewed as a complexification of $\big(\CP_\R,\tfrac{i}{2\hbar}\Omega_\R\big)$, where $\CP_\R$ is the slice of $\CP$ with $X,\,P\in \R$, and $\Omega_\R$ is the corresponding real restriction of the holomorphic symplectic form \eqref{Omega}. Quantization then produced a Hilbert space $\CH^{SL(2)}\simeq L^2(\R)$ containing functions $f(X;\hbar)$. Actual Chern-Simons wavefunctions had the additional property that they were locally \emph{analytic} in $X$ and $\hbar$. Such analytic ``$SL(2)$'' wavefunctions were identified with ellipsoid partition functions of theories $T_M$ in \cite{Yamazaki-3d,DG-Sdual,DGG}.
We recall from \cite{Dimofte-QRS} that the symplectic group $Sp(2,\Z)$ acted on wavefunctions $f(Z;\hbar)$ in the Weil representation \cite{Shale-rep, Weil-rep}. In particular, the generators $T$ and $S$ were implemented as
\begin{subequations} \label{SpSL}
\begin{align} T\;&:\quad f(X;\hbar)\;\mapsto\; (T\circ f)(X';\hbar)= e^{\frac{1}{2\hbar}X'^2}f(X;\hbar)\,, \\
S\;&:\quad f(X;\hbar)\;\mapsto\; (S\circ f)(X';\hbar)= \frac{1}{\sqrt{2\pi i\hbar}}\int_\R dX\, e^{\frac{1}{\hbar}XX'} f(X;\hbar)\,.
\end{align}
\end{subequations}
These generate a projective representation of $Sp(2,\Z)$ on $L^2(\R)$ --- projective because the group relations $(ST)^3=id$ and $S^2=C$ only hold up to a phase.

In the present case of $SL(2,\C)$ Chern-Simons at $k=0$, the complex phase space $(\CP,\omega_{0,s})$ can roughly%
\footnote{A precise statement is that the analytic continuation of $(\CP,\omega_{0,s})$, with independent complex coordinates $X,P$ and $\ol X,\ol P$, is isomorphic to $(\CP,\tfrac{i}{2\hbar}\Omega)\times(\CP,-\tfrac{i}{2\hbar}\Omega)$. We do \emph{not} want to analytically continue anything here, though.} %
be considered a product of holomorphic and antiholomorphic copies of $\big(\CP_\R,\pm\tfrac{i}{2\hbar}\Omega_\R\big)$. This is easy to see from the expression \eqref{ws1} for $\omega_{0,s}$. We can then guess what the appropriate symplectic action on $(\CP,\omega_{0,s})$ should be by combining holomorphic and antiholomorphic copies of \eqref{SpSL}.

For example, let us work in representation $(b)$ of the Hilbert space $\CH$, so that wavefunctions $f^{(b)}(x,\ol x;q)$ depend non-holomorphically on $x=\exp(X)$. The natural guess for the $T$ action is
\be T\;:\quad f(x,\ol x;q)\;\mapsto\; (T\circ f)(x',\ol x';q) = e^{\frac{1}{2\hbar}X'^2-\frac{1}{2\hbar}\ol X'^2}f(x',\ol x';q)\,, \label{Ts}\ee
with $X'=\log x'$.
It is important to check that the right-hand side of \eqref{Ts} is still an element of $\CH$, namely that it is periodic as $X'\to X'+2\pi i$, and that $\Re \,X' \in \tfrac\hbar2\Z$ is quantized. The latter condition is automatic, since $f(x,\ol x;q)$ was only initially defined for $\Re \,X = \log |x| \in \tfrac\hbar2\Z$; so let us set $\Re \,X' = m'\tfrac\hbar 2$. Then the exponential in \eqref{Ts} becomes
\be e^{\frac{1}{2\hbar}X'^2-\frac{1}{2\hbar}\ol X'^2} = e^{im'\,\Im\,X'}\,, \ee
enforcing the required periodicity in $\Im\, X'$.

Similarly, a first guess for the $S$ transformation would be
\be S\;:\quad f(x,\ol x;q)\;\overset{?}{\mapsto}\; (S\circ f)(x',\ol x';q)=\int dX\,d\ol X\,e^{\frac1\hbar XX'-\frac1\hbar \ol X\ol X'}f(x,\ol x;q)\,, \label{Ss1}\ee
with some appropriate normalization.
Now, setting $\Re\,X = m\tfrac\hbar2$ and $\Re\,X'=m'\tfrac\hbar2$, the exponent becomes $\tfrac1\hbar XX'-\tfrac1\hbar \ol X\ol X'=im'\,\Im\,Z+im\,\Im\,X'$\,. Therefore, quantization of $\Re\,X$ ensures periodicity in $\Im\,X'$, while requiring quantization of $\Re\,X'$ makes the entire integrand in \eqref{Ss1} periodic in $\Im\,X$. This is almost as desired, except that the integration measure $\int dX\,d\ol X$ does not quite make sense. The correct interpretation of $\int dX\,d\ol X$ is as an integration over a fundamental domain where $f(x,\ol x;q)$ is defined, \ie\ $\int dX\,d\ol X \,\to\, \sum_{m\in \Z}\int_0^{2\pi} d\,\Im X$\,. If we use the notation $x = q^{\frac m2}\zeta,\,\ol x'=q^{\frac {m'}2}\zeta'$, then the proper transformation becomes
\be S\;:\quad f(m,\zeta;q)\;\mapsto\; (S\circ f)(m',\zeta';q) = \sum_{m\in \Z} \int \frac{d\zeta}{2\pi i\zeta}\,\zeta^{m'}{\zeta'}^mf(m,\zeta;q)\,, \label{Ss2}\ee
with integration done over the unit circle. The normalizations in \eqref{Ts} and \eqref{Ss2} are just right to ensure that $T$ and $S$ generate a true representation of $Sp(2,\Z)$ acting on our Hilbert space $\CH$.

It should now be clear that the transformations \eqref{Ts}--\eqref{Ss2} coincide with the $Sp(2,\Z)$ action on the index, described in \cite{KW-index} and in Section \ref{sec:Sp} above. To make the correspondence even clearer, we could use a ``charge basis'' for wavefunctions in $\CH$, option $(d)$ of Table \ref{tab:pol}. Using Fourier transform to relate polarizations $(b)$ and $(d)$, we then recover the more familiar symplectic action
\be T\;:\; f(m,e;q)\;\mapsto\; f(m'-e',e';q)\,, \qquad S\;:\; f(m,e;q)\;\mapsto\; f(e',-m';q)\,,\ee
or more generally,
\be g\;:\; f(m,e;q)\;\mapsto\; f\big( g^{-1}\!\cdot\! \big(\begin{smallmatrix} m\\e\end{smallmatrix}\big);q)\,,\ee
\cf\ \eqref{Spem}.
This last expression generalizes nicely to an $Sp(2N,\Z)$ action on products of Hilbert spaces $\CH$.

Note that both the options $(a\!-\!d)$ in Table \ref{tab:pol} and the present $Sp(2,\Z)$ action can be understood as choices of basis in $\CH$, or of polarization for $\CP$. More precisely, it is a choice of element $g\in Sp(2,\Z)$ together with a choice $(a\!-\!d)$ from Table \ref{tab:pol} that fully determines a polarization. The two choices are largely independent. The element $g\in Sp(2,\Z)$ acts \emph{holomorphically} on the phase space $\CP$, viewed as $\C^*\times \C^*$, and makes a rough selection of ``position'' ($X'$) and ``momentum'' ($P'$) coordinates, with $\left(\begin{smallmatrix} X'\\ P'\end{smallmatrix}\right) = g\left(\begin{smallmatrix} X\\ P\end{smallmatrix}\right)$. The choice $(a\!-\!d)$ then splits the real and imaginary parts of $X'$ and $P'$ to provide a specific real polarization for $(\CP,\omega_{0,s})$. In terms of an index, the former comes from an $Sp(2,\Z)$ action on 3d SCFT's, while the latter is a choice of fugacity-vs-charge basis for writing the index. The one relation between the two choices is that switching $(b)\leftrightarrow(c)$ is equivalent to applying $S\in Sp(2,\Z)$.

In addition to a symplectic action, both Chern-Simons theory and the index require an action of affine shifts. In the holomorphic version of Chern-Simons theory discussed in \cite{Dimofte-QRS}, such shifts in position or momentum always came in multiples of $i\pi+\tfrac\hbar2$, with $i\pi$ the classical shift and $\hbar/2$ a quantum correction. Now, for full $SL(2,\C)$ Chern-Simons theory, we would expect a shift $\sigma_X$ in position $X$ to act as
\be \sigma_X\,:\,f(x,\ol x;q)\;\mapsto\; f(-q^{-\frac12}x,-q^{\frac12}\ol x;q)\,,\quad\text{or}\quad  \sigma_X\,:\,f(m,\zeta;q)\;\mapsto\; f(m,-q^{-\frac12}\zeta;q)\,,\ee
intertwining the operator algebra transformation $\big(\hat X,\hat{\ol X}\big)\mapsto \big(\hat X+i\pi+\tfrac\hbar2,\hat {\ol X}-i\pi-\tfrac\hbar2\big)$. Note that the quantum correction has a different sign for the holomorphic and antiholomorphic coordinates, due to the opposite signs appearing in the symplectic structure $\omega_{0,s}$ \eqref{ws1}. Similarly, a shift by $i\pi +\tfrac\hbar2$ in momentum $P$ acts as
\be \sigma_P\,:\,f(x,\ol x;q)\;\mapsto\; e^{\frac1\hbar X(i\pi+\frac\hbar 2)-\frac1\hbar X(-i\pi-\frac\hbar 2)}f(x,\ol x;q) = \big(-q^{\frac12}\big)^mf(x,\ol x;q)\,.\ee
Again, these transformations are consistent with the combined fermion number and R-charge shifts of the index defined in Section \ref{sec:aff}.

In contrast to analytically continued Chern-Simons theory, where the representation of the affine symplectic group on wavefunctions is only projective, and leads to normalization ambiguities for wavefunctions of 3-manifolds, the present actions in $SL(2,\C)$ Chern-Simons theory at $k=0$ define a true, faithful representation of $Sp(2,\Z)\ltimes \big[\big(i\pi+\tfrac\hbar2\big)\Z\big]^2$ on $\CH$. It is straightforward to check that group relations such as \eqref{affrels} hold on the nose. More generally, we have a true representation of $Sp(2N,\Z)\ltimes \big[\big(i\pi+\tfrac\hbar2\big)\Z\big]^{2N}$ on $\CH^{\otimes N}$. Therefore, wavefunctions obtained via a symplectic gluing procedure --- the procedure outlined in Section \ref{sec:rules} --- should be unambiguously defined. As already mentioned above, invariance of gluing under 2-3 moves completely fixes the normalization of the basic building block for any triangulated 3-manifold, the wavefunction of tetrahedron.

The final step in constructing wavefunctions for a triangulated 3-manifold is a ``symplectic reduction,'' the quantum equivalent to setting internal edge coordinates $C_I\to 2\pi i$. We finish this discussion by commenting briefly on it. In the analytically continued Chern-Simons theory, the reduction is implemented by transforming a wavefunction to a (holomorphic) polarization such that $C_I$ is a position coordinate, and simply setting $C_I \to 2\pi i +\hbar$, where the `$+\hbar$' is the usual quantum correction. For the non-holomorphic $SL(2,\C)$ wavefunction, this translates to setting $C_I\to 2\pi i+\hbar \simeq \hbar$ and $\ol C_I \to -2\pi i-\hbar\simeq -\hbar$; or, with $x_I = e^{C_I} = q^{\frac{m_I}{2}}\zeta_I$, setting $m_I\to 0$ and $\zeta_I \to q$. In a charge basis, this means
\be\text{reduction\,:}\qquad \CZ(m_I,m,e_I,e;q)\;\mapsto\; \sum_{e_I\in \Z} q^{e_I}\CZ(0,m,e_I,e;q)\,,\qquad\ee
just as in \eqref{indsum}, the last step of the index gluing rules. This completes the argument that the Chern-Simons wavefunction at $k=0$ is fully equivalent to the index.

\subsection{Chern-Simons theory from 6d}
\label{sec:6d}

As we argued in \cite{DGG}, a three-dimensional $\CN=2$ theory $T_M$ labeled by a 3-manifold $M$ can be thought of
as the result of compactification of the six-dimensional $(2,0)$ theory of type $A_1$ on a 3-manifold $M$.
Therefore, a 3d $\CN=2$ theory $T_M$ in the space-time $S^1 \times S^2$ relevant to the index computation
can be equivalently approached by studying the $A_1$ 6d theory that describes two coincident M5-branes on $S^1 \times S^2 \times M$.

This configuration of two five-branes on $S^1 \times S^2 \times M$ fits in a large web of dualities
that connects a number of closely related systems studied in the literature. Here, we briefly discuss some of
these dualities, providing a further piece of evidence for the interpretation of 3d index
in terms of the full $SL(2,\C)$ Chern-Simons theory (not just a holomorphic half of it).

In general, putting a 3d $\CN=2$ theory $\CT$ in different space-times with an $\Omega$-background
allows one to associate different quantities (partition functions) to $\CT$.
Below we list several examples of such partition functions and the corresponding 3d space-times,
illustrating their concrete form in a basic example of the tetrahedron theory $\CT = T_{\Delta}$,
which basically consists of a single chiral multiplet:
\begin{align}
S^1 \times S^2 &\; : \qquad \CI [T_{\Delta}]  \; = \;
\prod_{r=0}^\infty \frac{1- \q^{r+1}z^{-1}}{1-\q^{r}\bar z^{-1}} \notag \\
\text{``solid torus''} \quad S^1 \times \R^2_{\hbar} &\; : \qquad \CZ_{\text{vortex}} [T_{\Delta}] \; = \;
\prod_{r=0}^{\infty} (1 - q^{r +1} z^{-1}) \label{ZZZ} \\
\text{``ellipsoid''} \quad S^3_b \quad & \; : \qquad \CZ_b [T_{\Delta}] \; = \; \prod_{r=0}^{\infty}
\frac{1 - q^{r+1}\, z^{-1}}{1 - ({}^Lq)^r ({}^Lz)^{-1}} \notag
\end{align}
where, as usual, the $\Omega$-deformation parameter is proportional to $\hbar$, and in the case of $S^3_b$ there are dual deformations proportional to $\hbar=2\pi ib$ and ${}^L\hbar = 2\pi i b^{-1}$. In all cases, $z$ is related to a fugacity for the flavor $U(1)$ symmetry (and ${}^Lz = z^{\frac{2\pi i}{\hbar}}$).

By comparing the expressions of these partition functions for $\CT = T_{\Delta}$,
the reader will immediately notice that both the index $\CI [\CT]$ and the ellipsoid partition function $\CZ_{b} [\CT]$
roughly contain two copies of the vortex partition function $\CZ_{\text{vortex}} [\CT]$.
This observation has a simple heuristic explanation based on the geometry of the three-dimensional space-times
that lead to these partition functions.
Indeed, both a 3-sphere and $S^1 \times S^2$ consist of two copies of the solid torus glued together either ``back-to-back''
or with a twist, {\it i.e.} in a way where the A-cycle on the boundary torus of one copy
is glued to the A-cycle on the boundary of the second copy, or to the dual B-cycle.
These two ways of gluing solid tori produce $S^1 \times S^2$ and $S^3$, respectively,
which in turn lead to the partition functions~\eqref{ZZZ}.

\begin{wrapfigure}{r}{2in}
\centering
\includegraphics[height=2.3in]{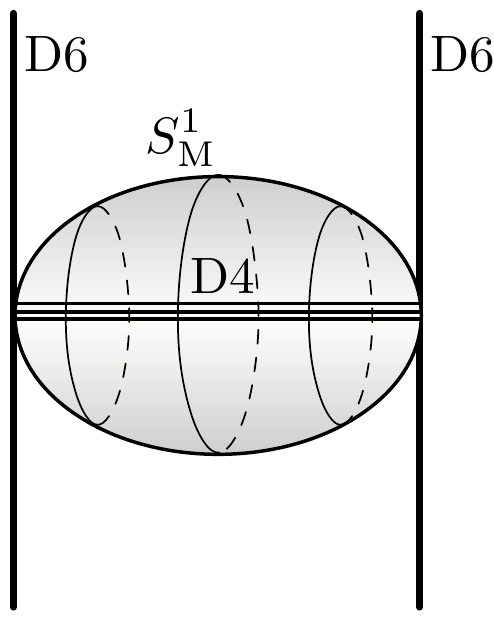}
\caption{Upon reduction on a circle fiber of $S^2$, a configuration of M5-branes wrapped on $S^2$
becomes a system of D4-branes stretched between two D6-branes in type IIA string theory.}
\label{D4D6}
\end{wrapfigure}

Another intuitive explanation of the relation between different partition functions \eqref{ZZZ}
and Chern-Simons theory with complex gauge group $SL(2,\C)$
can be seen directly in the system of two M5-branes supported on
$S^3_b \times M$ or $S^1 \times D \times M$, where $D = S^2$ leads to the superconformal index
and $D = \R^2_{\hbar}$ leads to the vortex partition function.
In each case, it is convenient to use a dual description of this five-brane configuration
by reducing it on a circle down to type IIA string theory.
There are several choices on can consider, which depend on the choice of the ``M-theory circle,''
and all of which are supposed to provide equivalent descriptions of the same physical system.
For example, if a 3-manifold $M$ admits a circle action, then one can find a dual description
of the five-brane system in terms of D4-branes wrapped on the quotient space $M/U(1)$.

In our discussion, the 3-manifold $M$ is generic\footnote{\emph{I.e.}, we do not make any special assumptions
about isometries of $M$.} and since the $S^1$ part of the 3d space-time
is used in the definition of the 3d index, our only choice is to take the M-theory circle be a part of the $D$ or $S^3$ geometry.
Luckily, all our 3d space-times in \eqref{ZZZ} admit such circle actions (aside from obvious isometries arising from the $S^1$ factor).
In particular, the sphere $D = S^2$ relevant for the 3d index can be represented as a circle fibration,
\be
\begin{matrix}
S^1_{\rm M} & \to & S^2 \\
& & \downarrow \\
& & I
\end{matrix}
\ee
with two degenerate fibers at the end-points of the interval $I = [0,1]$.
Therefore, if we identify the M-theory circle with $S^1_{\rm M}$,
we can relate the original five-brane system on $S^1 \times S^2 \times M$
with the system of D4-branes stretched between two D6-branes, as shown in Figure~\ref{D4D6}.
In other words, the 3d index (= partition function of the five-brane theory on $S^1 \times S^2 \times M$)
can be equivalently computed in a gauge theory that lives on the world-volume of the D4-branes,
\be
S^1 \times I \times M \,,
\ee
with suitable boundary conditions associated with the D6-branes at the end-points of the interval $I$.
Note, upon this reduction to type IIA theory the angular momentum for the $U(1)$ rotation symmetry
of $D = S^2$ (similarly for $D = \R^2$) becomes the instanton charge $k_{\text{inst}} = \frac{1}{16 \pi^2} \int_{I \times M} \Tr F \wedge F$
in the D4-brane gauge theory. One quick way to see this is to note that instantons in
the D4-brane theory can be interpreted as D0-branes which, in turn, are precisely the Kaluza-Klein modes
for the reduction on $S^1_{\rm M}$.

The resulting configuration of D4-branes ending on the D6-branes essentially comprises two (interacting) copies
of the D6-D4 brane system studied in \cite{Wfiveknots}, where it was argued that its partition
function equals the partition function of analytically continued Chern-Simons theory, see also~\cite{DGH}.
In the present case, we have two such systems coupled together. We claim that these are precisely the
two sectors (``holomorphic'' and ``anti-holomorphic'') of the full $SL(2,\C)$ Chern-Simons theory discussed in Section~\ref{sec:k0}.

Moreover, the coupling constants in these two sectors are related precisely as in Section~\ref{sec:k0}.
Indeed, in a single D6-D4 brane system the coupling constant of the analytically continued $SU(2)$ Chern-Simons theory
is identified with the fugacity for the instanton charge $k_{\text{inst}}$. If we denote this fugacity by $q = e^{\hbar}$,
then the fugacity of the second D6-D4 brane system is equal to $\tilde q = e^{- \hbar} = q^{-1}$ due to the opposite orientation.
This is exactly the relation between coupling constants $t = - \tilde t = is$ that we encountered in Section~\ref{sec:k0},
with the identification $\hbar = \frac{4 \pi}{s}$. In fact, we can even observe that the $\rho$ symmetry of Section \ref{sec:kappa}, lifted to the D6-D4 system, simply looks like a reflection of the geometry that exchanges the holomorphic and anti-holomorphic sectors, and the two fugacities $q$ and $\tilde q$. \\


\acknowledgments{We would like to thank Christopher Beem, Abhijit Gadde,
Daniel Green, Anton Kapustin, Sara Pasquetti, Leonardo Rastelli, Schlomo Razamat,
Nathan Seiberg, Cumrun Vafa, Roland van der Veen, and Edward Witten for illuminating discussions.
The work of TD is supported primarily by the Friends of the Institute for Advanced Study, and in part by DOE grant DE-FG02-90ER40542.
The work of DG is supported in part by NSF grant PHY-0503584
and in part by the Roger Dashen membership in the Institute for Advanced Study.
The work of SG is supported in part by DOE Grant DE-FG03-92-ER40701 and in part by NSF Grant PHY-0757647.
TD and SG thank the Simons Center for Geometry and Physics for their hospitality during the Simons Workshop in Mathematics and Physics, 2011, where part of this work was initiated.
Opinions and conclusions expressed here are those of the authors and do not necessarily reflect the views of funding agencies.} \\

\noindent\hrulefill

\appendix

\section{Proof of triality for the tetrahedron index}
\label{app:ST}

We would like to prove that the index is invariant under the affine $\sigma_eST$ transformation discussed in Section \ref{sec:tri}. We will prove it in the form
\be \label{appST}
\CI_\Delta(m,e;q) = \big(-q^{\frac12}\big)^m\,\CI_\Delta(-e-m,m;q)\,,
\ee
which immediately implies the middle equality of \eqref{tettri} as well.

There are several possible ways to show this, but one of the easiest (and most instructive) is to invoke the difference equations for the index. Recall from Section \ref{sec:tetdiff} that
\be (\hat p_+ + \hat x_+^{-1}-1)\CI_\Delta(m,e) = (\hat p_- + \hat x_-^{-1}-1)\CI_\Delta(m,e) = 0\,, \label{appTeqs}\ee
where $\hat p_\pm,\,\hat x_\pm$ are defined by
\be \hat x_\pm = q^{\frac m2}e^{\mp \pd_e}\,,\qquad \hat p_\pm = q^{\frac e2}e^{\pm \pd_m}\,. \label{appXP1}
\ee
The validity of \eqref{appTeqs} is trivial to show in a fugacity basis, using formula \eqref{Iztet2}, and the fugacity-basis operators \eqref{XPz}. We claim that the equations \eqref{appTeqs} determine the index completely as long as two of its values are fixed, on either the $m$ or $e$ axis (and also assuming that $\CI_\Delta(m,e)$ is a formal series in $q^{\frac 12}$ for every $(m,e)$).

To see this, first suppose that we knew $\CI_\Delta(m,e)$ everywhere on (say) the $m$-axis, at $e=0$. Then the equations \eqref{appTeqs} involve a single shift in $e$, in opposite directions, so they can be used to solve for $\CI_\Delta(m,e)$ everywhere else. To get $\CI_\Delta(m,e)$ on the $m$-axis, we can rewrite the operators $\hat x_\pm,\hat p_\pm$ in terms of the rotated operators
\be \hat\eta  = e^{\pd_e}\,,\quad \hat\epsilon=q^e\,;\qquad \hat \eta_m=e^{\pd_m}\,,\quad \hat \epsilon_m = q^m\,.\ee
Then, essentially by using elimination in a left ideal, we can eliminate the electric shift $\eta$ from \eqref{appTeqs}, and find a single equation only involving $\hat\epsilon,\hat\epsilon_m,\hat\eta_m$. At $e=0$ ($\hat \epsilon=1$), it reads
\be \big(\hat \eta_m+\hat \eta_m^{-1}+\hat\epsilon_m^{-1}-2\big)\CI_\Delta(m,0) =0\,,\ee
or
\be \CI_\Delta(m+1,0)+\CI_\Delta(m-1,0)+(q^{-m}-2)\CI_\Delta(m,0) = 0\,.\ee
This is a second-order difference equation in $m$, whose coefficients are nowhere vanishing. Therefore, knowing $\CI_\Delta(m,0)$ at two values of $m$ is sufficient for finding a unique solution along the entire $m$-axis. (A similar story holds for the $e$-axis, but we will not need it.)

Now, let's define $\CI'(m',e') \equiv \big(-q^{\frac12}\big)^{m'}\CI_\Delta(-e'-m',m';q) = \CI_{(\sigma_eST)^{-1}\circ T_\Delta}(m',e')$. The difference equations that $\CI'(m',e')$ obeys are the affine-symplectic $(\sigma_eST)^{-1}$--images of \eqref{appTeqs}. For example, from the inverse of \eqref{XPST}, we find that we should send
\be \hat x_\pm \to -\frac{1}{\hat p_\pm'}\frac{1}{\hat x_\pm'}\,,\qquad \hat p_\pm \to \hat x_\pm\,, \ee
so that the operators annihilating $\CI'(m',e')$ are
\be \hat p_\pm+\hat x_\pm-1\quad\to\quad \hat x_\pm'-\hat x_\pm'\hat p_\pm'-1 = -\hat x_\pm'(\hat p_\pm'+\hat x_\pm'{}^{-1}-1)\,. \label{appopsp}\ee
Therefore, since $\hat x_\pm'$ is invertible, the transformed index $\CI'(m',e')$ satisfied exactly the same difference equations, in the new variables:
\be (\hat p_+' + \hat x_+'{}^{-1}-1)\CI'(m',e') = (\hat p_-' + \hat x_-'{}^{-1}-1)\CI'(m',e') = 0\,. \ee
Then, to prove \eqref{appST}, it suffices to show that $\CI'(m',e')=\CI_\Delta(m',e')$ at two points on the $m'$-axis.

It is trivial that $\CI'(0,0)=\CI_\Delta(0,0)$. For our second point, we take $(m',e')=(-1,0)$, and need to show that $\CI'(-1,0)\equiv -q^{-\frac12}\CI_\Delta(1,-1) = \CI_\Delta(-1,0)$. But this follows from parity symmetry \eqref{tetPem} of the tetrahedron index! So we are done.

For completeness, since we never proved parity explicitly in the text, let us do so here. In an electric fugacity basis, parity symmetry takes the form
\be \CI_\Delta(m;\zeta,q) = \big(-q^{\frac12}\big)^m\zeta^{-m}\CI_\Delta(-m;\zeta,q)\,. \ee
Using formula \eqref{Iztet2} for the tetrahedron index, we then check
\begin{align*}
\frac{\CI_\Delta(-m,\zeta)}{\CI_\Delta(m,\zeta)}
&= \prod_{r=0}^\infty \frac{1-q^{r+\frac m2+1}\zeta^{-1}}{1-q^{r+\frac m2}\zeta} \frac{1-q^{r-\frac m2}\zeta}{1-q^{r-\frac m2+1}\zeta^{-1}} \\
&= \left\{ \begin{array}{l@{\quad}l}
   \prod_{r=0}^{m-1}\frac{1-q^{r-\frac m2}\zeta}{1-q^{r-\frac m2+1}\zeta^{-1}} & m \geq 0 \\[.2cm]
   \prod_{r=0}^{|m|-1}\frac{1-q^{r+\frac m2-1}\zeta^{-1}}{1-q^{r+\frac m2}\zeta} & m < 0
   \end{array}\right. \\
&= \big(-q^{\frac12}\big)^{-m}\zeta^m\,,
\end{align*}
by straightforward algebra, exactly as desired.

\section{Figure-eight index from six tetrahedra}
\label{app:6}

\begin{figure}[hb]
\centering
\includegraphics[width=5in]{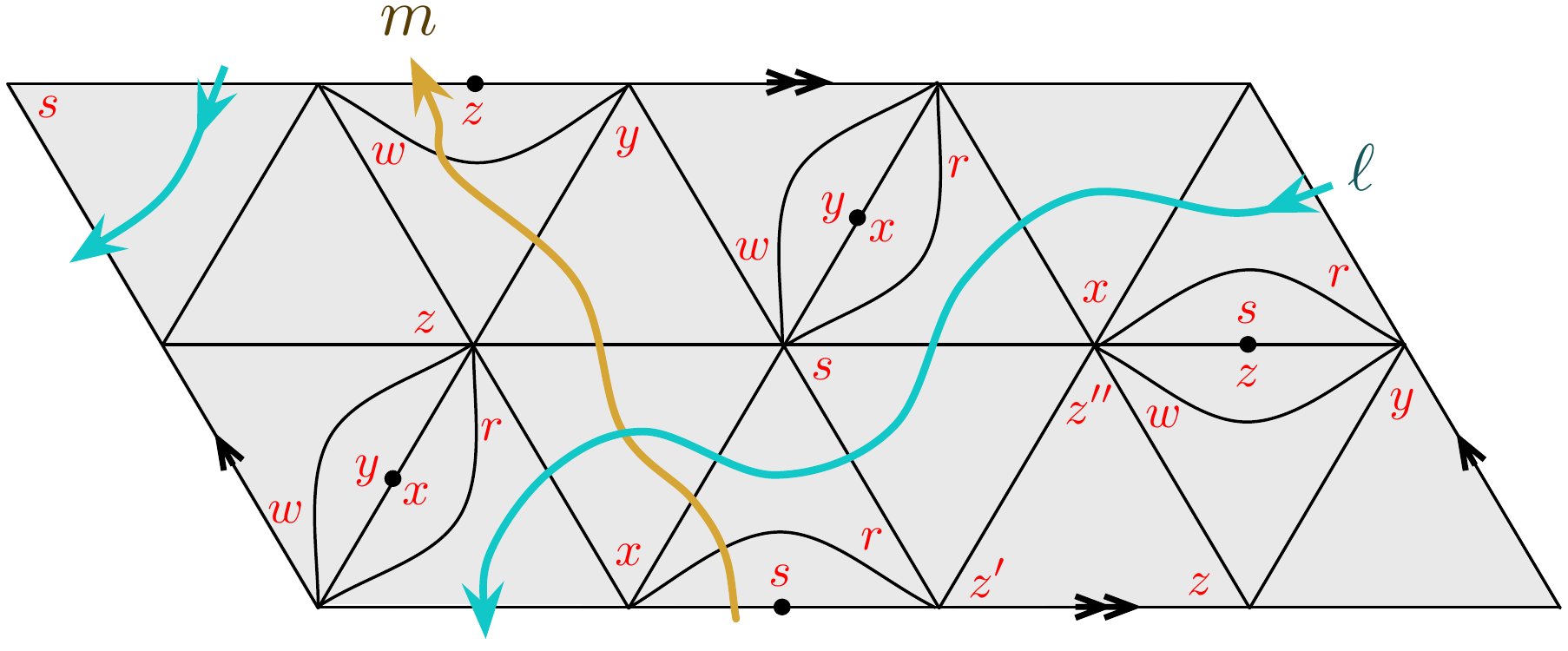}
\caption{Map of the torus boundary for the refined triangulation of the figure-eight knot.}
\label{fig:cusp6}
\end{figure}

The gluing data for the six-tetrahedron triangulation of the figure-eight knot complement was described in Section 4.6 of \cite{DGG}. A map of the torus boundary, which encodes all the necessary combinatorial data, is reproduced in Figure \ref{fig:cusp6}. The six tetrahedra have edge parameters $(Z,W,X,Y,R,S)$. Here we start out in a product polarization $\Pi_{Z}\times\cdots\times \Pi_{S}$, and find that we need to implement a change of polarization
\be \begin{pmatrix} U\\C_1\\C_2\\C_3\\C_4\\C_5\\ v \\ \Gamma_1\\\Gamma_2\\\Gamma_3\\\Gamma_4\\\Gamma_5 \end{pmatrix}
 = \left(\begin{array}{cccccccccccc}0 & 0 & 1 & 0 & -1 & -1 & 1 & -1 & 1 & -1 & -1 & -1 \\
 -2 & -2 & 0 & 1 & 1 & 0 & -2 & -2 & 0 & 0 & 0 & 2 \\
 0 & 1 & 0 & -2 & 0 & 1 & 0 & 0 & 2 & -2 & 2 & 0 \\
 1 & 0 & -2 & 0 & 1 & 0 & 0 & 2 & -2 & 2 & 0 & 0 \\
 0 & 0 & 1 & 1 & 0 & 0 & 0 & 0 & 0 & 0 & 0 & 0 \\
 1 & 0 & 0 & 0 & 0 & 1 & 0 & 0 & 0 & 0 & 0 & 0 \\
 0 & 0 & -1 & 0 & 1 & 1 & 0 & 0 & 0 & 0 & 2 & 0 \\
 -1 & -1 & 0 & 1 & 1 & 0 & -1 & -1 & 0 & 0 & 1 & 1 \\
 0 & 0 & 0 & 1 & 1 & 0 & 0 & 1 & 0 & 0 & 2 & 0 \\
 -1 & 0 & 1 & 1 & 1 & 0 & 0 & 0 & 1 & -1 & 3 & 0 \\
 -1 & 0 & -2 & 0 & 3 & 0 & -1 & 3 & 0 & 1 & 5 & 1 \\
 0 & 0 & 0 & -1 & -1 & 0 & 1 & -2 & 1 & -1 & -1 & 0
   \end{array}\right)
   \begin{pmatrix} Z_1\\Z_2\\Z_3\\Z_4\\Z_5\\Z_6 \\ Z_1''\\Z_2''\\Z_3''\\Z_4''\\Z_5''\\Z_6'' \end{pmatrix}
    + \begin{pmatrix} 0 \\ 2\pi i+\hbar \\ 2\pi i+\hbar \\ 2\pi i+\hbar \\ 2\pi i+\hbar \\ 2\pi i+\hbar  \\ 0 \\ 0 \\0\\0\\0\\0 \end{pmatrix}\,.
\label{pol6}
\ee
By applying the gluing rules of Section \ref{sec:rules}, we then obtain an index
\begin{align} \CI_{\mb{4_1}}(m,e) = &\sum_{e_i\in \Z}
 \big(-q^{\frac12}\big)^{m-e2+2e_5+2e_6}\,\CI_\Delta(2 e_2-e,-2 e_2+e_4+e_6)\,
 \CI_\Delta(e+2
   e_2-2 e_4,e_3-2 e_2) \notag \\ &\times \CI_\Delta(-e-2 e_3+2
   e_4,e-2 e_4+e_5+m)\,
 \CI_\Delta(e+2 e_3-2
   e_4,e_2-2 e_3+e_5) \notag \\ &\times \CI_\Delta(e-2 e_3+2
   m,-e+e_2+e_4-m)\,
 \CI_\Delta(e-2 e_2,-e+e_3+e_6-m)\,. \label{ind6}
\end{align}
As discussed in Section \ref{sec:knots}, we have checked computationally that this is equivalent to the much simpler expression \eqref{ind41} for the figure-eight knot index.

\section{Quantum Lagrangian calculations}
\label{app:L}

Here we calculate explicitly the quantum Lagrangian operators $\hat \CL_M$ that annihilate a few indices, as promised in Section \ref{sec:glueops}. We follow the computational framework delineated in Section \ref{sec:glueops}, or in \cite{Dimofte-QRS}.

\subsubsection*{Bipyramid}

Let's begin with the operators for the bipyramid theory. If we build the bipyramid from two tetrahedra, we can start with two pairs of operators
$ \hat r''_\pm+\hat r^{-1}_\pm -1,\; \hat s''_\pm+\hat s^{-1}_\pm -1 $
that both annihilate the product index $\CI_\Delta(m_R,e_R)\,\CI_\Delta(m_S,e_S)$, as in Section \ref{sec:bip}. We could write this somewhat more suggestively as
\be \hat r''_\pm+\hat r^{-1}_\pm -1 \simeq 0\,,\qquad \hat s''_\pm+\hat s^{-1}_\pm -1 \simeq 0\,,\ee
where by ``$\simeq$'' we mean ``annihilates an appropriate index when acting on the left.'' To be completely explicit, the fundamental generators $\hat r_\pm = \exp(\hat R_\pm)$, $\hat s_\pm =\exp(\hat S_\pm)$, etc. act as
\begin{align*} \hat R_\pm = \tfrac\hbar 2m_R\mp \pd_{e_R} &\qquad \hat S_\pm =\tfrac\hbar 2m_S\mp \pd_{e_S}\,,\\
\hat R_\pm'' = \tfrac\hbar 2e_R\pm \pd_{m_R} &\qquad \hat S_\pm'' =\tfrac\hbar 2e_S\pm \pd_{m_S}\,,
\end{align*}
\cf\ \eqref{XP}. (These $(\hat r,\hat r'')$ and $(\hat s,\hat s'')$ are taking the place of what was called $(\hat x,\hat p)$ in Step 1 above.) Now, in accordance with the change of polarization \eqref{pol2}, we  change variables to
\be \hat X_{1\,\pm} = \hat R_\pm+\hat S_\pm''\,,\qquad \hat X_{2\,\pm}=\hat R_\pm''+\hat S_\pm\,,\qquad \hat P_{1\,\pm}=\hat R_\pm''\,,\qquad \hat P_{2\,\pm}=\hat S_\pm''\,.\ee
Then the properly polarized index of the bipyramid, $\CI_{\rm bip}(m_1,m_2,e_1,e_2) = \CI_\Delta(m_1-e_2,e_1)\,\CI_\Delta(m_2-e_1,e_2)$ \eqref{ind2}, is annihilated by
\be \hat \CL_{\rm bip}^{(1)} = \hat p_{1\,\pm}+\hat x_{1\,\pm}^{-1}\hat p_{2\,\pm}-1 \simeq 0\,,\qquad \hat \CL_{\rm bip}^{(2)} = \hat p_{2\,\pm} + \hat x_{2\,\pm}^{-1}\hat p_{1\,\pm}-1\simeq 0\,. \label{bip2ops}
\ee
This is an obvious quantization of the bipyramid Lagrangian, given for example in Section 2 of \cite{DGG}. There are no internal edges, so we are done.

Alternatively, we can use three tetrahedra. We start with three equations%
\footnote{Remember from Section \ref{sec:tri} that the operator for a tetrahedron theory is cyclically invariant under permutations of the shape parameters. Thus we can freely write the $y$ equation using $(\hat y',\hat y)$ instead of $(\hat y,\hat y'')$, as we did here.}
\be \hat z''+\hat z^{-1}-1\simeq 0\,,\qquad
\hat w''+\hat w^{-1}-1\simeq 0\,,\qquad
\hat y+\hat y'{}^{-1}-1\simeq 0\,, \label{bip3ops0} \ee
where for clarity we just use the `$+$' operators and remove the subscript $\pm$. In parallel to the change of polarization \eqref{pol3}, we change variables to
\be \hat X_{1} = \hat Z\,,\qquad \hat X_{2}=\hat W\,,\qquad \hat C=\hat Z+\hat W+\hat Y\,, \notag \ee
\be \hat P_{1}=\hat Z''+\hat Y'\,,\quad
\hat P_{2}=\hat W''+\hat Y'\,,\quad
\hat \Gamma = -\hat Y'\,. \notag
\ee
Note that $\hat p_1\hat x_1 = q\hat x_1\hat p_1$, $\hat p_2\hat x_2=q\hat x_2\hat p_2$, and $\hat \gamma\hat c= q\hat c\hat \gamma$.
In terms of exponentiated operators, the equations \eqref{bip3ops0} become
\be \hat\gamma\,\hat p_{1}+\hat x_{1}^{-1}-1\simeq 0\,,\qquad
\hat\gamma\,\hat p_{2}+\hat x_{2}^{-1}-1\simeq 0\,,\qquad
\hat c\,\hat x_{1}^{-1}\hat x_{2}^{-1}+\hat\gamma -1 \simeq 0\,. \label{bip3ops1}\ee
By multiplying on the left and taking differences of these equations, we can eliminate $\hat \gamma$ from \eqref{bip3ops1}, leaving two equations that only involve $\hat x_i,\hat p_i,\hat c$. Setting $\hat c\to q$ in these equations recovers the Lagrangian operators \eqref{bip2ops}.

\subsubsection*{Figure-eight knot}

In order to build the operator for the figure-eight knot, it suffices to use the decomposition into two tetrahedra. Let's again just focus on `$+$' operators. We start with $\hat z''+\hat z^{-1}-1\simeq 0$ and $\hat w''+\hat w^{-1}-1\simeq 0$, and change variables to
\be \hat U = \hat Z-\hat W''\,,\qquad \hat C = \hat Z''+\hat W''-\hat Z-\hat W+2\pi i+\hbar\,,\qquad \hat v=\hat Z''-\hat Z\,,\qquad \hat \Gamma = -\hat W''\,. \notag
\ee
According to (semi)standard conventions in the mathematics literature, let us define $\hat M = e^{\hat U}$ and $\hat \ell = -e^{\hat v}$. The the two tetrahedron Lagrangians can then be rewritten as%
\footnote{One must be careful with factors of $q$ in these equations when de-exponentiating. For example $\hat z'' = \exp(\hat U+\hat v-\hat \gamma)=q^{\frac12}\hat M\hat \ell \hat \gamma^{-1}=q^{-\frac12}\hat \ell\hat M\hat \gamma^{-1}$.}
\be -q^{\frac12}\frac{\hat M \hat \ell}{\hat \gamma}+\frac{\hat \gamma}{\hat M}-1\simeq 0\,,\qquad \frac{1}{\hat \gamma}-q^{-\frac12}\frac{\hat c\,\hat \gamma}{\hat \ell}-1\simeq 0\,. \label{41ops1} \ee
Eliminating $\hat\gamma$ from these equations is not entirely trivial, but can be done by hand. After doing so and setting $\hat c\to 1$, what results is a single operator
\be \hat \CL_{\mb 4_1} = \big(q^{\frac12}\hat M-q^{-\frac12}\hat M^{-1}\big)\ell^{-1}-\big(\hat M-\hat M^{-1}\big)\big(\hat M^{-2}-\hat M^{-1}-q-q^{-1}-\hat M+\hat M^2\big)+\big(q^{-\frac12}\hat M-q^{\frac12}\hat M^{-1}\big)\hat\ell\,, \label{41op} \ee
which annihilates the figure-eight knot index $\CI_{\mb{4_1}}(m,e)$ in \eqref{ind41}. Here $\hat M$ and $\hat \ell$ act in the `$+$' representation, so that
\be \hat M = e^{\hat U}= \exp\big(\tfrac\hbar2m-\pd_e\big)\,,\qquad -\hat\ell=e^{\hat v}=\exp\big(\tfrac\hbar2e+\pd_m\big)\,. \label{MLact} \ee
The complementary `$-$' operator is obtained from \eqref{41op} by sending $\hat M\to \hat M_-$, $\hat \ell\to \hat \ell_-$, and $q\to q^{-1}$.

The quantum Lagrangian \eqref{41op} is well known in knot theory and Chern-Simons theory as the ``quantum A-polynomial'' of the figure-eight knot \cite{garoufalidis-2004, DGLZ, Dimofte-QRS}.%
\footnote{It is easiest to compare \eqref{41op} to the conventions of \cite{DGLZ} or \cite{Dimofte-QRS}. For example, \eqref{41op} is identical to Eqn. (1.8) of \cite{Dimofte-QRS} upon setting $\hat M\to\hat m^2$.}

\subsubsection*{Trefoil}

The quantum Lagrangian for the trefoil can be similarly obtained using our gluing rules, but we can also just take the known result from knot theory. In terms of longitude and meridian holonomy eigenvalues $\ell=-e^v$ and $M=e^U$, the classical Lagrangian (\ie\ the nonabelian A-polynomial) for the trefoil is simply $\ell+M^3$, which becomes quantized as
\be \hat\CL_{\mb{3_1}}=\hat \ell+q^{\frac32}\hat M^3\,. \label{31op} \ee
It is easy to check that this annihilates the delta-function index for the trefoil, $\CI_{\mb{3_1}}(m,e)=\delta_{e,3m}$, with $\hat \ell$ and $\hat M$ acting just as in \eqref{MLact}. The index is also annihilated by the `$-$' version of \eqref{31op}, namely $\hat\CL_{\mb{3_1}\,-}=\hat \ell_-+q^{-\frac32}\hat M_-^3$.

\section{Tentacles and difference equations}
\label{app:tent}

In this appendix, we seek to motivate the relation between tentacles of a classical amoeba of $\CL_M$ and ``linear growth'' in the index, as discussed in Section \ref{sec:tent}. We show how these two phenomena are connected via quantized difference equations $\hat \CL_M\CI_M = 0$. Our argument is heuristic, but indicates a path toward a more rigorous proof, given further information about quantization and/or indices.

For simplicity, suppose that a theory%
\footnote{This could actually be a general theory in class $\CR$; is still has a classical $\CL$ and quantum operators $\hat\CL$.} %
$T_M$ has a single $U(1)$ symmetry, so that the index $\CI_M(m,e;q)$ depends on a single pair of charges $(m,e)$. Correspondingly, the classical Lagrangian $\CL_M$ (or a component of it) can be described by a single equation, which we just write as $\CL_M(x,p)=0$, with $x,p\in \C^*$. Let us also suppose, without loss of generality, that the amoeba of $\CL_M$ has a tentacle in the negative $\Re\,P$ direction, \ie\ $\CL_M(x,p)=0$ has a solution with $|p|\to 0$ and $x$ finite. Otherwise, we can use the $Sp(2,\Z)$ action to rotate any other amoeba tentacle to this position.

Let us write
\be \CL_M = a_0(x) + a_1(x)\,p+\ldots +a_d(x)p^d\,. \label{LMpoly} \ee
This polynomial is only well-defined up to multiplication by monomials in $x$ and $p$. The fact that $\CL_M(x,p)=0$ has a solution at $|p|\to 0$ means that the coefficient $a_0(x)$, multiplying the lowest power of $p$, has a nontrivial root. In particular, $a_0(x)\neq 1$. If the tentacle is directly on the negative $\Re\,P$ axis (as we would expect for character varieties of 3-manifolds, \cf\ \cite{cooper-1994}), then the root must actually be on the unit circle; \ie\ $a_0(x)=0$ for some $|x|=1$. The simplest example would be $a_0(x)=1-x$.

This property of the polynomial \eqref{LMpoly} is also familiar from the relation between amoebas and Newton polygons. Namely, every tentacle of the amoeba is perpendicular to an edge of the Newton polygon of $\CL_M$. A tentacle in the $-\Re\,P$ direction means that the Newton polygon has a ``vertical'' ($x$-direction) edge at the lowest power of $p$, which is the same thing as saying that $a_0(x)$ is nontrivial.%
\footnote{Meaning not a monomial. All equations and operators are defined up to (left) multiplication by monomials.}

Now, the index is annihilated by quantized versions of \eqref{LMpoly},
\be \hat\CL_M(\hat x_+,\hat p_+;q)\,\CI_M = \hat\CL_M(\hat x_-,\hat p_-;q^{-1})\,\CI_M = 0\,. \ee
Sometimes the degree of $\hat \CL_M$ in $\hat p$ or $\hat x$ jumps upon quantization, but we will assume here that it does not (the jumping does not affect the general story). Then we could write
\be \hat \CL_M(\hat x_+,\hat p_+;q) = \sum_{n=0}^d \hat a_n(\hat x_+;q)\hat p_+^n\,,\ee
with $\hat a_n(\hat x_+;q) \to a_n(x)$ as $q\to 1$, and similarly for the $(-)$ operators. In particular, $\hat a_0(\hat x_{\pm};q^\pm)$ is a nontrivial (classical!) polynomial in $\hat x_\pm$, with $q$-dependent coefficients.

We want to consider the behavior of the index at $m=0$ and large positive electric charge $e$. For this, the operators $\hat x_\pm, \hat p_\pm$ are not convenient, because they combine shifts in $e$ and $m$. Recall from Section \ref{sec:ops} that $(\hat x_\pm,\hat p_\pm)=(e^{\hat X_\pm},e^{\hat P_\pm})$ with
\be \label{appXP2}
\begin{array}{l@{\qquad}l}
\hat X_+ = \tfrac{\hbar}{2} m - \partial_e\,, & \hat P_+ = \tfrac{\hbar}{2} e + \partial_m\,, \\[.1cm]
\hat X_- = \tfrac{\hbar}{2} m + \partial_e\,, & \hat P_- = \tfrac{\hbar}{2} e - \partial_m\,.
\end{array}
\ee
We can rotate \eqref{appXP2} to define a more useful set of operators
\be \hat \eta = \sqrt{\frac{\hat x_-}{\hat x_+}} = e^{\pd_e}\,,\quad
\hat \epsilon = \hat p_+\hat p_- = q^e\,;\qquad
\hat \eta_m = \sqrt{\frac{\hat p_+}{\hat p_-}} = e^{\pd_m}\,,\quad
\hat \epsilon_m = \hat x_+\hat x_- = q^m\,. \label{rotatee}
\ee
Quantum-mechanically, it should then be possible to work in the left ideal generated by $\CL_M(\hat x_+,\hat p_+;q)$ and $\CL_M(\hat x_-,\hat p_-;q)$, and eliminate the combination $\hat\eta_m$, leaving a single operator
\be \hat E_M(\hat \eta,\hat \epsilon,\hat \epsilon_m;q)\,\CI_M = 0 \ee
that does not shift $m$. This operator must still annihilate the index.

In the classical limit $q\to 1$, $\hat E_M$ reduces to a classical polynomial $E_M(\eta,\epsilon,\epsilon_m)$. This polynomial is the result of taking (redundant) complex conjugate equations
\be \CL_M(x,p) = 0\,,\qquad \CL_M(\bar x,\bar p) = 0\,, \ee
setting $\eta = \sqrt{\bar x/x} = e^{-i\,\Im\,X}$, $\epsilon = |p|^2$, $\eta_m = \sqrt{p/\ol p}= e^{i\,\Im\,P}$, $\epsilon_m = |x|^2$, and eliminating the phase $\eta_m$. Thus, $E_M(\eta,\epsilon,\epsilon_m)$ is half-way along to defining the amoeba of $\CL_M$! Let us write
\be E_M(\eta,\epsilon,\epsilon_m) = E_M\Big(\sqrt{\frac{\bar x}{x}},|p|^2,|x|^2\Big) = \sum_{n=0}^{d'} b_n(\eta,\epsilon_m)\epsilon^n\,.\ee
It is not hard to see, by analyzing the limit $|p|\to 0$, that our ansatz about the amoeba of $\CL_M$ implies that $b_n$ is a nonconstant (non-monomial) polynomial. This is because any solution to $\CL_M=0$ must be a solution to $E_M=0$. In fact, if we also fix $\epsilon_m =1$ (this is the classical equivalent of setting $m=0$ in the index), then $\eta = x^{-1}$, and the same root of $a_0(x)$ in \eqref{LMpoly} must be root of $b_0(x) \equiv b_0(x^{-1},1)$. In particular, if $a_0(x)$ has a root on the unit circle, then $b_0(x^{-1},1)$ should contain a cyclotomic factor.

What does this finally imply for the index? The operator $\hat E_M$ defines a recursion relation for $\CI_M$ that only shifts $e$. We are therefore free to set $m=0$, obtaining
\be \hat E_M(\hat \eta,\hat\epsilon,1;q)\,\CI_M(0,e;q) = 0\,, \ee
or
\be \sum_{n=0}^{d'}  \hat\epsilon^n\, \hat b_n(\hat\eta,1;q)\,\CI_M(0,e;q) = \sum_{n=0}^{d'} q^{ne}\,\hat b_n(e^{\pd_e},1;q)\,\CI_M(0,e;q)=0\,, \label{Ebeq}\ee
for some $q$-deformed polynomials $\hat b_n(e^{\pd_e},1;q)$ that reduce to $b_n(\eta,1)=b_n(x^{-1},1)$ as $q\to 1$. Now, suppose that the powers of $q$ appearing in $\CI_M(m,e;q)$ at finite $(m,e)$ are bounded from below, as they are for theories $T_M$; and let us see what happens if the minimum power grows linearly at $m=0$, \ie\
\be \CI_M(0,e;q) \sim q^{\alpha e}\big(1+O(q)\big) \label{linind} \ee
for some $\alpha$. After plugging \eqref{linind} into \eqref{Ebeq}, we can try to take the limit $q\to 0$. Since the operator $e^{\pd_e}$ only modifies the ansatz \eqref{linind} by constant factors of $q$ at leading order, only the $n=0$ term of the sum \eqref{Ebeq} will contribute as $q\to 0$. This term must vanish by itself. That is:
\be \hat b_0(e^{\pd_e},1;q)\,\CI_M(0,e;q) = \hat b_0(e^{\pd_e},1;q) q^{\alpha e}\big(1+O(q)\big)  = \hat b_0(q^\alpha,1;q)q^{\alpha e}\big(1+O(q)\big) = 0 \label{leadb01} \ee
at leading order in $q$. One way to satisfy \eqref{leadb01} is to have
\be \boxed{\hat b_0(q^\alpha,1;q)=0}\,, \label{b0const} \ee
which is \emph{only} possible if $\hat b_0(x^{-1},1;q)$ is a nontrivial polynomial in $x$. But we know from our analysis that when the amoeba of $\CL_M$ has a tentacle along the negative $\Re\,P$ axis, $b_0(x^{-1},1)$ has a cyclotomic (or nontrivial) factor. The $q$-deformed version of this factor could then allow a solution to \eqref{b0const}. For the moment, this is the closest we can get to the mathematical claim/conjecture that tentacles of the amoeba allow linear growth.

We note that \eqref{leadb01} does not universally imply \eqref{b0const}. For this, a bit more must be known about the $O(q)$ corrections to the index --- for example, that they vanish (involve higher and higher powers of $q$) as $e\to \infty$. This is true in every calculated example of an index $\CI_M$, and there is likely a physical argument to justify it. When \eqref{b0const} is true, the leading polynomial $\hat b_0$ determines the growth rate $\alpha$.

If there is no tentacle of the amoeba lying on the negative $\Re\,P$ axis, then $a_0(x)$ and $b_0(x,1)$ are monomials, which generally implies that $\hat b_0(x^{-1},1;q)$ is a monomial. Then it is absolutely impossible to solve \eqref{b0const}, and (again, with some assumption about $O(q)$ corrections) the linear ansatz \eqref{linind} is excluded. The generic growth of the index is not linear but quadratic. Having $\CI_M(0,e;q)\sim q^{\alpha e^2}\big(1+O(q)\big)$ means that the operator $e^{\pd_e}$ can produce $q^e$ factors, and then all the terms in the sum \eqref{Ebeq} can mix at leading order. This allows much more general cancellations to occur.

\bibliographystyle{JHEP_TD}
\bibliography{toolbox}

\end{document}